\documentclass[usenatbib]{mnras}
\usepackage{graphicx}
\usepackage{hyperref}
\usepackage[usenames]{color}
\usepackage{rotating} 
\usepackage{amsmath} 
\usepackage{soul}
\usepackage{xargs}
\usepackage{color}

\newcommand{\lsim}{{\lower0.8ex\hbox{$\buildrel <\over\sim$}}}
\newcommand{\gsim}{{\lower0.8ex\hbox{$\buildrel >\over\sim$}}}
\def\simge{\mathrel{%
  \rlap{\raise 0.511ex \hbox{$>$}}{\lower 0.511ex \hbox{$\sim$}}}}
\def\simle{\mathrel{
  \rlap{\raise 0.511ex \hbox{$<$}}{\lower 0.511ex \hbox{$\sim$}}}}

\def\apj{ApJ}
\def\apjl{ApJL}
\def\aap{A\&A}

\def\mnras{MNRAS}

\def\aj{AJ}
\def\apjs{ApJ Supp}

\def\Chandra{\emph{Chandra}}

\def\XMM{{\it XMM-Newton}}

\newcommand{\Msun}{\ifmmode {M_{\odot}}\else${M_{\odot}}$\fi}
\newcommand{\Lsun}{\ifmmode {L_{\odot}}\else${L_{\odot}}$\fi}
\newcommand{\Rsun}{\ifmmode {R_{\odot}}\else${R_{\odot}}$\fi}

\definecolor{ForestGreen}{rgb}{0.148,0.414,0.180}

\begin{document}

\title[Constraining the Mass and Radius of Neutron Stars]{Constraining
  the Mass \& Radius of Neutron Stars in Globular Clusters}

\author[Steiner et al.]{A.~W. Steiner$^{1,2}$\thanks{awsteiner@utk.edu},
  C.~O. Heinke$^{3}$, S. Bogdanov$^{4}$, C. Li$^{3,5}$,
  W.~C.~G. Ho$^{6}$, \newauthor A. Bahramian$^{3,7}$, and S. Han$^{1}$ \\
  \vspace*{0.1cm} \\
  $^{1}$ Department of Physics and
  Astronomy, University of
  Tennessee, Knoxville, TN 37996, USA \\
  $^{2}$ Physics Division, Oak Ridge National Laboratory, Oak
  Ridge, TN 37831, USA \\
  $^{3}$ Dept. of Physics, University of Alberta, CCIS 4-183,
  Edmonton, AB T6G 2E1, Canada\\
  $^{4}$ Columbia Astrophysics Laboratory, Columbia University,
  550 West 120th Street, New York, NY 10027, USA\\
  $^{5}$ Key Laboratory for Particle Astrophysics, Institute of
  High Energy Physics, \\
  \hspace*{0.2cm} Chinese Academy of Sciences, 19B Yuquan Road,
  100049 Beijing, PR China\\
  $^{6}$ Mathematical Sciences, Physics \& Astronomy and
  STAG Research Centre, University of Southampton, \\
  \hspace*{0.2cm} Southampton, SO17 1BJ, UK\\
  $^{7}$ Department of Physics and Astronomy, Michigan State
  University, East Lansing, MI 48824, USA
}


\maketitle

\begin{abstract}
  We analyze observations of eight quiescent low-mass X-ray binaries
  in globular clusters and combine them to determine the neutron star
  mass-radius curve and the equation of state of dense matter. We
  determine the effect that several uncertainties may have on our
  results, including uncertainties in the distance, the atmosphere
  composition, the neutron star maximum mass, the neutron star mass
  distribution, the possible presence of a hotspot on the neutron star
  surface, and the prior choice for the equation of state of dense
  matter. We find that the radius of a 1.4 solar mass neutron star is
  most likely from 10 to 14 km and that tighter constraints are only
  possible with stronger assumptions about the nature of the neutron
  stars, the systematics of the observations, or the nature of dense
  matter. Strong phase transitions are preferred over other models and
  interpretations of the data with a Bayes factor of 8 or more, and in
  this case, the radius is likely smaller than 12 km. However, radii
  larger than 12 km are preferred if the neutron stars have uneven
  temperature distributions.
\end{abstract}


\section{Introduction}

With the exception of small corrections from rotation and magnetic
fields, the neutron star mass-radius relation is expected to be
universal~\citep{Lattimer01}. All neutron stars in the universe lie on
the same curve, and the determination of that curve informs the study
of many neutron star phenomena. The
Tolman-Oppenheimer-Volkov equations provide a one-to-one
correspondence between the mass-radius relation and the equation of
state (EOS) of dense matter, a quantity directly connected to quantum
chromodynamics, the theory of strong interactions. In particular, the
mass-radius curve is connected to the relationship between pressure
and energy density. Since neutron star temperatures are expected to be
much smaller than the Fermi momentum of the particles which comprise
the neutron star core, neutron stars probe the EOS at zero
temperature. There is strong interest in both determining the
mass-radius curve from observations and determining the equation of
state of cold and dense matter from nuclear experiments and theory.

A critical parameter is the nuclear saturation density, the central
density of matter inside atomic nuclei. Nuclear experiments and
nuclear theory are extremely successful at determining the nature of
matter at and below the saturation density. On the other hand,
experiments which probe matter more dense than the saturation density
are limited by the fact that they do so at the cost of introducing a
large temperature~\citep[see, e.g.,][]{Tsang12}. Theory is also
currently limited to lower densities: uncertainties are only
well-controlled where the Fermi momentum is small enough to employ
chiral effective theory or accurate phenomenological interactions
calibrated to nuclei~\citep[see recent reviews][]{Gandolfi15,Hebeler15nf}.

The central density of all neutron stars, however, is likely to be
larger than four times the nuclear saturation density
\citep{Steiner15un,Watts16mt}. Thus, unless there is a dramatic
advance which enables one to construct the cold EOS from experiments
which probe hot and dense matter, or there is an unexpected dramatic
improvement in nuclear theory-based calculations of dense matter,
observations of neutron star masses and radii are likely to be the
best probe of cold and dense matter.


\subsection{The quiescent LMXB method to constrain NS radii}

One promising method to constrain the NS radius is spectral fitting of
NSs in low-mass X-ray binaries during periods of little to no
accretion, called ``quiescence''. Low-mass X-ray binaries (LMXBs) are
binary systems containing a NS (or a black hole; we will not discuss
those systems here) and a low-mass star (less than 1-2 times the mass
of our Sun), where the orbit is tight enough that material can be
pulled from the low-mass star down onto the NS. In the majority of
LMXBs, the material falling from the companion star piles up in an
accretion disk around the NS, where it builds up for months to years
until the disk becomes dense and hot enough to become partly ionized,
leading to increased viscosity and flow of matter down onto the NS
\citep[e.g.][]{Lasota01}. During these ``outbursts'', the falling
material converts its potential energy into radiation, primarily in
X-rays where the LMXBs typically radiate many 1000s of times the
bolometric luminosity of our Sun. These outbursts are detectable
across the Galaxy with the use of (low-sensitivity) all-sky X-ray
monitors.

Between outbursts, as the disk builds up, the NS is much dimmer,
radiating $\sim$1-100\% of the Sun's bolometric luminosity
($10^{31}$--$10^{33}$ erg/s). During quiescence, the NS emits heat
deposited in the crust and core during outbursts as blackbody-like
radiation \citep{Brown98}. The (ionized) atmosphere quickly
stratifies, with the lightest accreted element on top; this topmost
layer (typically hydrogen) determines the details (spectrum, angular
dependence) of the emitted radiation field
\citep{Zavlin96,Rajagopal96}. In addition to this thermal,
blackbody-like radiation, quiescent NS LMXBs often also produce
nonthermal X-rays, which can typically be modelled (in the 0.5-10 keV
band) with a power-law of photon index 1-2. The nature of the
nonthermal X-rays are not clear, though they appear to generally be
produced by low-level accretion in quiescence
\citep{Campana98a,Chakrabarty14}. Continued accretion can also produce
thermal blackbody-like emission
\citep{Zampieri95,Deufel01,Rutledge02b,Cackett10}.

By measuring the X-ray flux and temperature of an object at a known
distance, the radius of the emitting object can be calculated.
Including the redshifting effects of general relativity means that the
quantity actually measured is the radius as seen at infinity,
$R_{\infty}$=$R(1+z)$=$R/\sqrt{1-2GM/(R c^2)}$, such that the outcome
is a constrained strip across the mass-radius plane. Since the final
spectrum has different dependences on the surface gravity in the
atmosphere and on the redshift, it is possible that future, larger
effective-area missions may tightly constrain both mass and radius.


Work in this direction has concentrated on quiescent LMXBs in globular
clusters for three reasons. First, quiescent LMXB radius measurements
depend on knowing the distance, since to first order we constrain the
quantity (R/D). Typical LMXBs in our galaxy have poorly known
distances (factors of 2 are not uncommon), while globular cluster
distances can be known as well as $\sim$6\% \citep[e.g.][]{Woodley12}. Dense
globular clusters produce close accreting binaries in dynamical
interactions \citep[e.g.][]{Benacquista13}, making them excellent targets to
search for quiescent LMXBs (identifiable through their unusual soft
spectra, \citealt{Rutledge02a}). Finally, unlike the majority of
quiescent LMXBs found outside globular clusters (through recent
outbursts), quiescent LMXBs identified in globular clusters tend to
have relatively simple spectra, dominated by thermal surface emission
with little or no power-law component \citep{Heinke03d}.

A number of quiescent LMXBs have been studied in some depth with the
\Chandra\ and/or XMM-Newton observatories, of which several provide
potentially useful constraints on mass and radius. These include
quiescent LMXBs in the globular clusters $\omega$ Cen
\citep{Rutledge02a,Webb07,Heinke14}, NGC 6397
\citep{Grindlay01b,Guillot11,Heinke14}, M28
\citep{Becker03,Servillat12}, M13 \citep{Gendre03b,Webb07,Catuneanu13},
NGC 6304 \citep{Guillot09a,Guillot13}, and M30
\citep{Lugger07,Guillot14}. Deep observations of the relatively bright
(few $10^{33}$ erg/s) quiescent LMXB X7 in 47 Tuc gave apparently
tight constraints and a large inferred radius \citep{Heinke06a}, but
suffered significantly from an instrumental systematic uncertainty,
pileup. Pileup occurs at relatively high count rates, when the energy
deposited from two photons is incorrectly recorded as coming from one
photon \citep{Davis01}; some combined photons are interpreted as
signals from cosmic rays, and rejected. Although it is possible to
model the effects of pileup, this modeling introduces systematic
errors that are difficult to quantify; \citet{Guillot13} pointed out
that due to the large fraction ($\sim$20\%) of piled-up photons in X7
in these data, these systematics were quite large. Recently, new
\Chandra\ observations of 47 Tuc X7 in a mode using a shorter frame
time to dramatically reduce pileup have provided more reliable radius
constraints \citep{Bogdanov16}.

\subsection{Previous combined constraints using quiescent LMXBs}
\label{s:1p2}

Since each quiescent LMXB provides a constraint covering a large range
of mass and radius, several groups have sought to combine constraints
from several systems to constrain the locus of mass and radius points
for neutron stars. These works have made different assumptions about
the composition of the NS atmospheres, and used different methods to
combine the constraints from different systems. \citet{Guillot13}
simultaneously fit five quiescent LMXB spectra, forcing all five to
have the same radius, arguing for a NS radius of $9.1^{+1.3}_{-1.5}$
km for the set. The choice of a single radius for all was motivated by
the finding that the equation of state curves for most plausible NS
structures are nearly vertical for much of the range of measured NS
masses \citep{Lattimer01}. This analysis used a Markov-chain Monte
Carlo method to sample the parameter space, and allowed uncertainties
in the distances to the globular clusters, variation in the extinction
column to each source, and for the possible presence of a hard
power-law spectral component (even if not confidently detected).

A key assumption in Guillot et al.'s work was that all quiescent LMXBs
have pure hydrogen atmospheres. This is reasonable if the donor stars
are hydrogen-rich, since the accreted elements will stratify in less
than a minute \citep{Alcock80,Hameury83}, unless accretion continues
at such a rate as to replace the photosphere in this time
\citep{Rutledge02b} Such an accretion rate would produce an
accretion-derived X-ray luminosity of order $10^{33}$ erg/s; since
accretion is fundamentally a variable process, the lack of detected
variability on timescales of years to decades in most globular cluster
quiescent LMXBs (including the objects used in these analyses) argues
that the thermal emission in these objects comes from stored heat in
the NS, and thus that the accretion rate is low enough that the
atmosphere is stratified \citep{Heinke06a,Walsh15,Bahramian15}.

However, the donor stars may not be hydrogen-rich. Between 28 and 44\%
of luminous globular cluster LMXBs have orbital periods less than 1
hour \citep{Bahramian14}, indicating that these NSs accrete from white
dwarfs, and suggesting that a similar fraction of quiescent LMXBs may
also have white dwarf donors. Detailed calculations of plausible
evolutionary scenarios show that the transferred mass will be devoid
of hydrogen, and dominated by helium, or carbon and oxygen, depending
on the composition of the donor star \citep{Nelemans10}. (Note that
binary evolution starting with a slightly evolved secondary star,
which would be likely to contain some hydrogen in the transferred
mass, cannot explain the observed period distribution of short-period
LMXBs in globular clusters; e.g. \citealt{vanderSluys05}.) It is
possible that accreted matter could spallate nuclei on impact,
releasing protons\citep{Bildsten93}, though spallation might require
infalling protons \citep{intZand05}, and there is no evidence of
spallation-produced H in observed thermonuclear bursts in extremely
short-period LMXBs \citep{Cumming03,Galloway08}. Finally, diffusive
nuclear burning may also consume the hydrogen at the photosphere
\citep{Chang04}.

Since helium and carbon atmospheres shift the emitted X-ray spectra to
slightly higher energies with respect to hydrogen atmospheres, the
inferred radii (if fit with hydrogen atmospheres) would be smaller
than the true radii \citep{Rajagopal96,Ho09}. For carbon atmospheres,
the radius difference (a factor of $\sim$two) is large enough that
identification should be immediate (none have yet been seen), but the
effects of helium atmospheres are more subtle. Several works have
considered fits of specific quiescent LMXBs to either H or He
atmospheres \citep{Servillat12,Catuneanu13,Heinke14}, finding
increases of the radius from the fits of $\sim$20-50\%.

Other works have combined the individual results for quiescent LMXBs
in a Bayesian formalism. \cite{Steiner10} combined mass-radius
constraints from three thermonuclear burst
systems~\citep{Ozel09,Guver10a,Guver10b} with results from three
quiescent LMXBs~\citep{Heinke06,Webb07}. Steiner et al. used a
Bayesian framework to combine the results, introducing a parametrized
equation of state (incorporating causality constraints, the minimum NS
maximum mass, and the low-density nuclear equation of state), and
preferred radii (for 1.4 $\mathrm{M}_{\odot}$ NSs) between 11 and 12
km. \citet{Lattimer14} used a similar Bayesian framework (accounting
for the discovery of a 2 $\mathrm{M}_{\odot}$ NS, which increased the
minimal NS maximum mass) to reconsider the results of
\citet{Guillot13}. Lattimer \& Steiner estimated an average bias in
the inferred radius of a He-covered NS applied to H atmosphere models
of 33\%. They took estimates of the probability distribution functions
of \citet{Guillot13}, altered them by allowing He atmospheres and
alternative choices of interstellar extinction (both alterations
performed by analytically estimating the impact), and combined them,
finding a preferred radius range of 10.45$-$12.66 km.

\citet{Ozel16} also combined thermonuclear burst constraints with
quiescent LMXB constraints, using an alternative Bayesian formalism
that maps the measured masses and radii to the pressures at three
fiducial densities. Ozel et al. combined individual constraints on six
quiescent LMXBs, assuming a hydrogen atmosphere for all systems except
the NGC 6397 system, for which they assumed a helium atmosphere,
finding a preferred radius between 10.1 and 11.1 km. Finally,
\citet{Bogdanov16} obtained new constraints on the mass and radius of
the quiescent LMXBs X7 and X5 in 47 Tuc, and combined these
constraints with the results of \citet{Ozel16} to find a preferred
radius in the range 9.9-11.2 km.

The goal of this work is to carefully analyze the quiescent LMXB
sample, allowing each quiescent LMXB to have either a hydrogen or
helium atmosphere, except when independent evidence indicates a
particular composition. We combine these measurements in a Bayesian
framework, producing results for different assumptions about the NS
equation of state, and different assumptions about the quiescent LMXB
population.
  

\section{Data on individual quiescent LMXBs}\label{sec:data}

We briefly describe the data used to study each quiescent LMXB, and
our best estimates of the distance to each globular cluster. We
summarize the key information about these sources in Table
\ref{tab:sources}.

\begin{table}
  \begin{tabular}{cccc}
    Cluster & Distance & $N_H$ & References \\
     & (kpc) &  (cm$^{-2}$) & \\
    \hline
    47 Tuc (X7,X5) & 4.53$\pm0.06$ & $3.5\times10^{20}$ & 1,2,3,4 \\
    $\omega$ Cen & 5.22$\pm0.17$ & $1.3\times10^{21}$  & 5,6,4 \\
    NGC 6397 & 2.47$\pm0.07$  & $1.2\times10^{21}$ & 3,4,5 \\
    M13  & 7.8$\pm0.36$ & $1.1\times10^{20}$ & 7,8,4 \\
   NGC 6304 & 6.22$\pm0.26$ & $4.6\times10^{21}$  & 7,9,10\\
   M30  &  8.78$\pm0.33$ & $2.6\times10^{20}$ & 7,11,4\\
  \end{tabular}
  \caption{List of clusters containing the quiescent LMXBs we analyze,
    with the best measurements of their distances and $N_H$ columns,
    and references pertaining to those. References: 1)
    \citet{Bogdanov16}, 2) \citet{Bergbusch09}, 3) \citet{Hansen13},
    4) \citet{Harris96}, 2010 update, 5) \citet{Heinke14}, 6)
    \citet{Bono08}, 7) \citet{Recio-Blanco05}, 8) \citet{Sandquist10},
    9) \citet{Piotto02}, 10) \citet{Guillot13}, 11) \citet{Dotter10} }
  \label{tab:sources}
\end{table}

\subsection{47 Tuc: X7 and X5}


47 Tuc is an excellent target due to its relatively short distance,
relatively bright quiescent LMXBs, and relatively low $N_H$ (the
amount of interstellar gas absorbing soft X-rays). The core is
sufficiently dense with X-ray sources that only \Chandra\ observations
can fully resolve the sources. The two relatively bright quiescent
LMXBs in 47 Tuc, X5 and X7, are bright enough that deep
\Chandra\ observations in 2000 and 2002 suffered substantial
($\sim$20\%) pileup. We therefore obtained 181 kiloseconds of new
\Chandra\ observations in 2014/2015, in an observing mode designed to
reduce pileup to $\sim$1\%. \citet{Bogdanov16} described these
observations (plus five shorter observations from 2000 and 2002 in
this mode), and the analysis to derive M-R constraints for X7 and X5;
we use those M-R constraints in our combined analysis.

X5 shows eclipses with an 8.67 hour period, and dips, stronger at
lower energies, due to varying local extinction \citep{Heinke03a}.
These indicate that X5 is viewed nearly edge-on, so that the companion
star eclipses X-rays from the NS, and that material from the dynamic
accretion disk is often present along the line of sight.

\citet{Bogdanov16} also compiled a list of literature determinations
of the distance to 47 Tuc, by a variety of methods, and compiles a
weighted mean of the distance determinations. To limit the effect of
flaws in any one method, they conducted ``jackknife'' tests, where
they removed all measurements taken with one method, to assess the
systematic errors, finding a final distance of 4.53$^{+0.08}_{-0.04}$
kpc for 47 Tuc.\footnote{\citet{Bogdanov16} also addressed the
  discrepancy with the dynamical distance estimate of
  \citet{Watkins15}, showing why Watkins' value is discrepant, and
  noting that when Watkins et al. use a larger dataset, the
  discrepancy disappears.} We use Bogdanov et al's distance estimate,
but assume symmetric errors, taking 0.06 kpc as the 1-sigma error
uncertainty.

\subsection{$\omega$ Cen}\label{sec:wCen}

$\omega$ Cen is a relatively nearby and low-density globular cluster,
for which either \Chandra\ or \XMM\ can resolve the known quiescent
LMXB. We use \Chandra\ data on $\omega$ Cen from 2000 (69 kiloseconds)
and 2012 (225 kiloseconds), along with the \XMM\ data from 2001 (40
kiloseconds), reduced as described by \citet{Heinke14}.

It is often easier to establish the relative distance scale of several
globular clusters than an absolute scale. We use the well-established
relative difference in distances between $\omega$ Cen and 47 Tuc
($\omega$ Cen is 16($\pm3$)\% farther than 47 Tuc, \citealt{Bono08})
to calculate the distance to $\omega$ Cen as 5.22$\pm0.17$ kpc. This
is consistent with the horizontal branch distance of 5.2 kpc of
\citet{Harris96}\footnote{http://physwww.physics.mcmaster.ca/$\sim$harris/mwgc.dat},
the dynamical distance estimate of 5.19$^{+0.07}_{-0.08}$ kpc of
\citet{Watkins15}, and in the middle of the other distance estimates
discussed by \citet{Heinke14}.

\subsection{NGC 6397}

NGC 6397 is the second nearest globular cluster, with a very dense
core. We use \Chandra\ observations taken in 2000 (49 kiloseconds),
2002 (55 kiloseconds), and 2007 (240 kiloseconds), reduced as
described by \citet{Heinke14} (similar to \citet{Guillot11}).

As for $\omega$ Cen, we use the relative distance measurements to NGC 6397
and 47 Tuc (that NGC 6397 is at 54.5$\pm$2.5\% of 47 Tuc's distance,
\citet{Hansen13}) to calculate NGC 6397's distance at 2.47$\pm$0.07
kpc. This is in agreement with the dynamical distance estimate of
2.39$^{+0.13}_{-0.11}$ kpc of \citet{Watkins15}, and with most other
recent distance estimates discussed in \citet{Heinke14}.

\subsection{M13}

M13 is a relatively low-density cluster with very low extinction. It
was the subject of extensive ROSAT observations (46 kiloseconds) in
1992 with the PSPC camera, which accurately measured the absorption of
the low-energy spectrum (though ROSAT has much poorer spectral
resolution in general). We use this ROSAT PSPC spectrum, 2002
\XMM\ observations (34.6 kiloseconds), and 2006 \Chandra\ observations
(54.7 kiloseconds), reduced as described by \citet{Catuneanu13}
(similar to \citet{Webb07}).

The distance to M13 has been extensively discussed by
\citet{Sandquist10}, who estimate 7.65$\pm$0.36 kpc (reconciling
distance estimates using the tip of the red giant branch with
horizontal branch estimates), while the homogeneous relative distance
estimates of \citet{Recio-Blanco05} (which perfectly matches our
distance estimate to 47 Tuc) gives a distance of 7.8$\pm$0.1 kpc. We
use Recio-Blanco's distance, but conservatively use the larger
distance uncertainty, 7.8$\pm$0.36 kpc.


\subsection{NGC 6304}

The quiescent LMXB in NGC 6304 is located in the core of its cluster
near other sources \citep{Guillot09b}, so we use only the deep
\Chandra\ observation of 2010 (98.7 kiloseconds), extracted following
\citet{Guillot13}, using CIAO 4.7 and CALDB 4.6.9.

The distance to NGC 6304 is uncertain due to its relatively high
extinction, and thus high uncertainty on its extinction. Using the
distance modulus of \citet{Recio-Blanco05} and the extinction estimate
of \citet{Piotto02} (following \citealt{Guillot13}) gives a distance
of 6.22$\pm$0.26 kpc.



\subsection{M30}

The quiescent LMXB lies in the extremely dense core of the distant,
but low-extinction, globular cluster M30, and can only be resolved by
\Chandra. We use the 2001 \Chandra\ observation (49 kiloseconds), and
extract the data following \citet{Lugger07} \citep[also][]{Guillot14},
using CIAO 4.7 and CALDB 4.6.9.

\citet{Recio-Blanco05} measure a distance of 8.78$\pm$0.33 kpc; since
their distance estimate aligns with ours for 47 Tuc, we adopt this. We
note that it is also in agreement with other estimates by, e.g., 
\citet{Dotter10} (8.8 kpc).
 
\subsection{M28}

The quiescent LMXB in M28 (source 26 of \citealt{Becker03}) lies in
the core of this relatively nearby, dense cluster. We use three
\Chandra\ observations from 2002 (42 kiloseconds) and two from 2008
(199.6 kiloseconds), reduced as described in \citet{Servillat12},
using CIAO 4.7 and CALDB 4.6.9.

We use the estimate of the distance to M28 of 5.5$\pm$0.3 kpc, derived
from the brightness of the horizontal branch \citep{Testa01}
calibrated by \citet[][2010 revision]{Harris96}.

\section{X-ray Spectral Fitting}\label{sec:result}

For the spectral fitting, we used the XSPEC software \citep{Arnaud96}.
We experimented with merging spectra taken with the same detector
close in time, or fitting them simultaneously; with grouping the data
into bins of $>$20, $>$50, or other numbers of counts; with analyses
using $\chi^2$ or C-statistics. In general, such changes do not make
substantial differences to the final results.

For all spectra, our spectral fits included a neutron star atmosphere,
either NSATMOS (hydrogen, \citealt{Heinke06a}) or NSX (helium,
\citealt{Ho09}). We included $N_H$ through the TBABS model (with the
extinction free to vary), using element abundances from
\citet{Wilms00} and photoelectric cross-sections from
\citet{Verner96}. We include a power-law in the spectral fitting,
although it is not required for any quiescent LMXB, with the photon
index fixed to 1.5, as typical for power-law components in quiescent
LMXBs \citep{Campana98a,Cackett10,Chakrabarty14}. We add systematic
errors, of magnitude 3\%, to all spectra, accounting for instrumental
calibration uncertainties, following \citet{Guillot13,Bogdanov16}. Our
spectral fits were performed assuming the nominal best-fit distances,
with distance uncertainties convolved with the probability density
functions (from the spectral fits) during the Bayesian MCMC
calculation (see below).

\citet{Bogdanov16} found that the inclusion of pileup in spectral
modelling for \Chandra\ observations of X7 in 47 Tuc made a
significant difference to the final radius contours, even at pileup
fractions as low as 1\%. For this reason, we include pileup in all
\Chandra\ spectral fits; this is particularly relevant for the NGC
6397 spectral fits, since previous fits \citep{Guillot13,Heinke14} did
not include pileup for this source. The M28 quiescent LMXB has the
highest fraction of piled-up events, with about 5\% of photons piled
up \citep{Servillat12}.

A crucial uncertainty is the chemical composition of
the atmosphere of the quiescent LMXB. In the Bayesian MCMC analysis
below, we analyze the spectra of each quiescent LMXB with both
hydrogen and helium atmospheres with two exceptions. The first
exception is the quiescent LMXB $\omega$ Cen which has a firm
detection of hydrogen in its spectrum \citep{Haggard04}. The second is
X5, which has a long orbital period \citep{Heinke03a} suggesting a
hydrogen-rich donor (see section~\ref{s:1p2}).

\subsection{Systematic Uncertainties}
\label{s:su}

There are remaining systematic uncertainties which we have not
robustly controlled. Perhaps the largest uncertainty is the possible
effect of hot spots upon the inferred radius. Many neutron stars show
pulsations, implying the presence of hotter regions on their surface;
examples include young pulsars \citep{deLuca05}, old millisecond
pulsars \citep{Bogdanov13}, young neutron stars without pulsar
activity \citep{Gotthelf10}, and accreting neutron stars
\citep{Patruno09b}. Hot spots may be produced by the accretion of
material onto a magnetic pole, collision of relativistic electrons and
positrons with the pole during pulsar activity, or preferential
leakage of heat from the core along paths with particular magnetic
field orientations \citep{Potekhin01}. \citet{Elshamouty16} explored
the effect of hot spots on quiescent LMXB spectra, focusing on the
cases of X7 and X5 in 47 Tuc. Elshamouty et al. used deep
\Chandra\ observations at high time resolution to search for
pulsations from these two sources. Considering a range of possible
magnetic latitudes and inclinations, they derived limits on the hot
spot temperature, and then derived limits on the bias that might be
inferred by fitting a single-temperature neutron star atmosphere to a
neutron star that has hot spots. Unfortunately, existing data poses
only weak limits, such that the spectroscopically inferred radius
could be biased downwards up to 28\% smaller than the true radius. We
will consider biases up to this level in some of our analyses.
 
Another concern is variability in the absorbing column. As mentioned
above, X5 in 47 Tuc is a nearly edge-on system that shows varying
photoelectric absorption. \citet{Bogdanov16} shows that if periods of
enhanced absorption (signified by dips in the count rate) are removed,
the inferred radius of X5 grows significantly. When time periods
including different absorption values are combined and fitted with a
single absorber, the spectrum appears intrinsically more curved (and
thus at a hotter temperature). However, we cannot be sure that we have
removed all the periods of enhanced photoelectric absorption; short
periods of enhanced absorption would not supply enough counts to
enable unambiguous determination of a dip. We thus suspect that the
spectral fits to X5 may be biased downwards by varying photoelectric
absorption. This problem might affect other quiescent LMXBs as well.
Though no other quiescent LMXB among our sample shows eclipses,
absorption dips have been seen in LMXBs that are not seen edge-on (cf.
\citealt{Galloway16} and \citealt{MataSanchez16}), although to date
these dips have only been seen during outburst.

\section{Bayesian Inference for Neutron Star Masses and Radii}
\label{s:4}

Ideally, one directly connects the probability distributions of the
quantities of interest to the observables. In this case, that would
imply directly connecting neutron star masses and radii to the flux of
photons at every energy. In practice, however, this would require an
integral over several energy bins for each neutron star data set.
In this work, we convert the X-ray spectrum for each source into a
probability distribution for the neutron star with radius
$R$, mass $M$, at distance $D$ with atmosphere composition $X$
\begin{equation}
  {\cal D}(R,M,D,X) = \exp \left[ - \chi^2(R,M,D,X)/2 \right]
  \label{eq:red}
\end{equation}
In essence, this exponential of a sum is equivalent to
a product over Gaussians for each energy bin for the X-ray spectrum
which is fit.

Note that this assumes that the uncertainties between energy bins in
the spectrum and uncertainties between objects are both uncorrelated.
This may not be the case, as systematic uncertainties in a particular
observation may affect a number of high-energy X-ray bins. Also,
\XMM~observations may have uncorrected systematics which are different
than those from \Chandra. Such systematics are beyond the scope of the
current work, except for the 3\% uncertainty which we added to all of
the spectra as described above (see, e.g., \citealt{Lee11,Xu14} for
potential future directions).

From the eight probability distributions for each neutron star, we can
directly apply a generalization of the approach first described in
\citet{Ozel09rt}, \citet{Steiner10}, and \citet{Ozel10}. Following
\citet{Steiner10} we employ Bayesian inference and we assume uniform
prior distributions unless otherwise specified below. Upper and lower
limits for uniform distribution for the model parameters, $\{p_i\}$,
are chosen to be extreme enough that they do not affect the final
results. The posterior probability distribution for quantity $Q$ is
then
\begin{eqnarray}
  P_Q(q) &\propto& \int \left\{ \prod_{i=1}^{M}
  {\cal D}_i[R(M_i,\{p_j\}),M_i,D_i,X_i] \right\} \nonumber \\
  && \times \delta[q-Q(p_1,\ldots,p_N,M_1,\ldots,M_N,
    \nonumber \\ && D_1,\ldots,D_N,X_1,\ldots,X_N)]
  \nonumber \\
  && \times d p_1~\ldots dp_N~dM_1 \ldots dM_N
  \nonumber \\ && \times
  dD_1 \ldots dD_N~dX_1 \ldots dX_N \, ,
  \label{eq:marg}
\end{eqnarray}
where $N$ is the number of neutron stars being considered. 
Several examples of relevant posteriors are reviewed briefly in
\citet{Lattimer14co}.

We assume the composition, $X_i$ of the atmosphere of each neutron
star (except for the neutron star in $\omega$ Cen and for X5 in 47
Tuc) is a discrete binary variable: either H or He. As a prior
distribution we assume a 2/3 probability of H and a 1/3 probability of
He, following the observed ratio of H-rich to He-rich donors in bright
LMXBs in globular clusters (\citealt{Bahramian14}, and references
therein). This method is superior to that employed in
\citet{Lattimer14} because it allows us to quantitatively predict the
posterior probability that any particular neutron star has a H or He
atmosphere. On the other hand, it has been argued by some authors that
helium atmospheres are expected to be unlikely~\citep{Guillot14}, so
we also try models where the prior probability for hydrogen is 90\% or
100\%. We remove the neutron star in X5 from our baseline data set
because of the varying absorption described in section~\ref{s:su}. We
assume the maximum mass must be larger than $2~\mathrm{M}_{\odot}$ as
implied by the recent observations of high mass neutron
stars~\citep{Demorest10,Antoniadis13}. We allow neutron stars to have
masses as low as $0.8~\mathrm{M}_{\odot}$ (which may be conservative),
but increasing this number has little effect on our results.

As discussed in \citet{Bogdanov16} (see also \citealt{Beznogov15}),
there are strong reasons to believe that these neutron stars are not
close to 2~$\mathrm{M}_{\odot}$ in mass, since that would likely
produce extremely rapid cooling (by processes such as direct Urca,
e.g. \citealt{Yakovlev03}) so that we would not observe strong thermal
radiation from their surfaces. This method is potentially powerful to
constrain the masses of neutron stars, but at this time the mass at
which rapid cooling turns on is not well-constrained. The neutron star
in SAX J1808.4-3658 is extremely cold, indicating that rapid cooling
processes are active \citep[e.g.][]{Heinke09a}. This neutron star
seems not to have a particularly large mass (the 2$\sigma$ upper limit
on the mass is $\sim$1.7~$\mathrm{M}_{\odot}$), though current
constraints are not very strong \citep{Wang13}.

\subsection{Distance and Hotspot Correction}

In this work, we choose to marginalize over the distance as a nuisance
variable instead of producing a separate fit for each distance. So
long as all of the probability distributions of interest (all of the
quantities $P_Q$ in eq.~\ref{eq:marg} above) are independent of
distance, we can perform the distance integrations first. This means
that we cannot generate any posterior distributions for the distance,
but we expect other methods to provide superior distance measurements
anyway.

In units where $c=1$, there is a bijection between $(R,M)$ and
$(R_{\infty},z)$ when $R>2 G M$ (the composition will be
considered later and is thus suppressed for now):
\begin{equation}
  R_{\infty}(R,M) = R \left(1 - \frac{2 G M}{R}\right)^{-1/2}
\end{equation}
and
\begin{equation}
  z(R,M)= \left(1 - \frac{2 G M}{R}\right)^{-1/2} -1 \, .
\end{equation}
The opposite transformation is given by
\begin{equation}
  M(R_{\infty},z) = \frac{R_{\infty}}{G} \left[ \frac{z(2+z)}{2(1+z)^3} \right]
\end{equation}
and
\begin{equation}
  R(R_{\infty},z) = \frac{R_{\infty}}{(1+z)} \, .
\end{equation}

To first order (adequate for small fractional uncertainties),  
distance scales with $R_{\infty}$. Thus to correct
for a variation in distance, $ D=D_{\mathrm{new}} \pm \delta D$ kpc
given an original probability distribution ${\cal D}_{\mathrm{old}}(R,M)$
based on a fixed distance, $D_{\mathrm{old}}$, we compute
\begin{eqnarray}
  {\cal D}_{\mathrm{new}}(\hat{R},\hat{M}) &=&
  \int_0^{\infty}~d\hat{R}_{\infty}
  \left[ \frac{D_{\mathrm{old}}}{ R_{\infty}(\hat{R},\hat{M}) \delta
      D \sqrt{2 \pi} } \right] \nonumber \\
&& \exp \left\{ - \frac{\left[\hat{R}_{\infty}-R_{\infty}(\hat{R},\hat{M})
      D_{\mathrm{new}}/D_{\mathrm{old}} \right]^2}
       {2 \left[ R_{\infty}(\hat{R},\hat{M}) \delta
           D/D_{\mathrm{old}} \right]^2}\right\} \nonumber \\
       && \times {\cal D}_{\mathrm{old}} \{
       R[\hat{R}_{\infty},z(\hat{R},\hat{M})],
       \nonumber \\ &&
       M[\hat{R}_{\infty},z(\hat{R},\hat{M})] \} \, .
       \label{eq:rinf_trans}
\end{eqnarray}
To see that this expression makes sense, take the limit
$\delta D \rightarrow 0$, then the exponential and associated
normalization is equal to a delta function, 
\begin{eqnarray}
  &\left[ \frac{D_{\mathrm{old}}}{ R_{\infty}(\hat{R},\hat{M}) \delta
      D \sqrt{2 \pi} } \right] \times & \nonumber \\
  & \exp \left\{ - \frac{\left[\hat{R}_{\infty}-R_{\infty}(\hat{R},\hat{M})
      D_{\mathrm{new}}/D_{\mathrm{old}} \right]^2}
  {2 \left[ R_{\infty}(\hat{R},\hat{M}) \delta
      D/D_{\mathrm{old}} \right]^2}\right\}
  \xrightarrow[\delta D \rightarrow 0]{} & \nonumber \\
  & \delta \left[\hat{R}_{\infty}-R_{\infty}(\hat{R},\hat{M})
    D_{\mathrm{new}}/D_{\mathrm{old}} \right] & \, , 
\end{eqnarray}
so in this limit
\begin{eqnarray}
  {\cal D}_{\mathrm{new}}(\hat{R},\hat{M}) &=&
  {\cal D}_{\mathrm{old}} \{ R[R_{\infty}(\hat{R},\hat{M})
    D_{\mathrm{new}}/D_{\mathrm{old}},z(\hat{R},\hat{M})],
  \nonumber \\ &&    
  M[R_{\infty}(\hat{R},\hat{M})
    D_{\mathrm{new}}/D_{\mathrm{old}},z(\hat{R},\hat{M})] \} \, .
  \label{eq:after_delta}
\end{eqnarray}
If $D_{\mathrm{new}} = D_{\mathrm{old}}$, the term on the right-hand
side of Eq.~\ref{eq:after_delta} just becomes
${\cal D}_{\mathrm{old}}(\hat{R},\hat{M})$ (i.e. the probability distribution
is unchanged). Finally, Eq.~\ref{eq:rinf_trans} also ensures that the
new relative uncertainty in $R_{\infty}$ is equal to $\delta
D/D_{\mathrm{new}}$, since
\begin{equation}
  \frac{\delta \hat{R}_{\infty}}{\hat{R}_{\infty}} \rightarrow
  \frac{R_{\infty}(\hat{R},\hat{M}) \delta D/D_{\mathrm{old}} }
       {R_{\infty}(\hat{R},\hat{M}) D_{\mathrm{new}}/D_{\mathrm{old}} }
       = \frac{\delta D}{D_{\mathrm{new}}}
\end{equation}

\begin{figure}
  \includegraphics[width=1.6in]{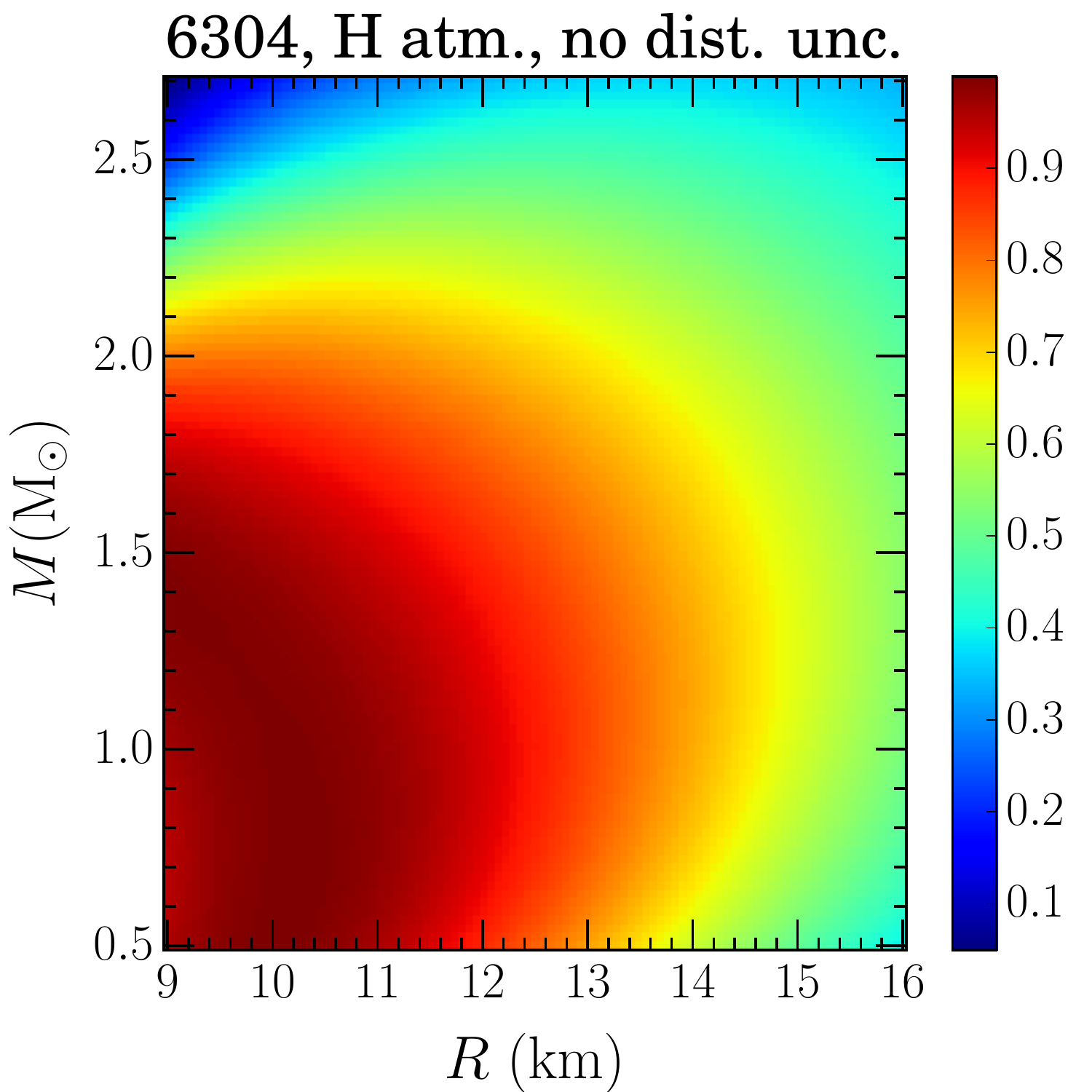}
  \includegraphics[width=1.6in]{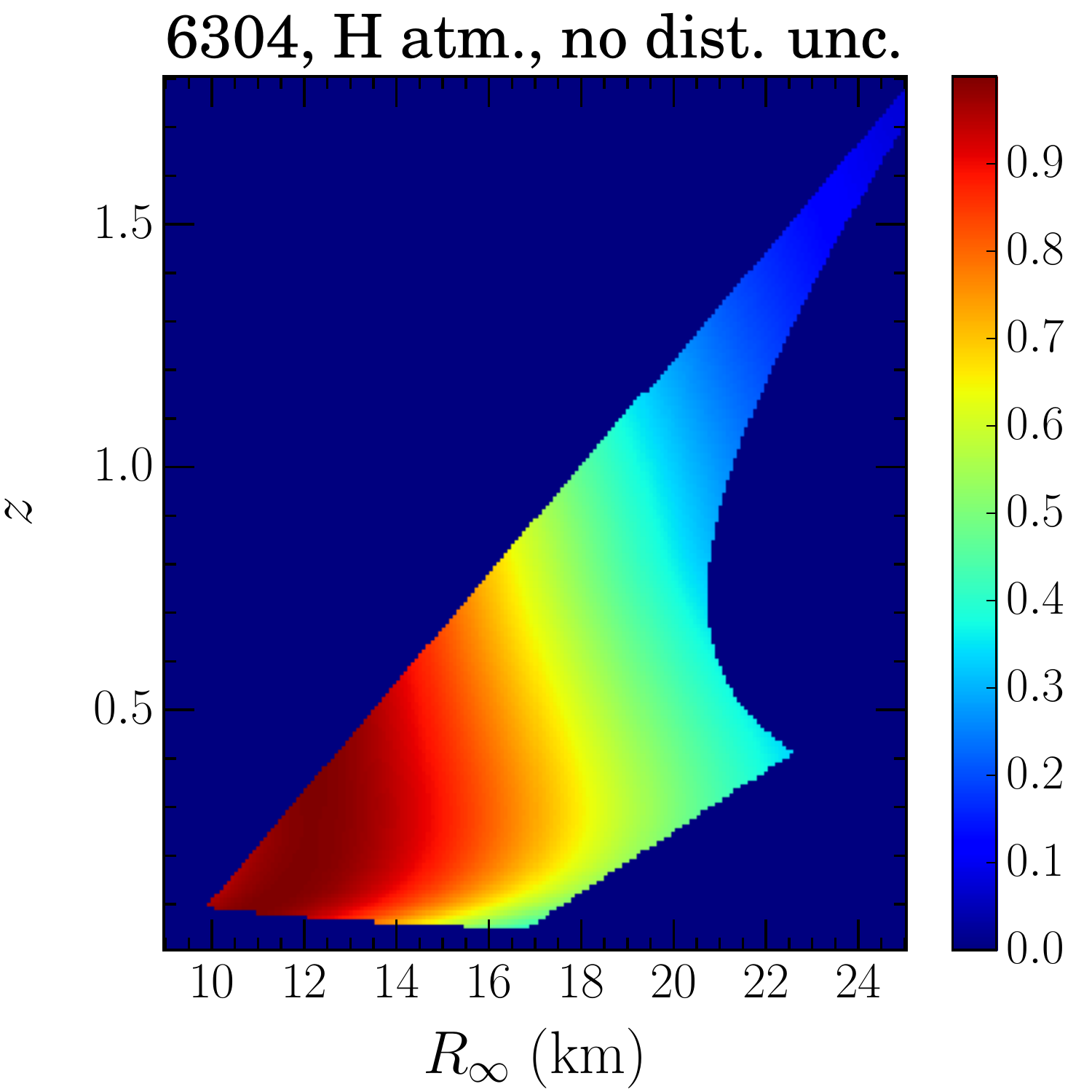}
  \includegraphics[width=1.6in]{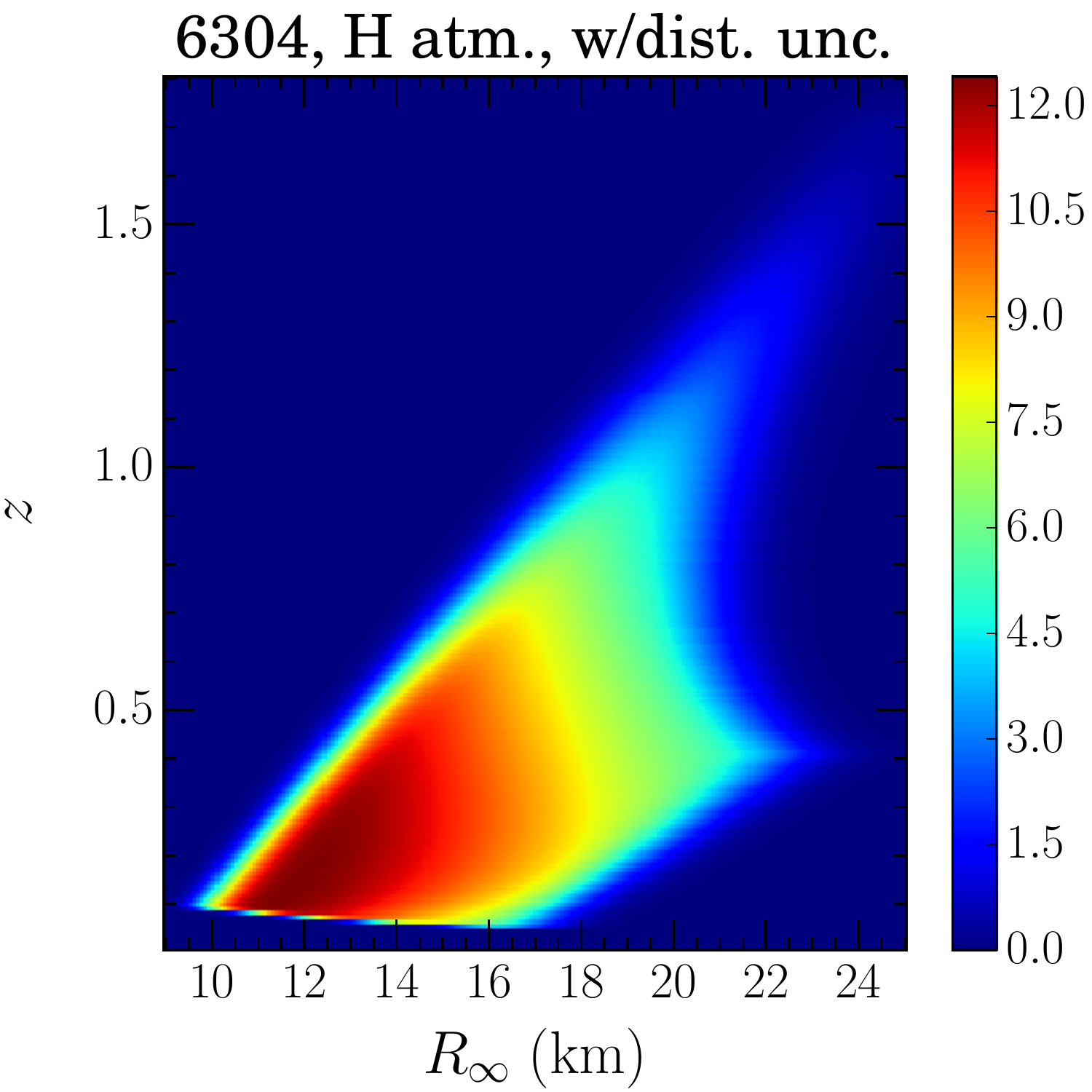}
  \includegraphics[width=1.6in]{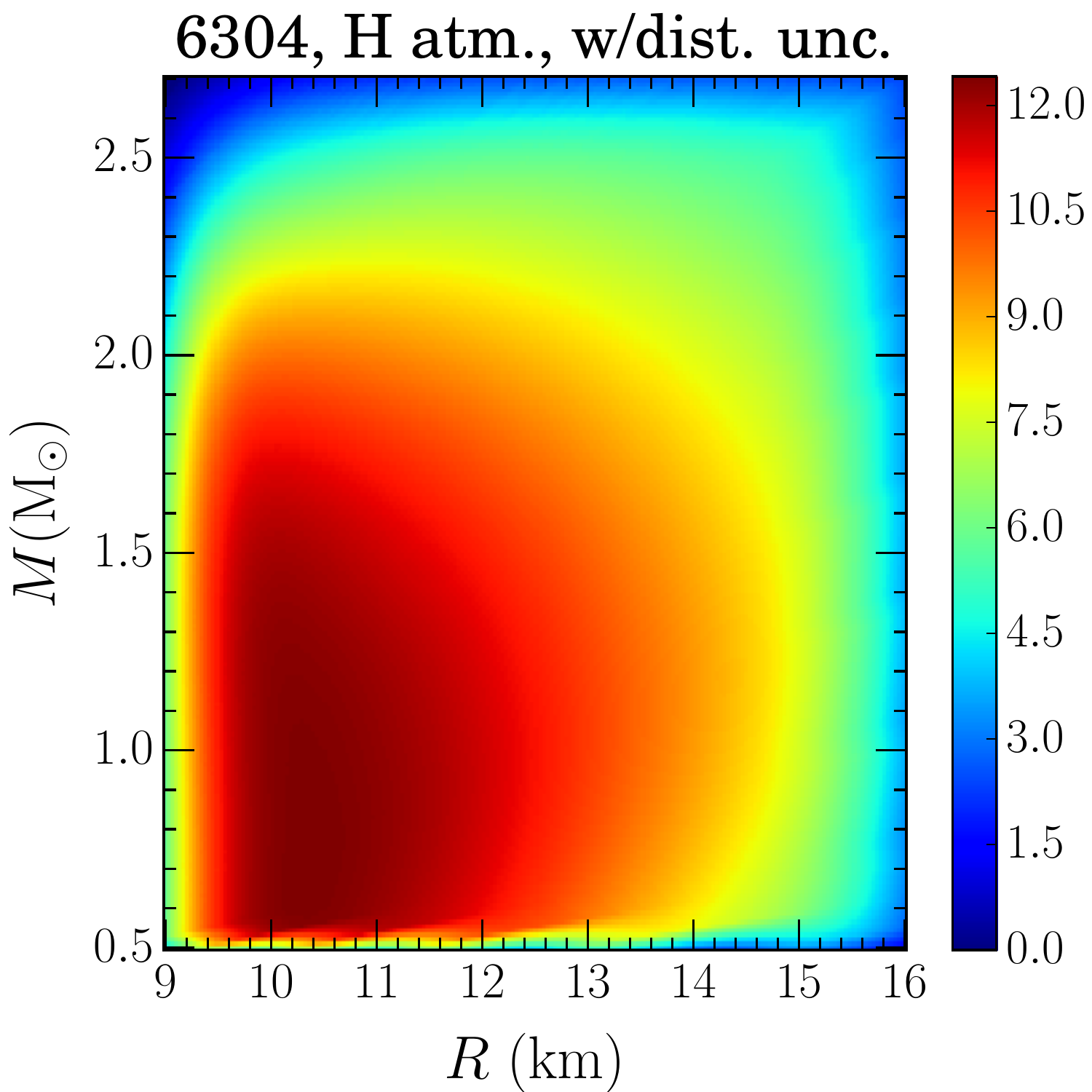}
  \caption{This figure is a demonstration of the incorporation of the
    distance uncertainty as in Eq.~\ref{eq:rinf_trans}. The original
    mass-radius probability distribution (arbitrary normalization) for
    a neutron star in NGC 6304 including a 3\% systematic uncertainty
    and X-ray absorption assuming abundances from \citet{Wilms00} is
    in the upper-left panel. The probability distribution is converted
    to $(R_{\infty},z)$ space (upper-right panel), and then shown
    again in the lower-left panel after an integration over a Gaussian
    distance uncertainty. The lower-right panel shows the result after
    the conversion back to $(M,R)$ space. The probability distribution
    in the lower right panel smoothly falls off at the edges, but this
    is will not affect our final results.}
  \label{fig:demo_Rinfz}
\end{figure}

Fig.~\ref{fig:demo_Rinfz} gives a demonstration of the method. The
distribution is transformed to $(R_{\infty},z)$ space (upper-right
panel) where the distance uncertainty is applied (lower-left panel),
and then transformed back to $(R,M)$ space (lower-right panel).

Alternatively, we can rescale $R$ at constant $z$ and obtain the same
distribution ${\cal D}_{\mathrm{new}}$. To show this, create a new dummy
integration variable, ${\cal R} \equiv
\hat{R}_{\infty}/[1+z(\hat{R},\hat{M})]$. The value
$z(\hat{R},\hat{M})$ is independent of the
integration variable, ${\cal R}$, so $ d{\cal R}/d \hat{R}_{\infty} =
1/[1+z(\hat{R},\hat{M})]$. Since
$R_{\infty}(\hat{R},\hat{M})/[1+z(\hat{R},\hat{M})]=\hat{R}$,
the integral on the right-hand side of Eq.~\ref{eq:rinf_trans}
can be rewritten to give
\begin{eqnarray}
  {\cal D}_{\mathrm{new}}(\hat{R},\hat{M}) &=&
  \int_0^{\infty}~d{\cal{R}}
  \left[ \frac{1}{ \hat{R} \delta D/D_{\mathrm{old}}
      \sqrt{2 \pi} } \right] \nonumber \\
  && \exp \left\{ - \frac{\left[{\cal{R}} -\hat{R}
      D_{\mathrm{new}}/D_{\mathrm{old}} \right]^2}
       {2 \left[ \hat{R} \delta
           D/D_{\mathrm{old}} \right]^2}\right\}
       \nonumber \\ &&
       {\cal D}_{\mathrm{old}} \{ {\cal{R}} ,
       M[{\cal{R}},z(\hat{R},\hat{M})] \} \, .
\label{eq:r_trans}
\end{eqnarray}

Note that, in the figures below, the rescaled results from either
Eqs.~\ref{eq:rinf_trans} or \ref{eq:r_trans} shows that the probability
distribution vanishes at the extremes in $M$ and $R$. This is because
we must assume the input probability distributions from the X-ray
fits are step functions (e.g. they drop immediately to zero
probability for $R<9$ km) and these step functions are softened by
the additional distance uncertainty.

\begin{figure}
  \includegraphics[width=1.6in]{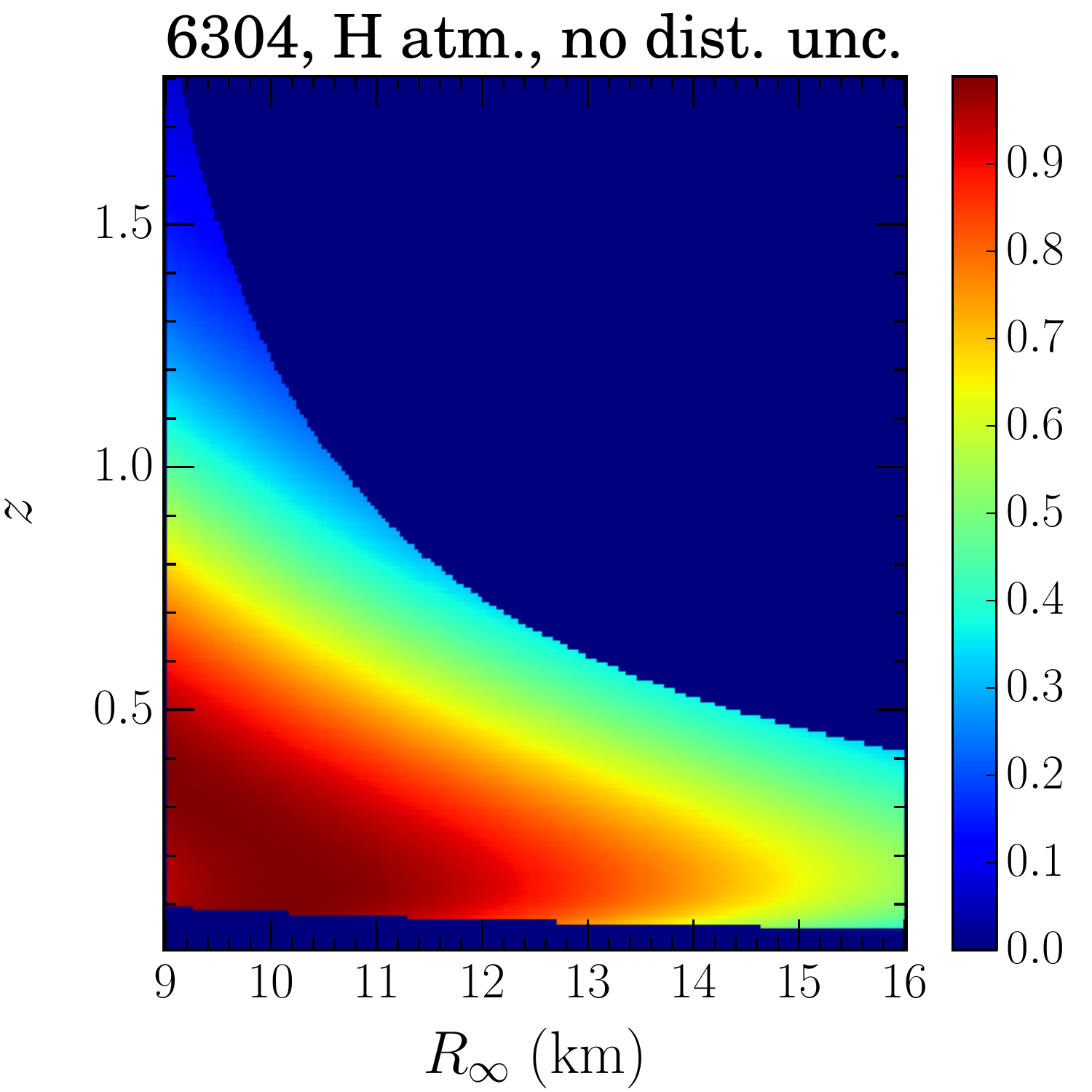}
  \includegraphics[width=1.6in]{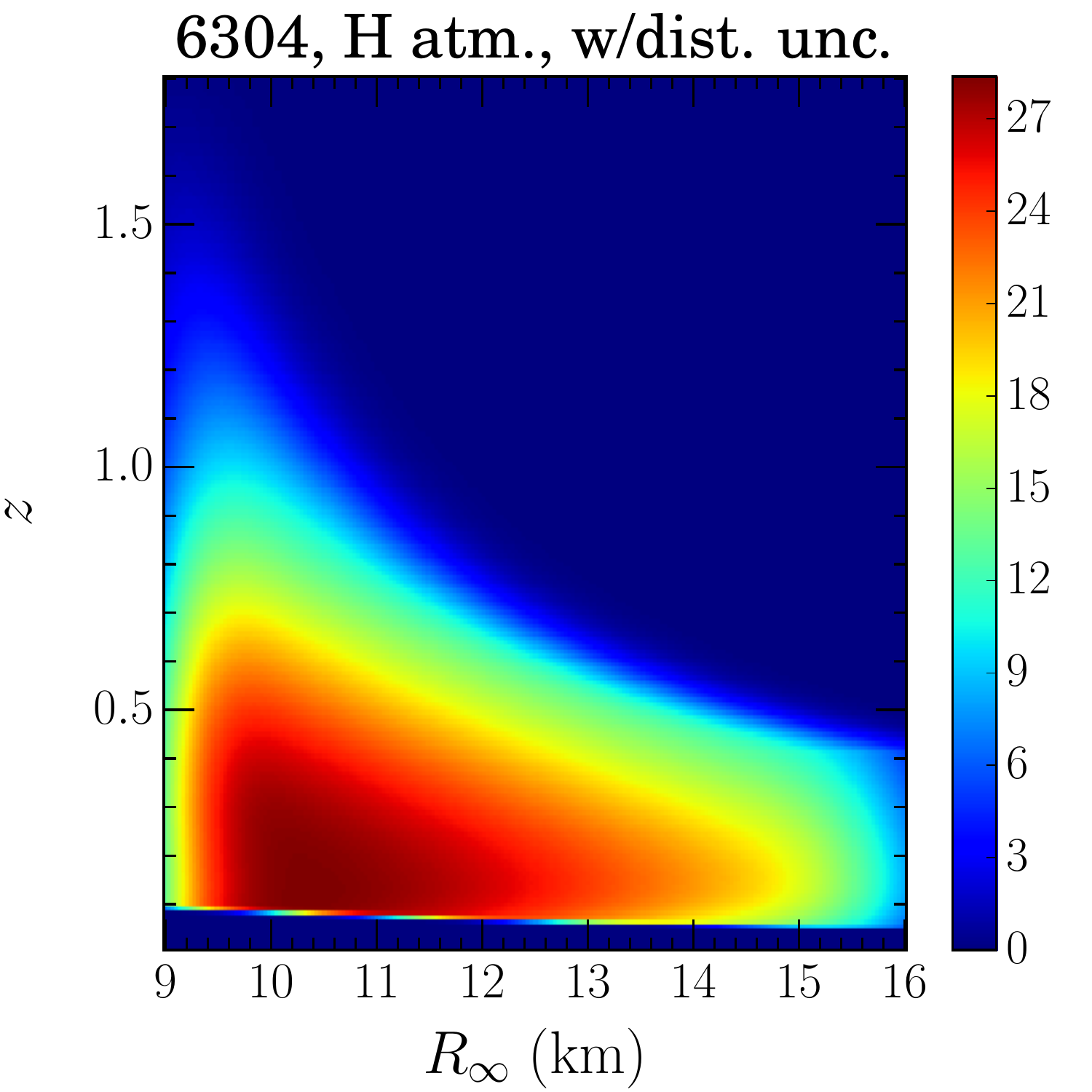}
  \includegraphics[width=1.6in]{demo4}
  \includegraphics[width=1.6in]{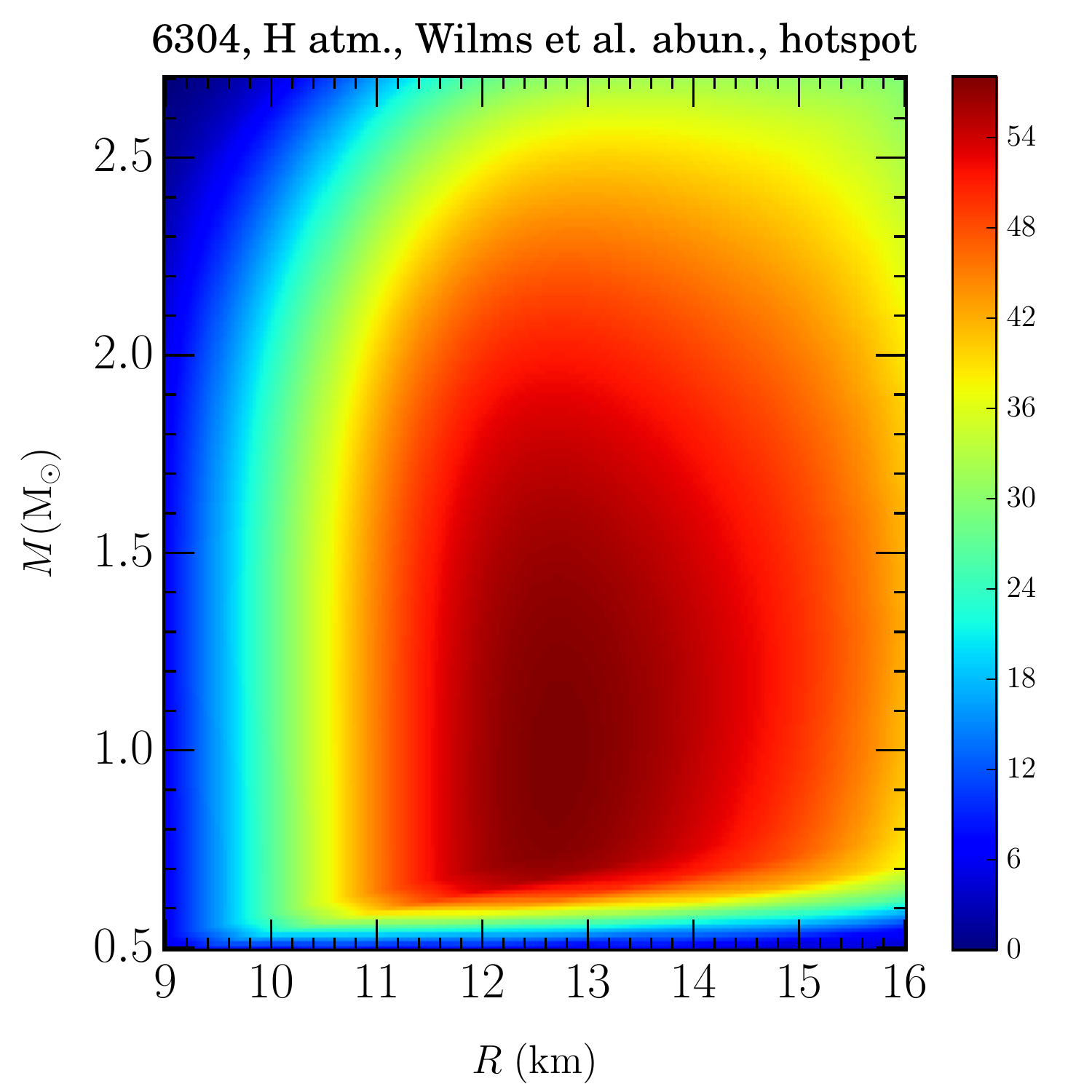}
  \caption{Upper left and upper right panels: A demonstration of the
    distance uncertainty having been applied in $(R,z)$ space as
    implied by Eq.~\ref{eq:r_trans}. The final result, in the lower
    left panel, is the same as
    that in the lower-right panel of Fig.~\ref{fig:demo_Rinfz}. Lower
    right panel: A demonstration of the effect of a hotspot, to be
    compared with the lower right panel in Fig.~\ref{fig:demo_Rinfz}.}
  \label{fig:demo_Rz}
\end{figure}

\begin{figure}
  \includegraphics[width=1.6in]{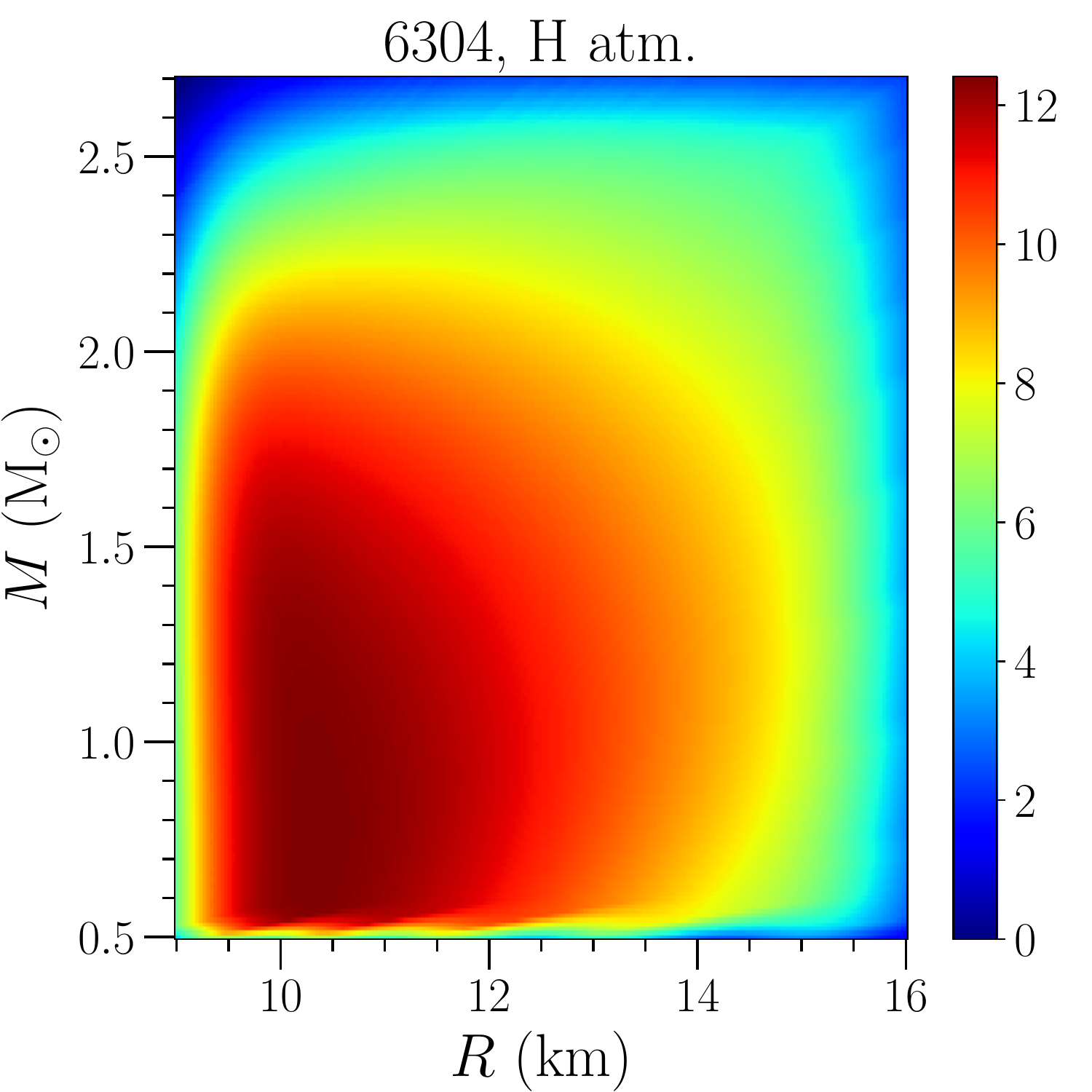}
  \includegraphics[width=1.6in]{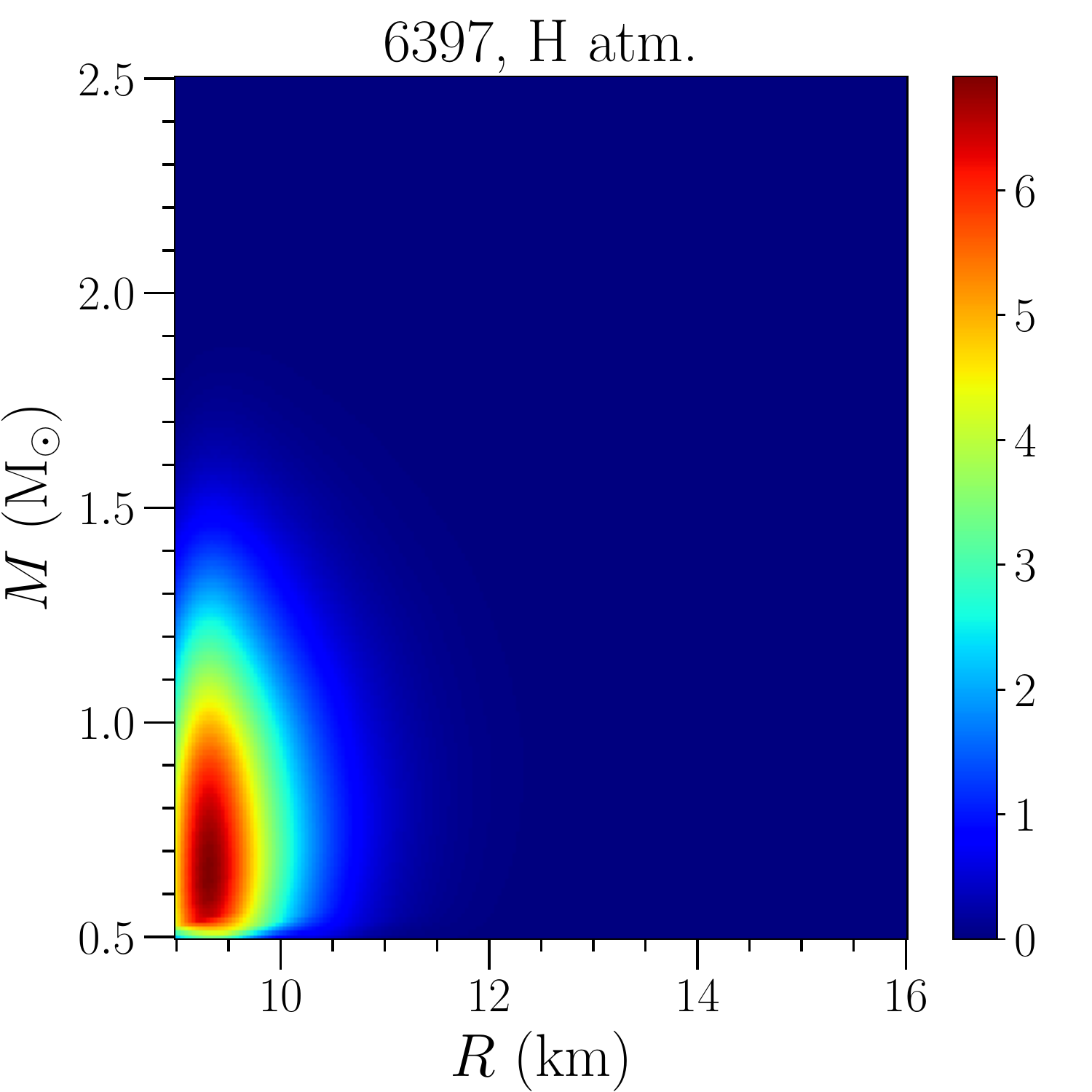}
  \includegraphics[width=1.6in]{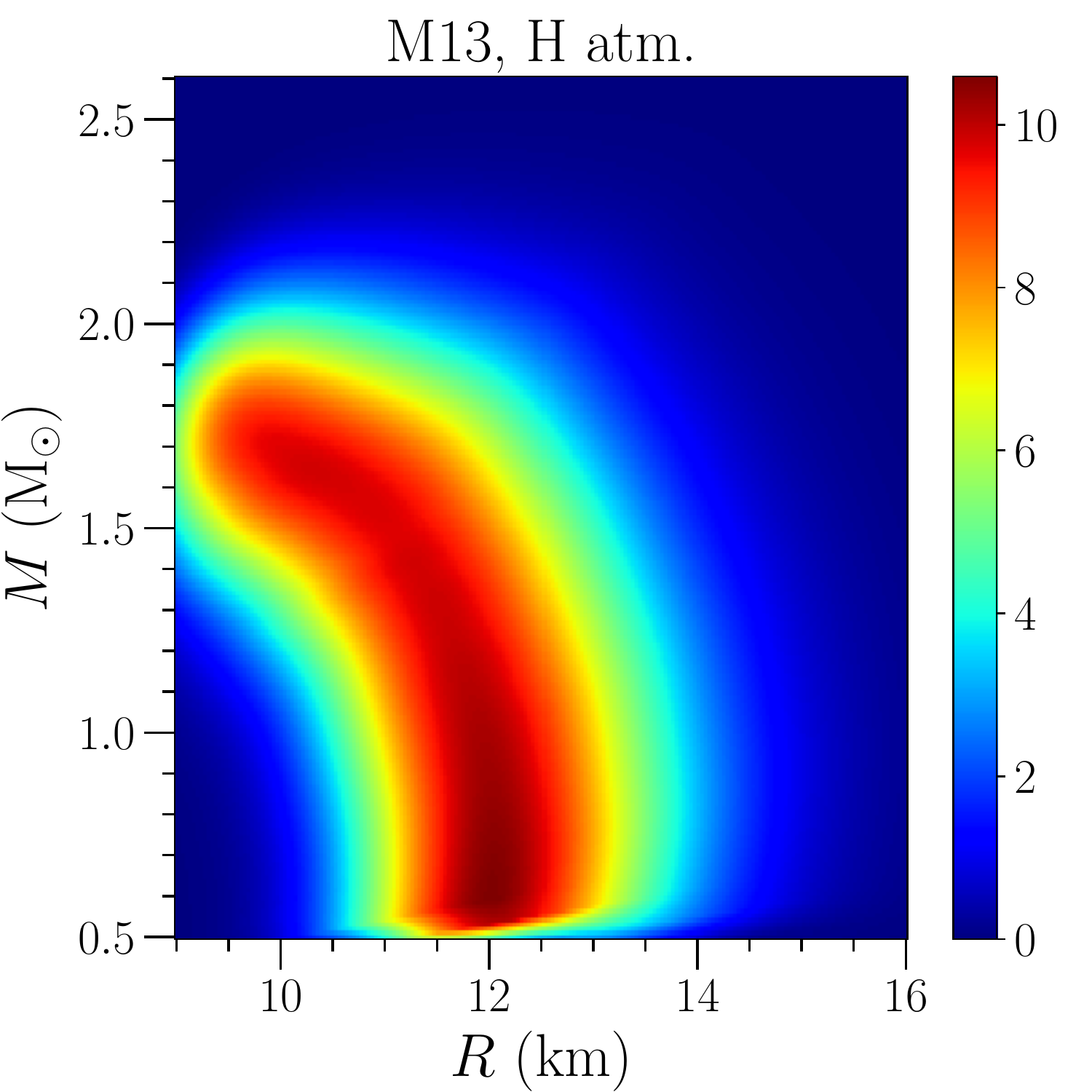}
  \includegraphics[width=1.6in]{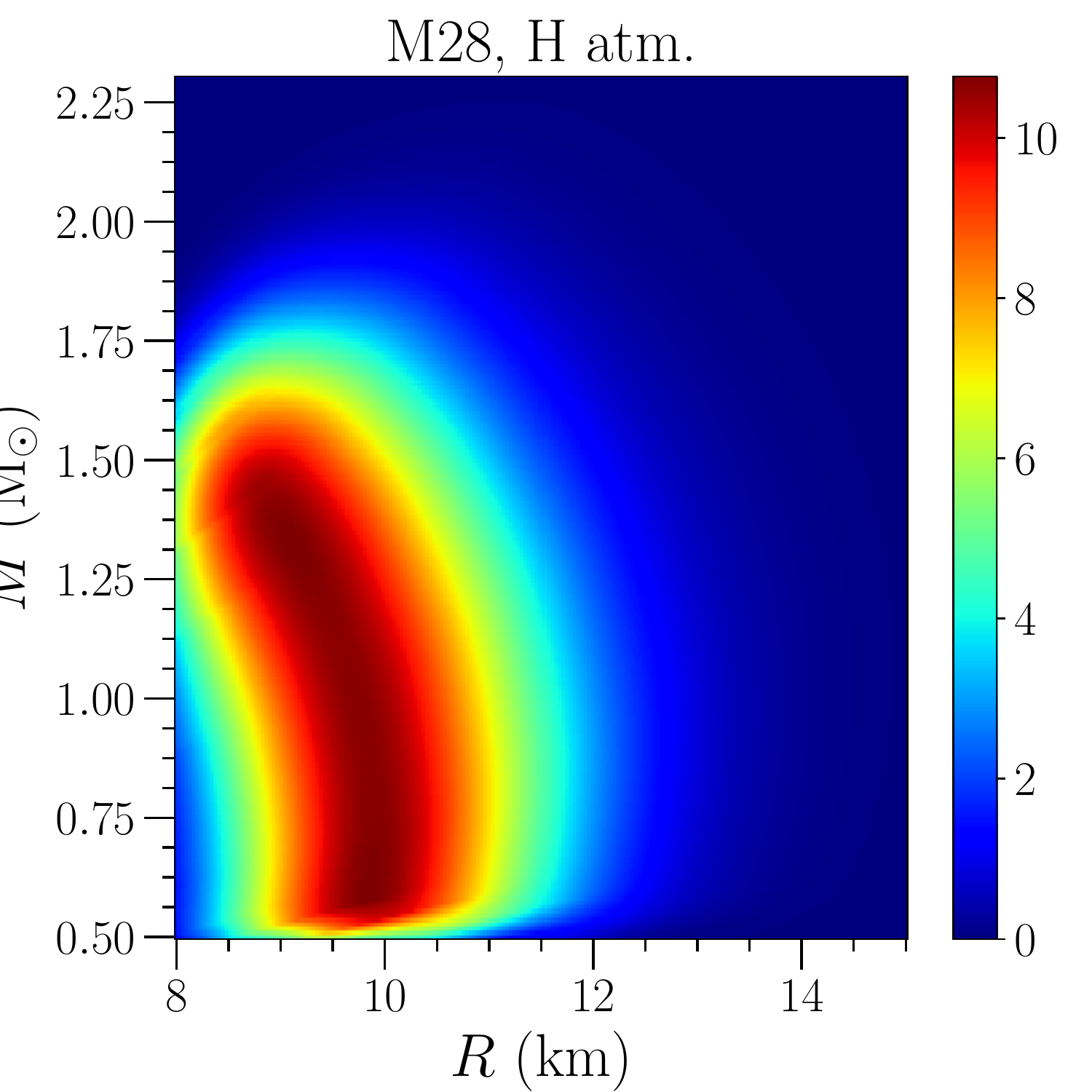}
  \includegraphics[width=1.6in]{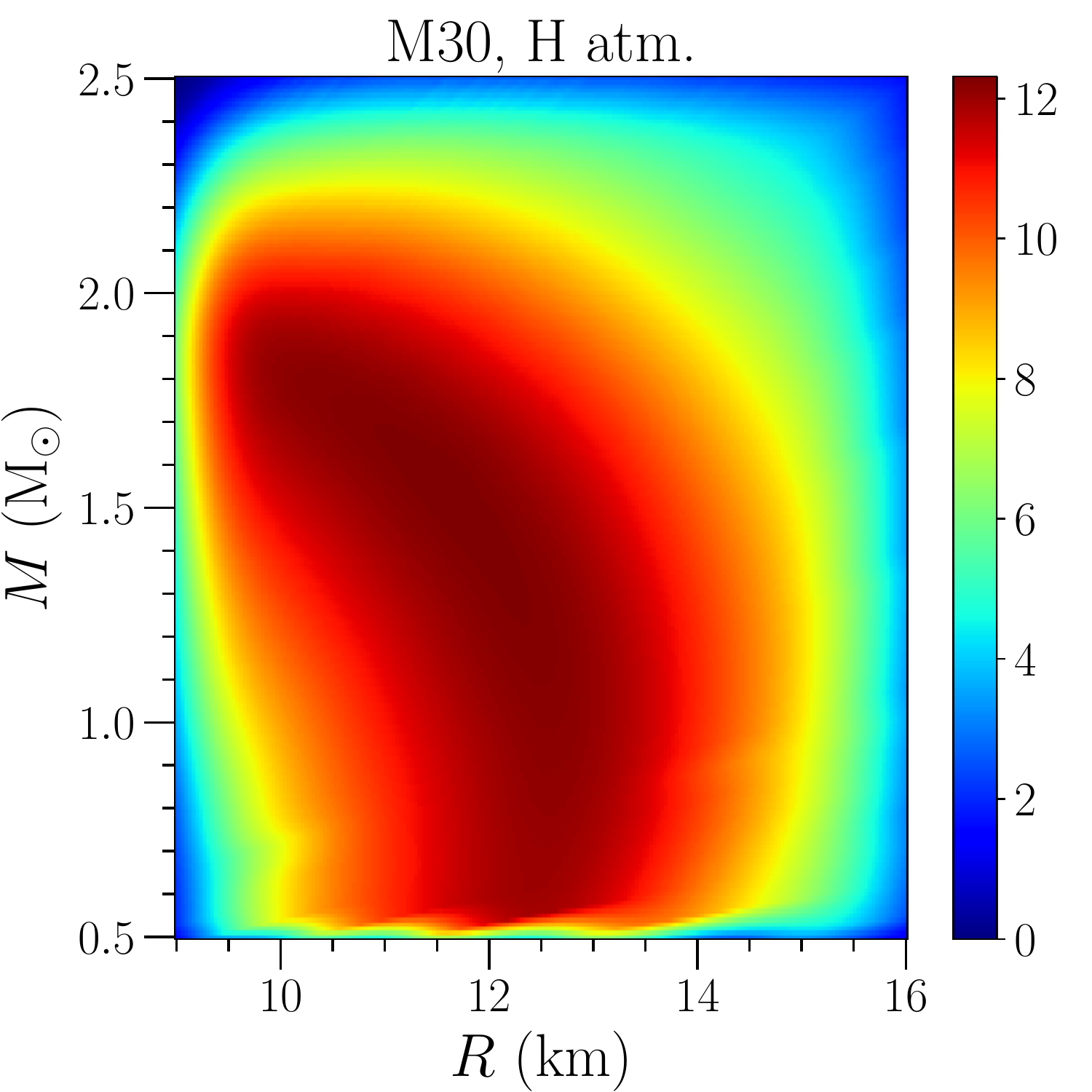}
  \includegraphics[width=1.6in]{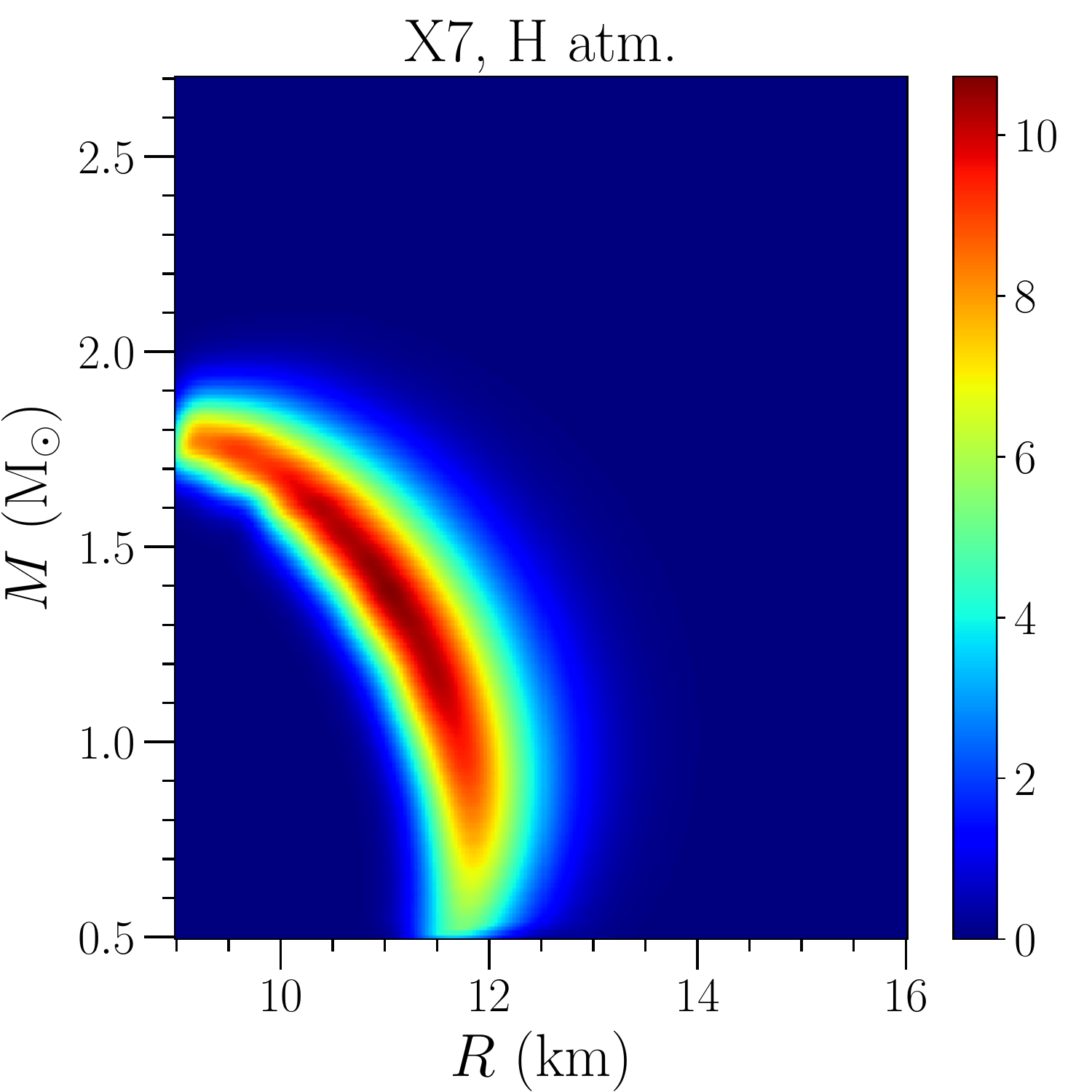}
  \includegraphics[width=1.6in]{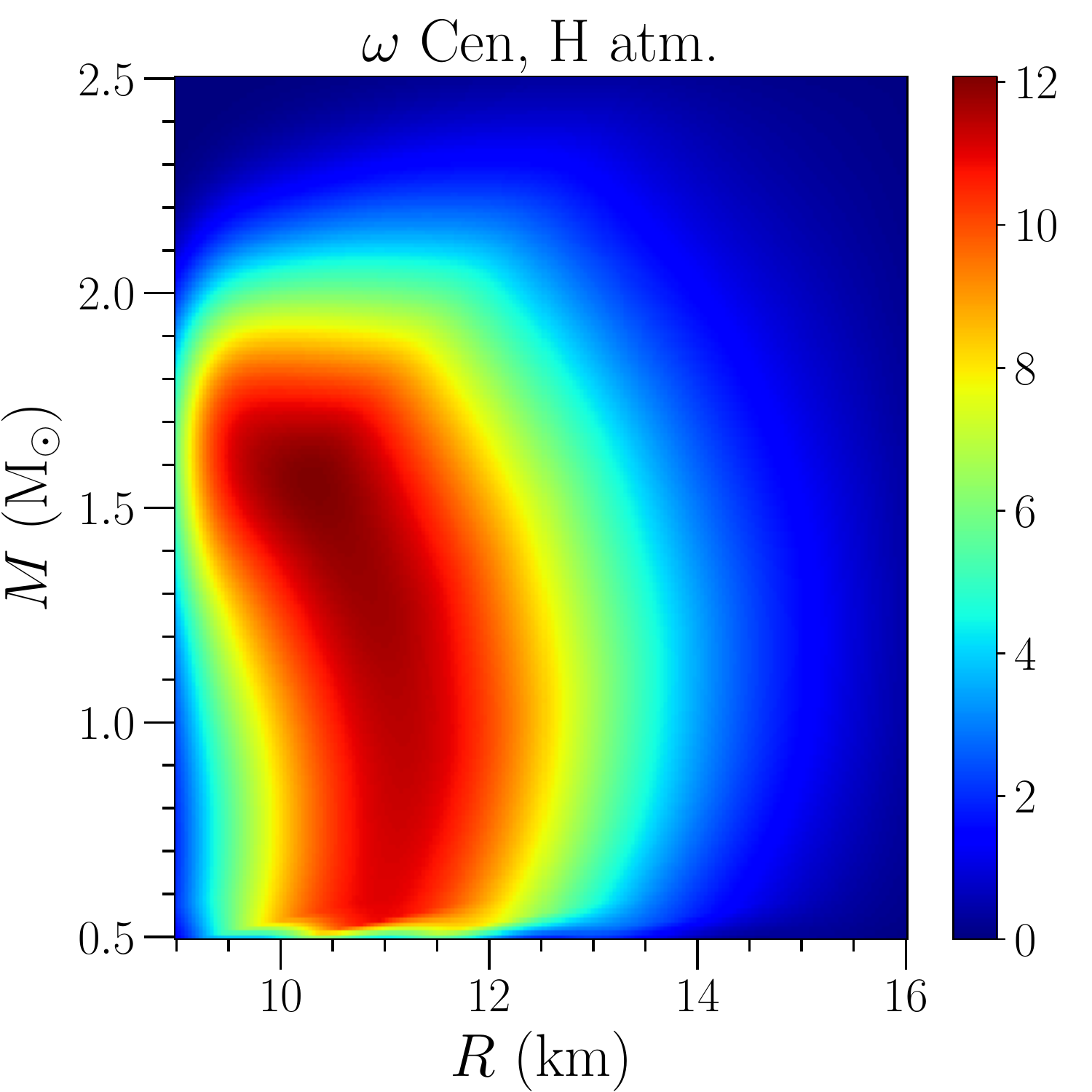}
  \includegraphics[width=1.6in]{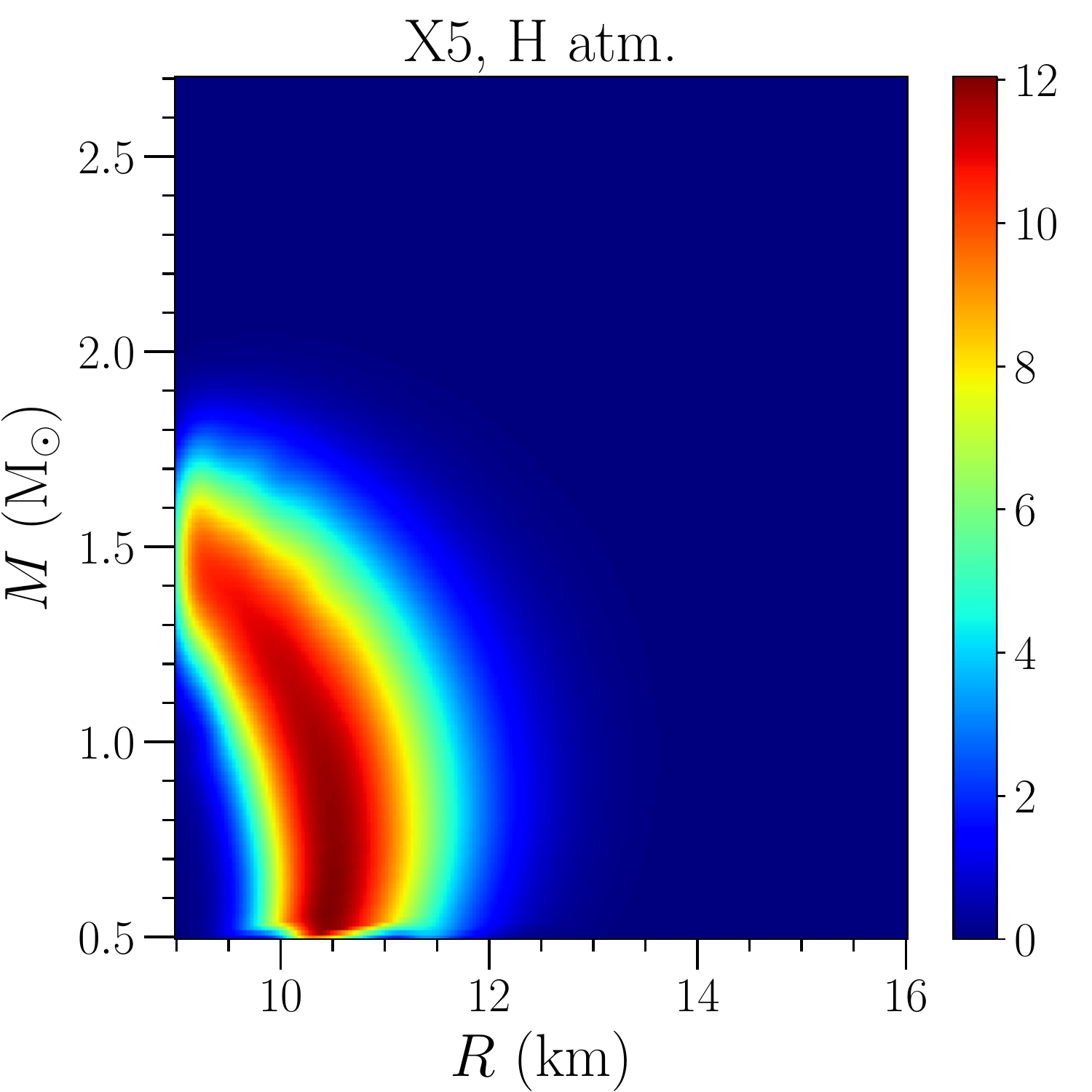}
  \caption{The H atmosphere part of our baseline data set plus the
    neutron star X5 in 47 Tuc. The \citet{Wilms00} abundances are used
    to correct for X-ray absorption in all cases, the normalization is
    arbitrary, and a distance uncertainty has been added following the
    prescription described in section \ref{s:4}. }
  \label{fig:data1}
\end{figure}

\begin{figure}
  \includegraphics[width=1.6in]{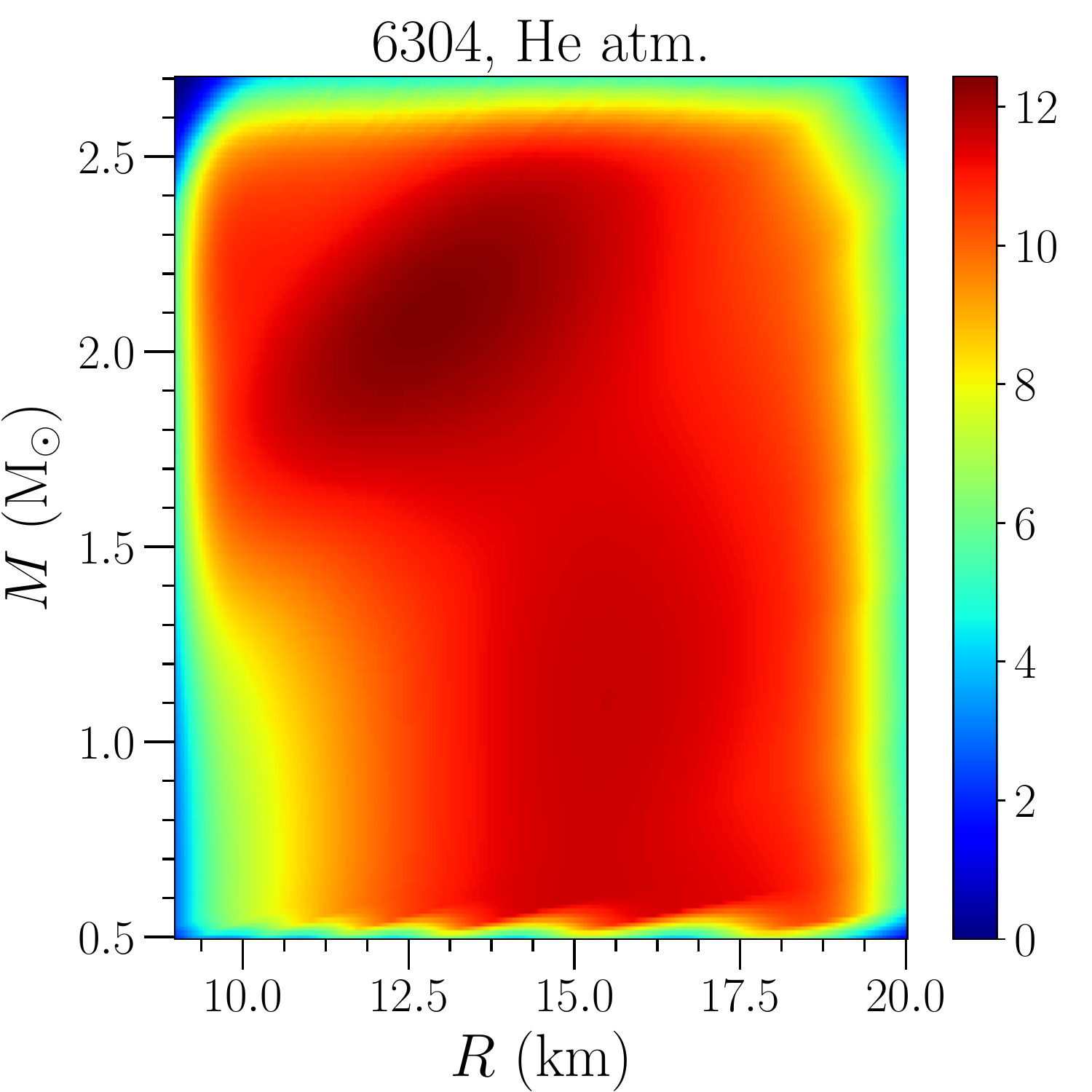}
  \includegraphics[width=1.6in]{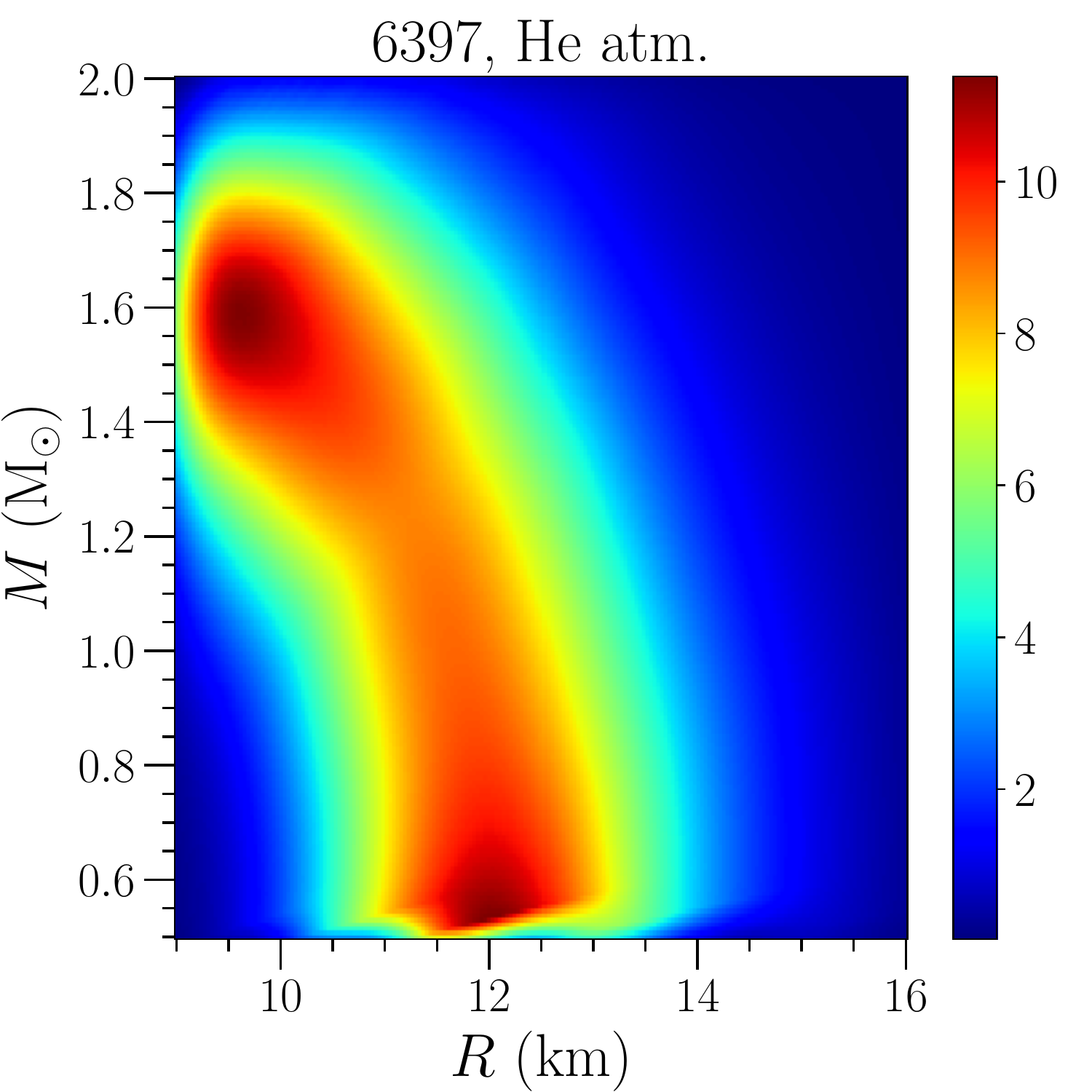}
  \includegraphics[width=1.6in]{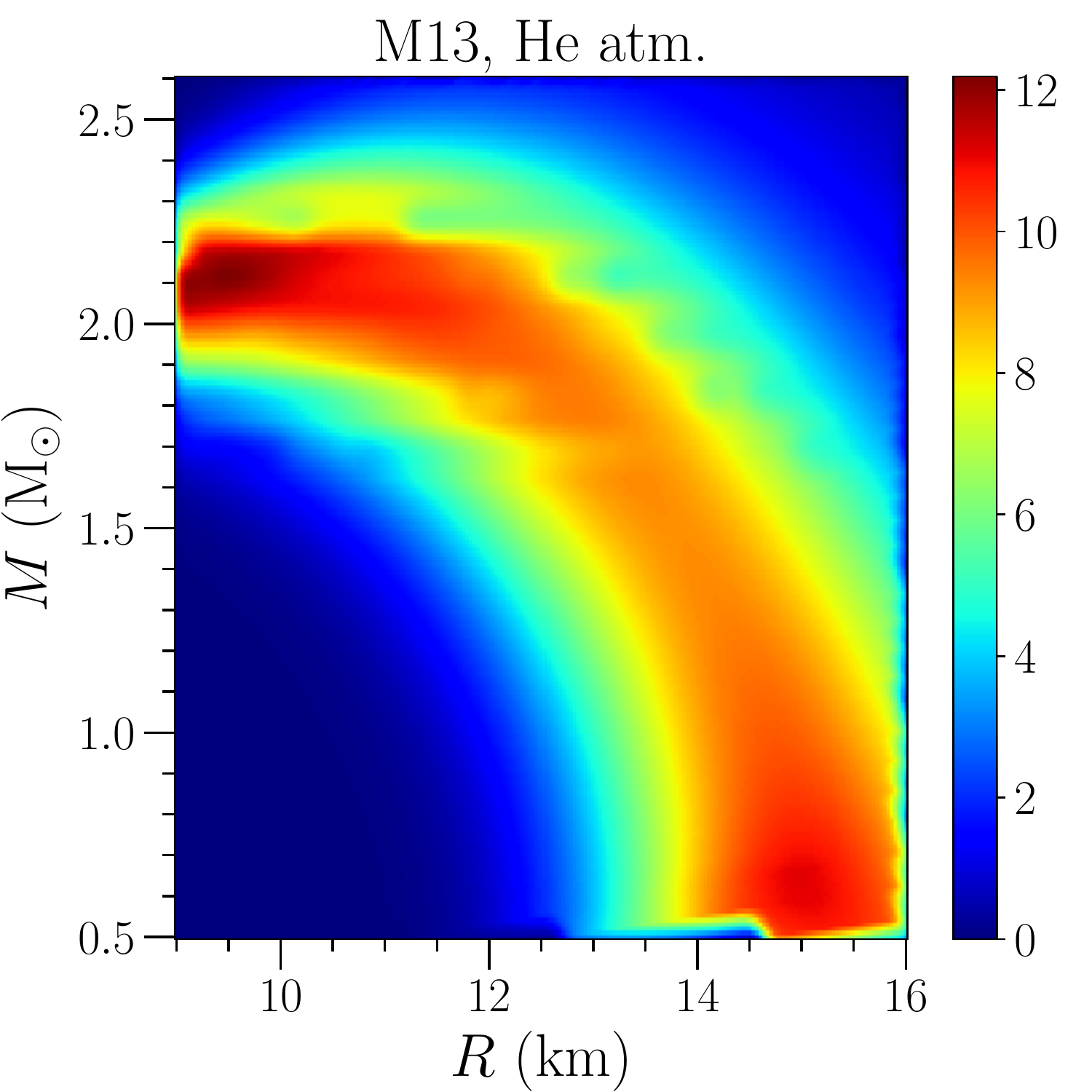}
  \includegraphics[width=1.6in]{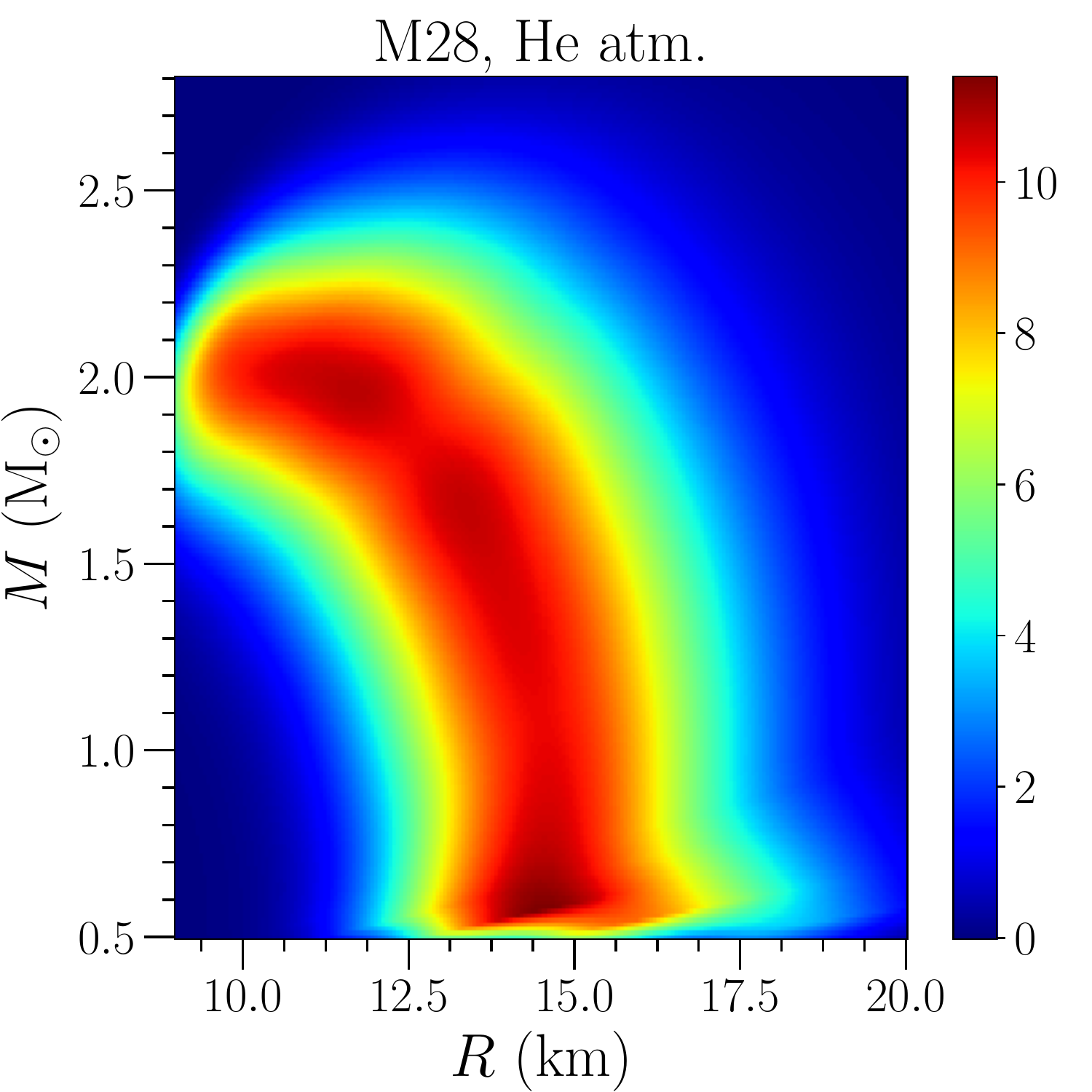}
  \includegraphics[width=1.6in]{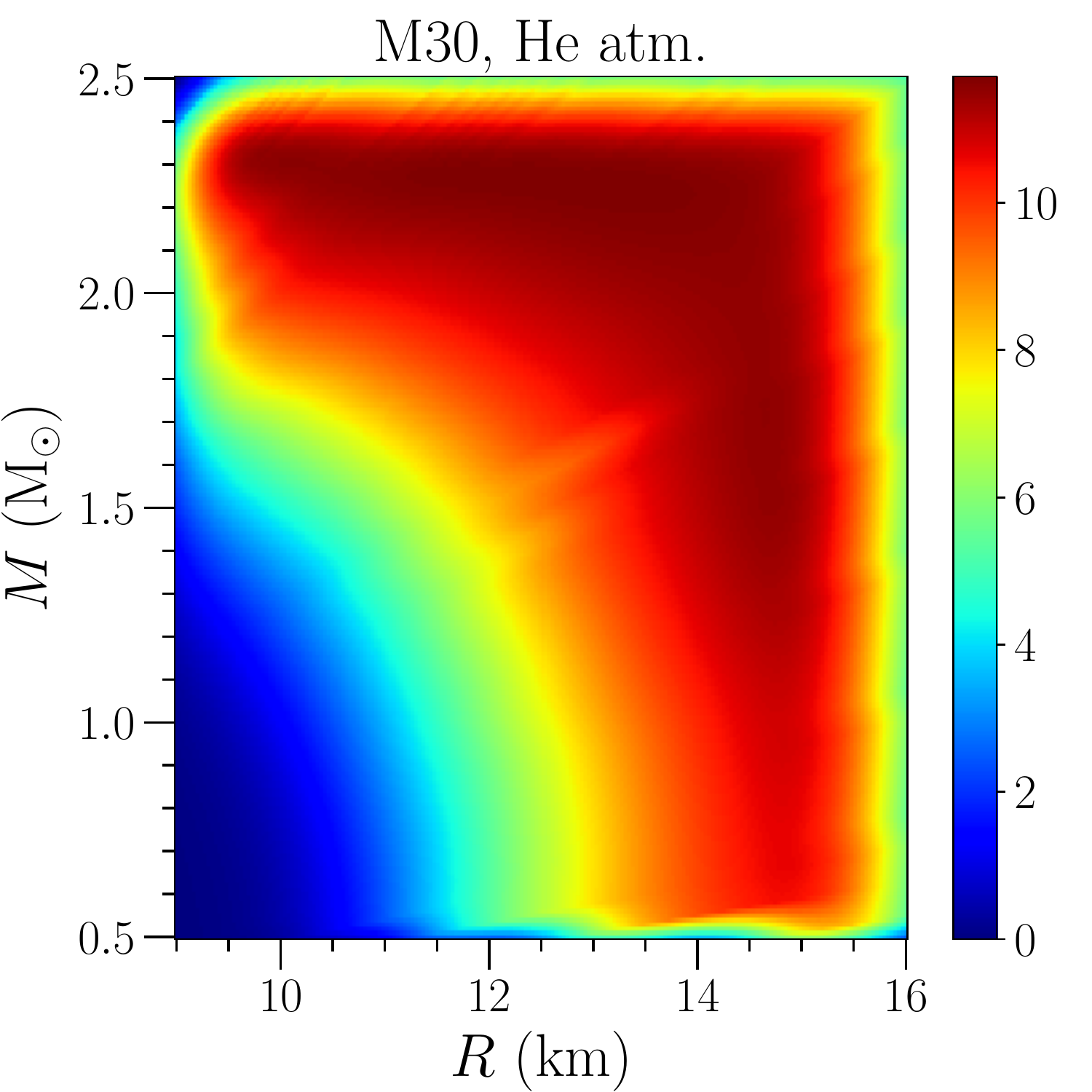}
  \includegraphics[width=1.6in]{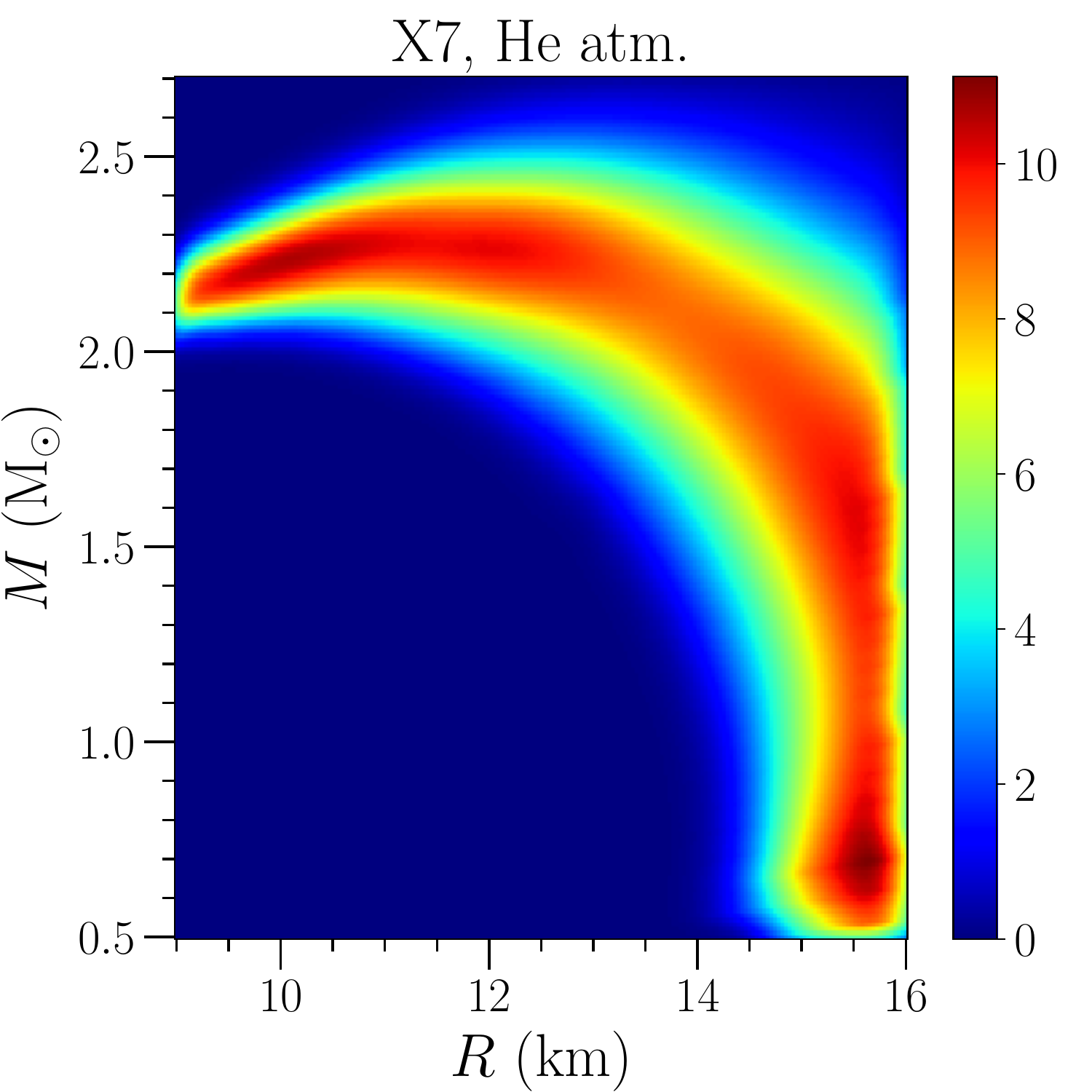}
  \caption{Left panel: The mass and radius constraints for the neutron
    stars in our data set when a He atmosphere is assumed (compare
    with Fig.~\ref{fig:data1}). Our baseline model includes He
    atmospheres for all neutron stars except those in $\omega$ Cen and
    X5.}
  \label{fig:data2}
\end{figure}

A demonstration of the method implied by Eq.~\ref{eq:r_trans}, applied
to the neutron star in NGC 6304 is given in the upper panels of
Fig.~\ref{fig:demo_Rz}. The final result (not shown) is
indistinguishable from the lower right panel in
Fig.~\ref{fig:demo_Rinfz}. The final results, given by
Eq.~\ref{eq:rinf_trans}, for the baseline data set and for X5 in 47
Tuc assuming a H atmosphere are presented in Fig.~\ref{fig:data1}. The
effect of an He atmosphere on the neutron stars (except for X5 and the
neutron star in $\omega$ Cen) in Fig.~\ref{fig:data2}.

\begin{figure}
  \includegraphics[width=1.6in]{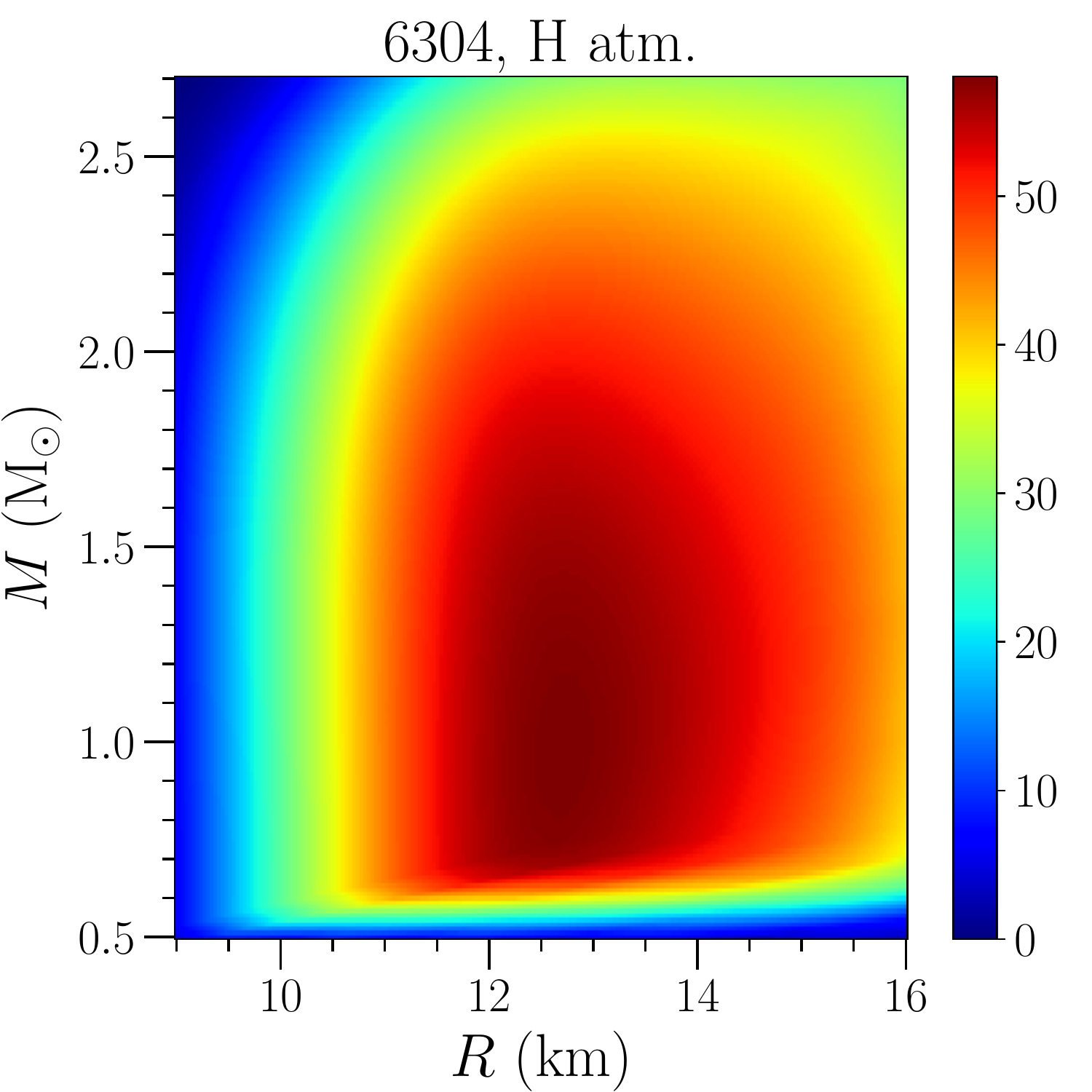}
  \includegraphics[width=1.6in]{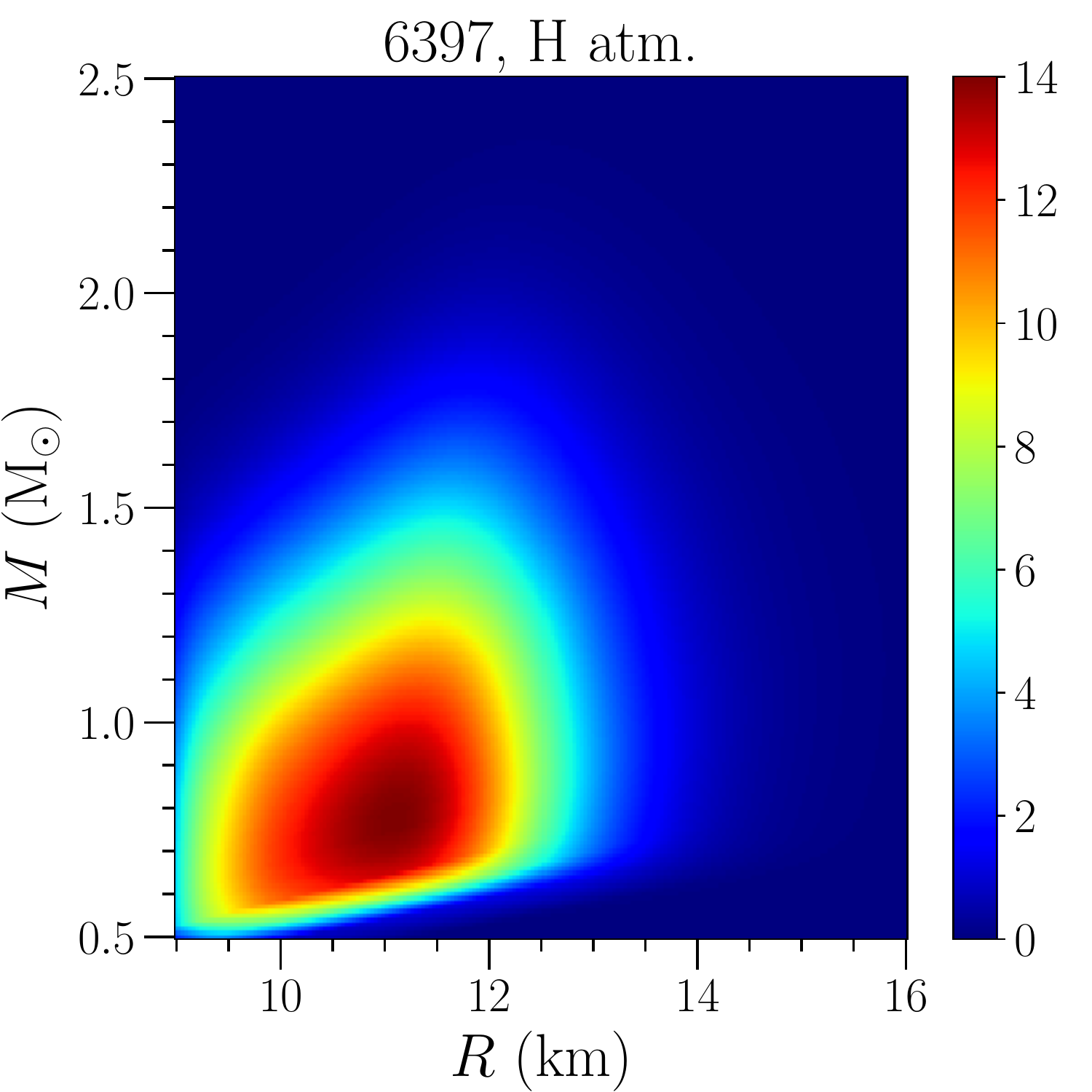}
  \includegraphics[width=1.6in]{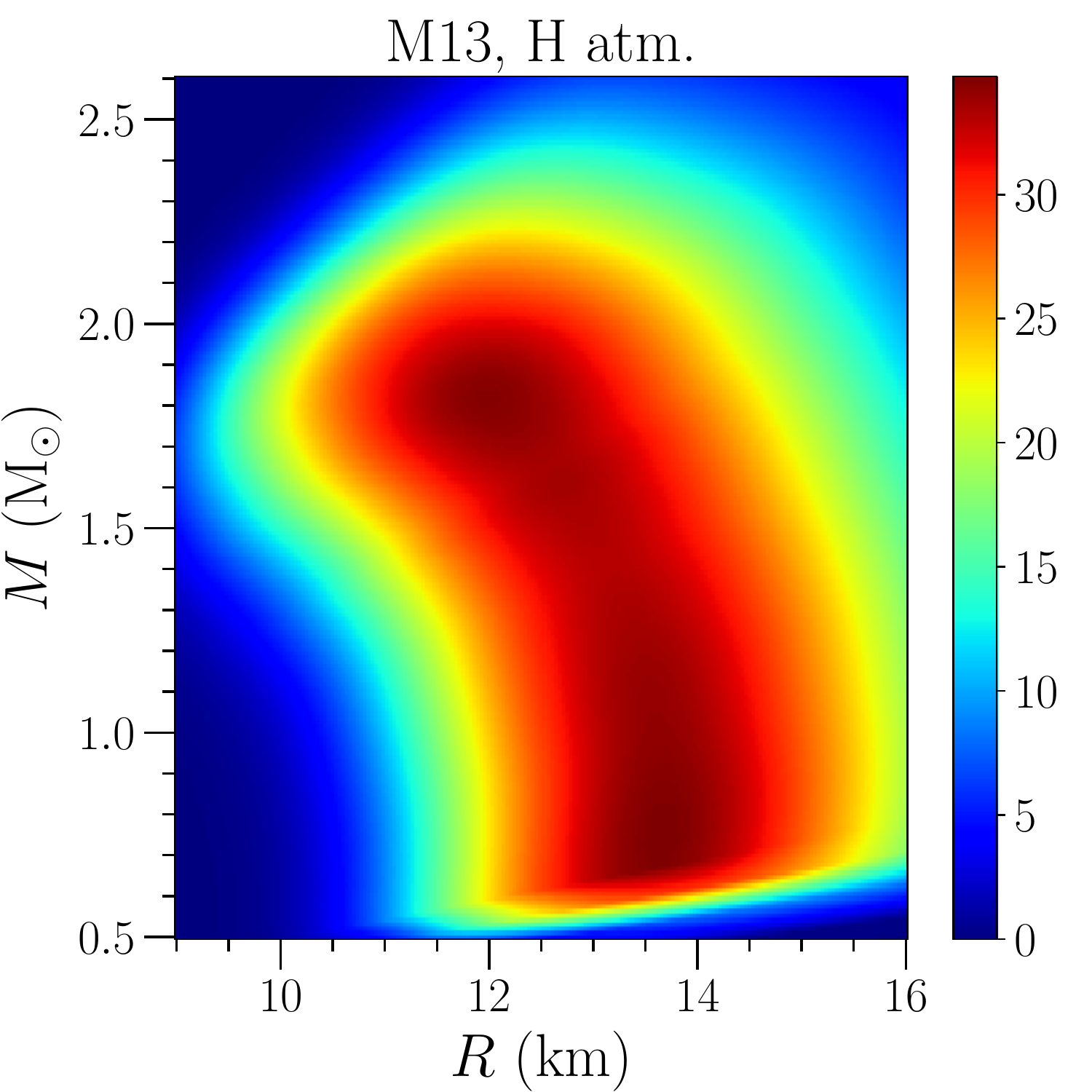}
  \includegraphics[width=1.6in]{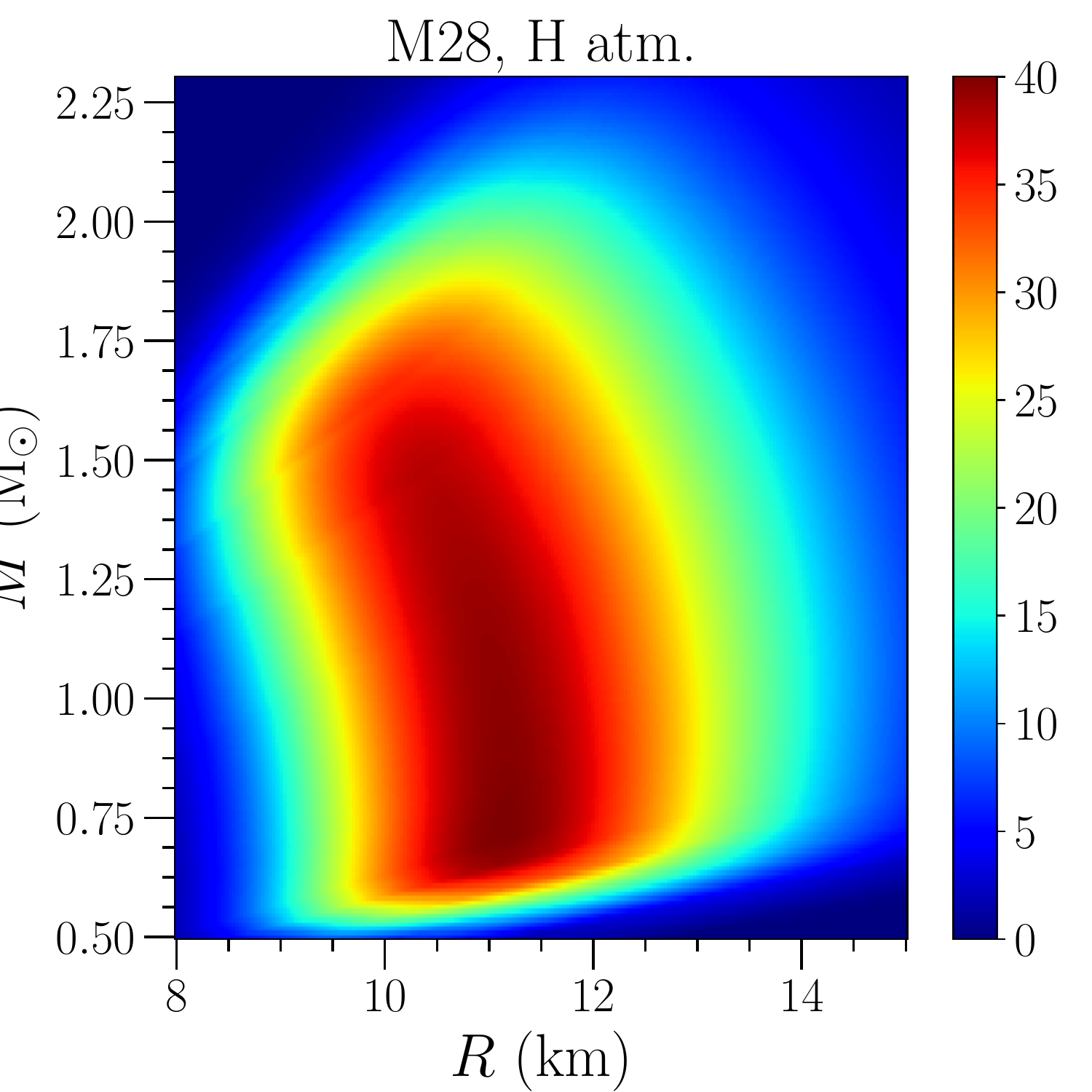}
  \includegraphics[width=1.6in]{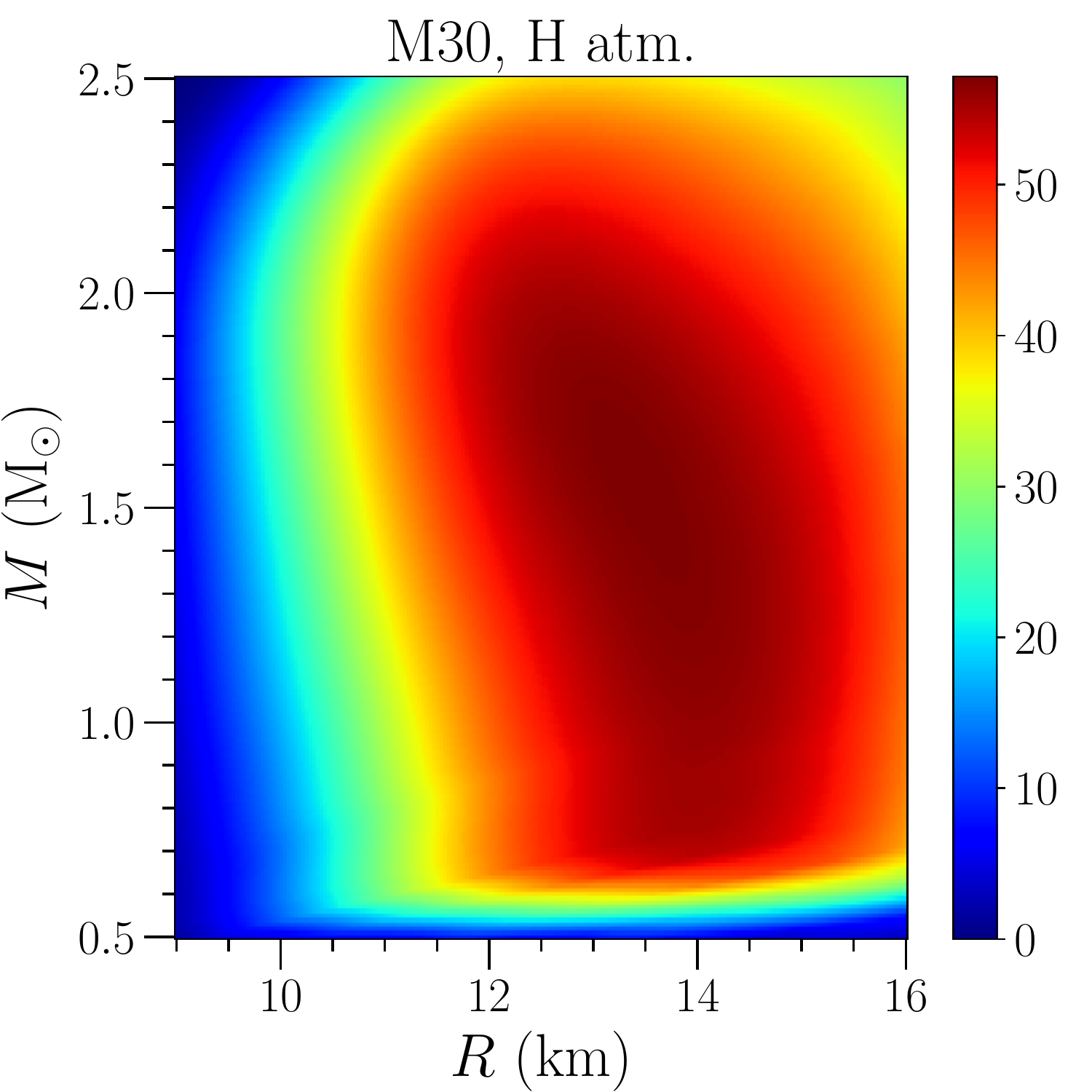}
  \includegraphics[width=1.6in]{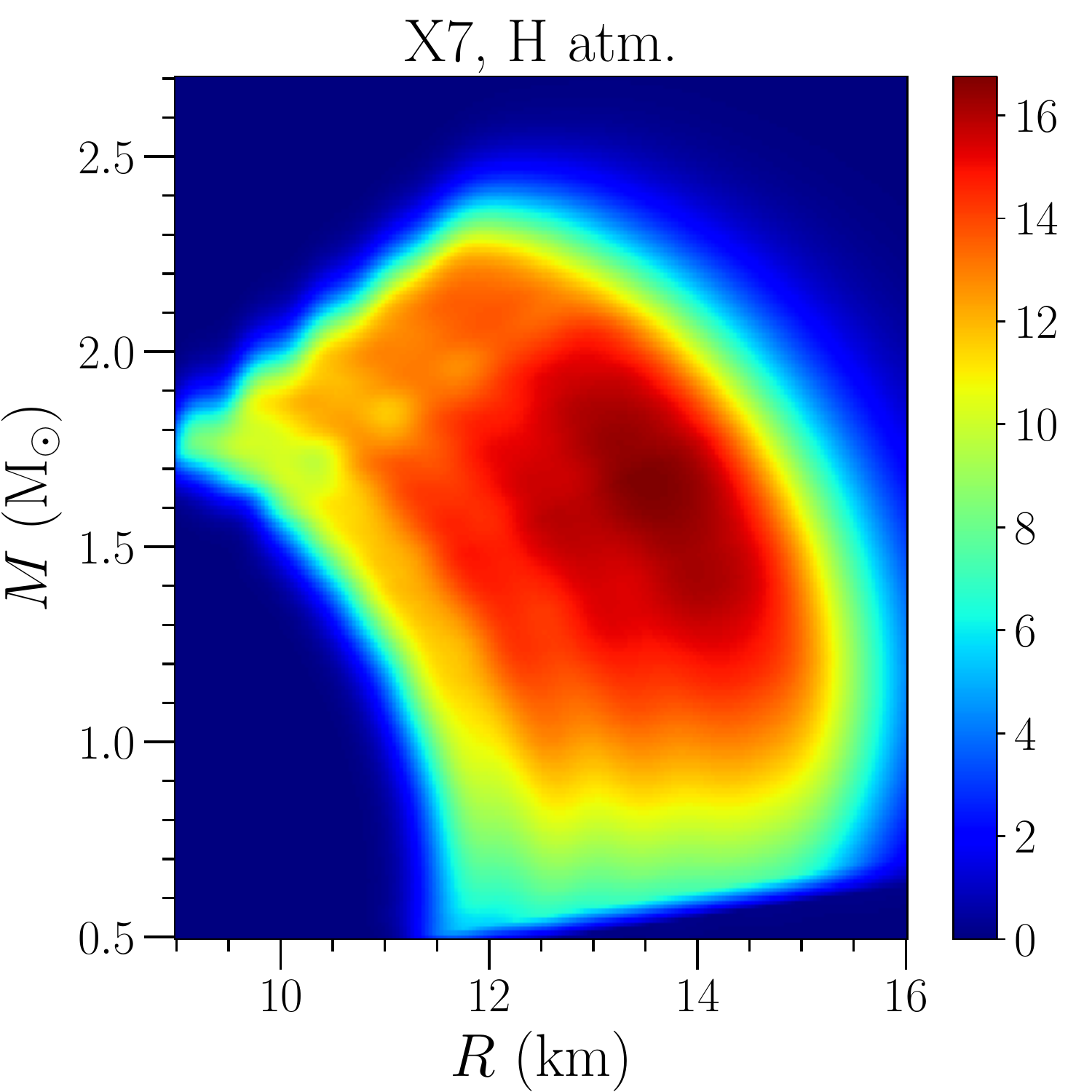}
  \includegraphics[width=1.6in]{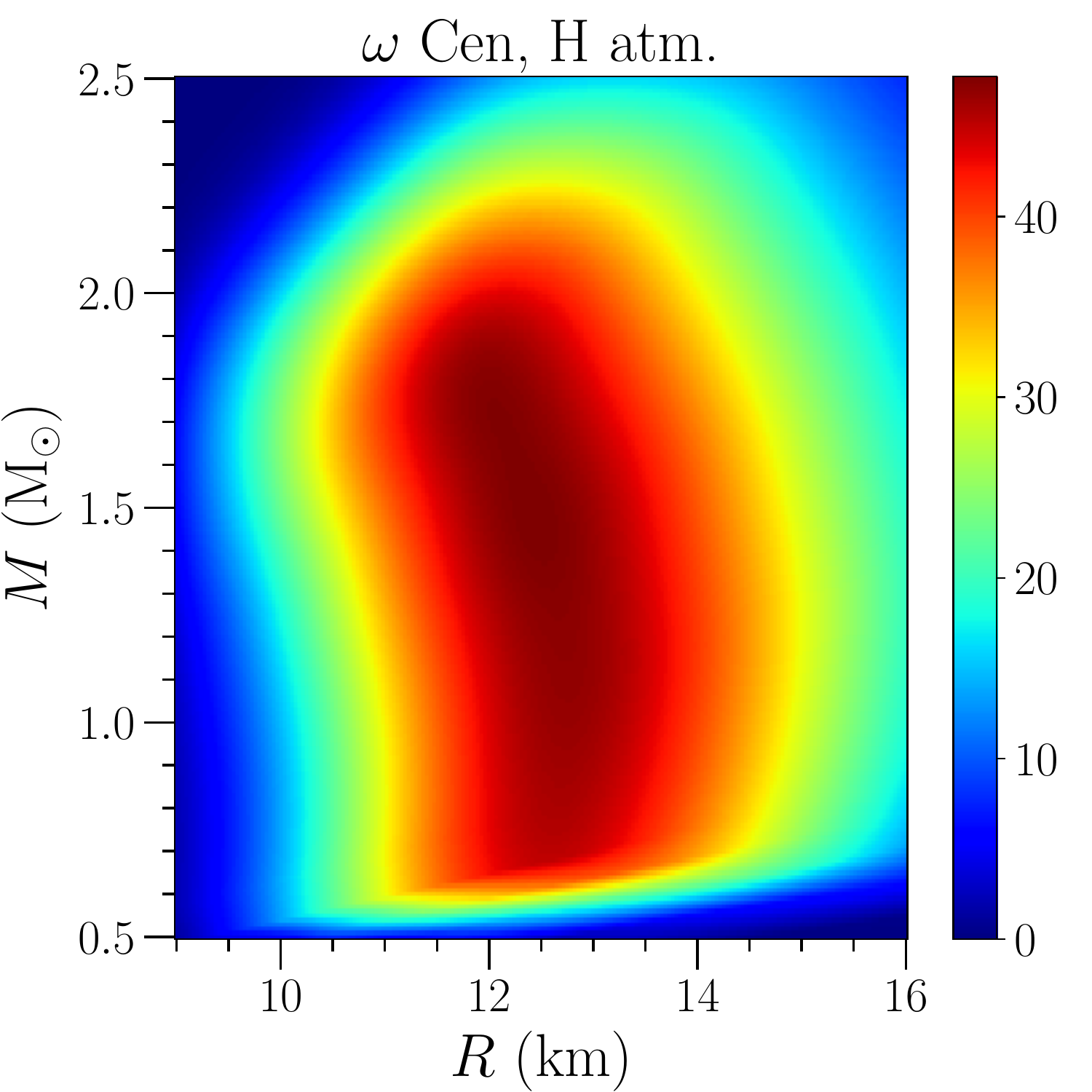}
  \includegraphics[width=1.6in]{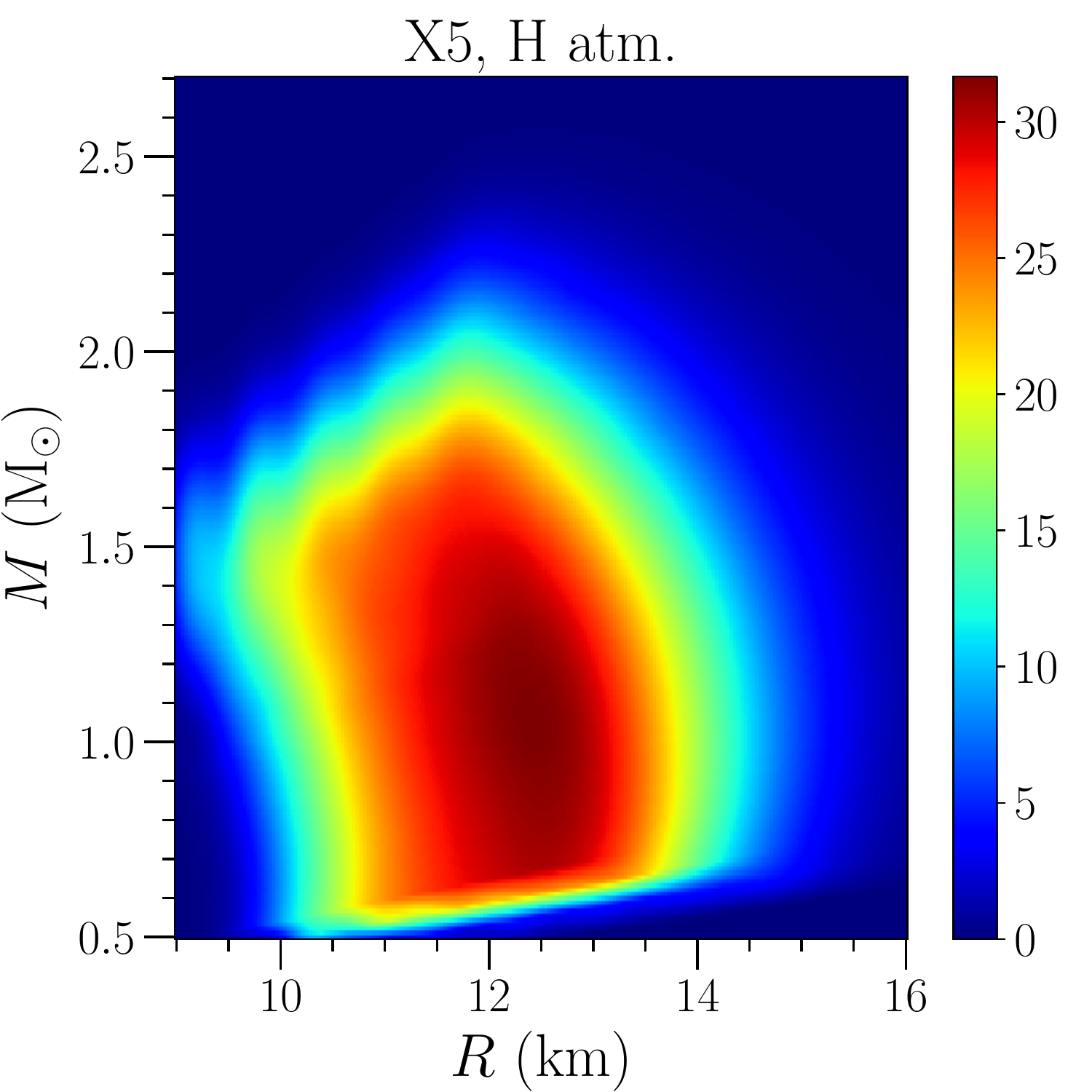}
  \caption{The H atmosphere part of our baseline data set plus the
    neutron star X5 in 47 Tuc assuming a hotspot may be present. The
    \citet{Wilms00} abundances are used to correct for X-ray
    absorption in all cases, the normalization is arbitrary, and a
    distance uncertainty has been added following the prescription
    described in section \ref{s:4}.}
  \label{fig:hot1}
\end{figure}

\begin{figure}
  \includegraphics[width=1.6in]{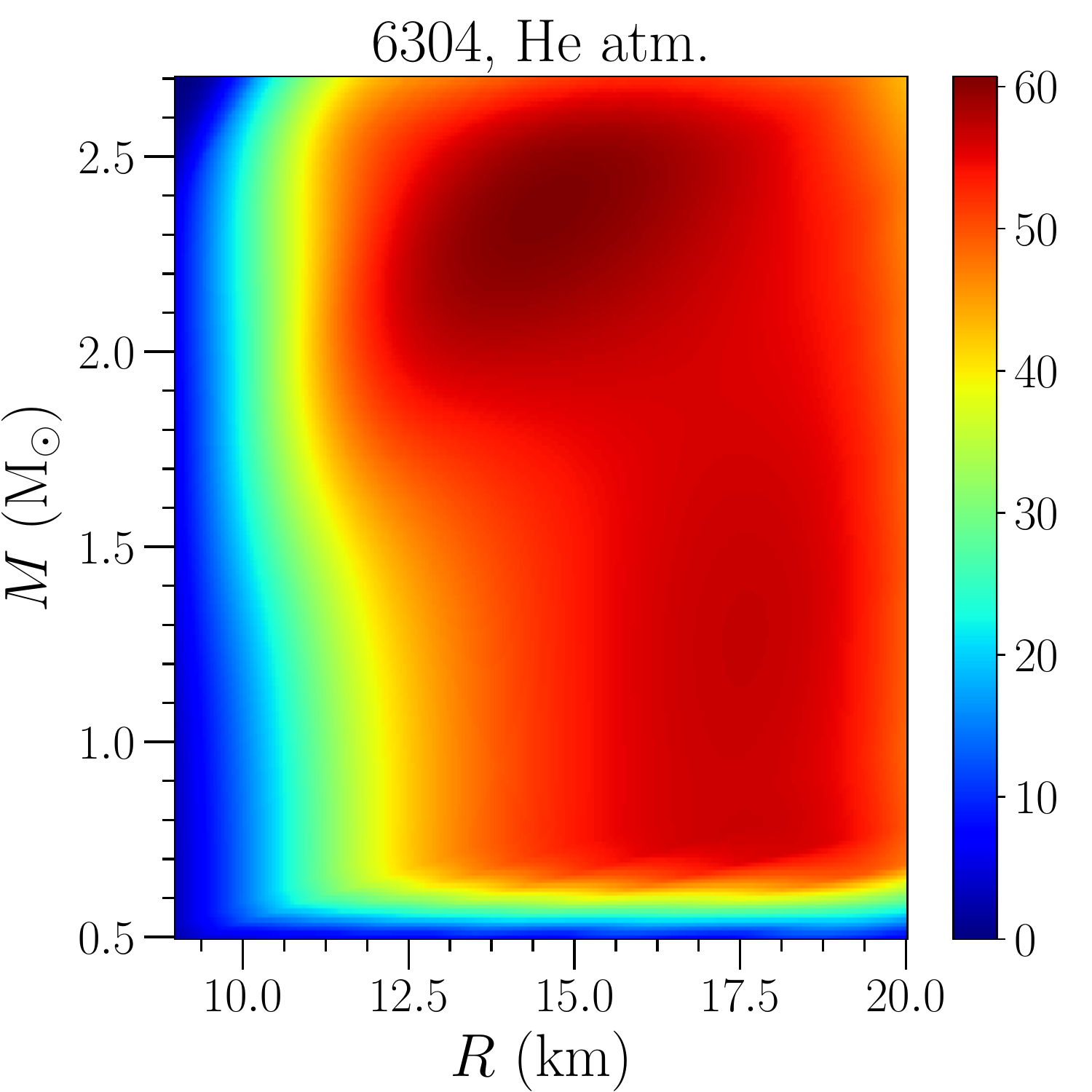}
  \includegraphics[width=1.6in]{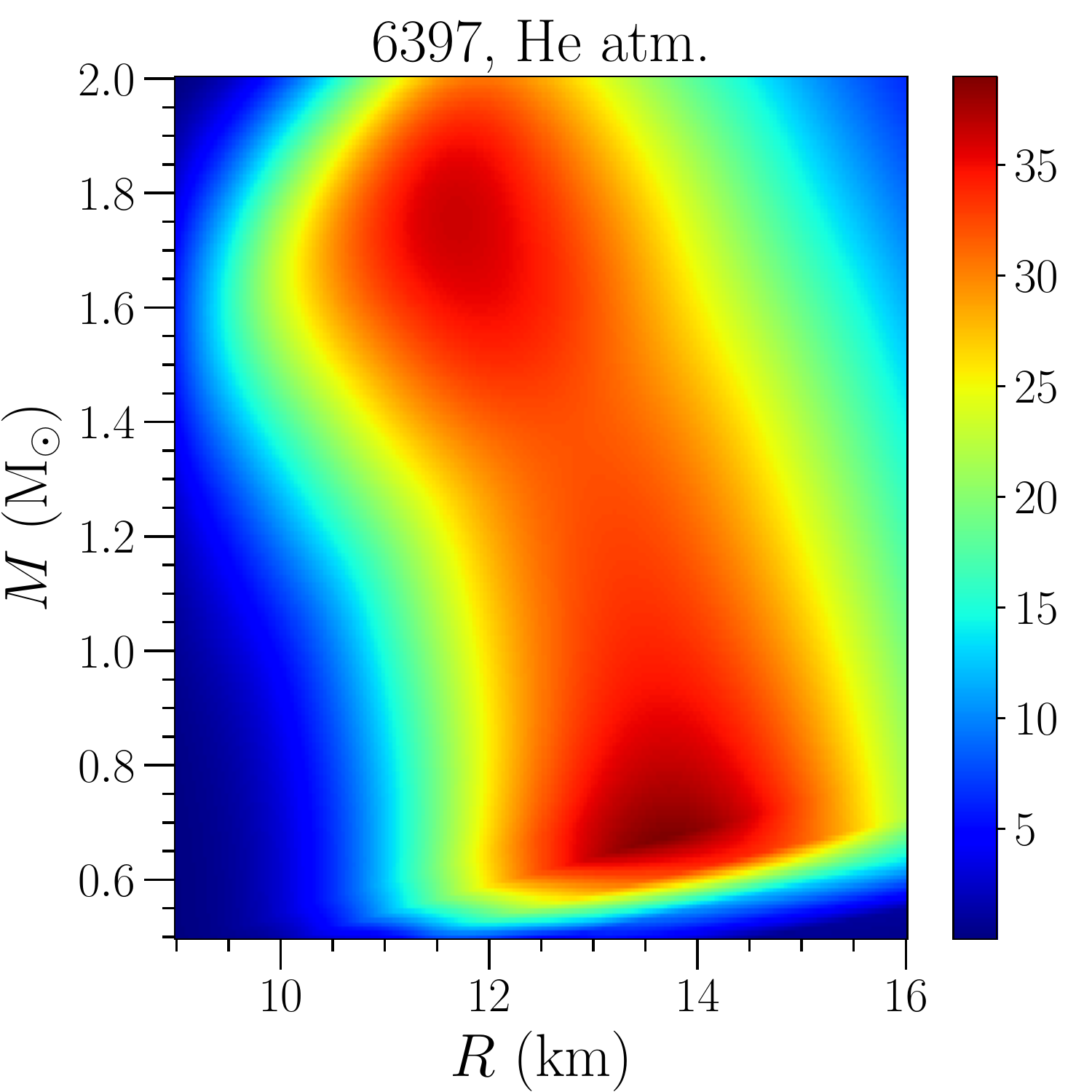}
  \includegraphics[width=1.6in]{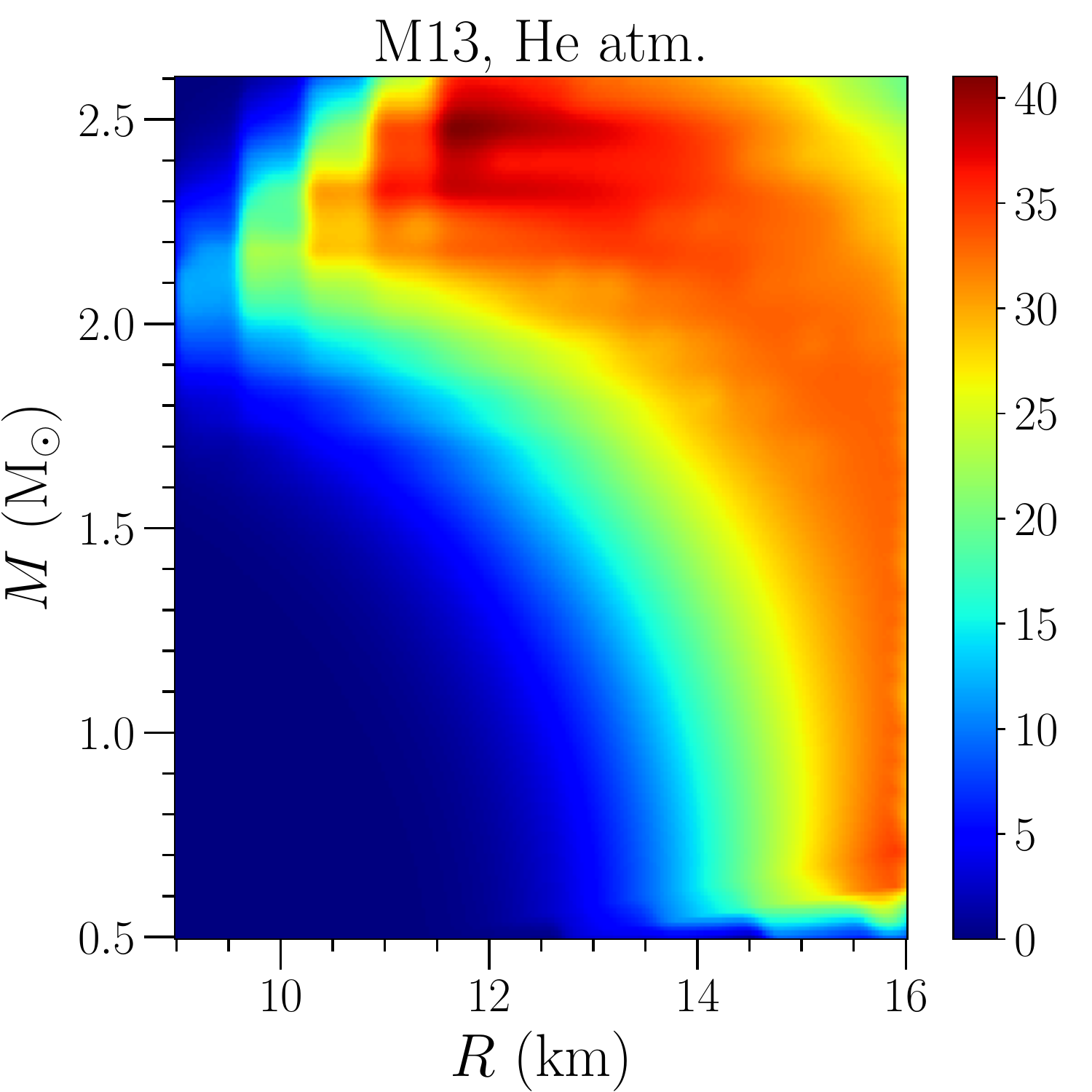}
  \includegraphics[width=1.6in]{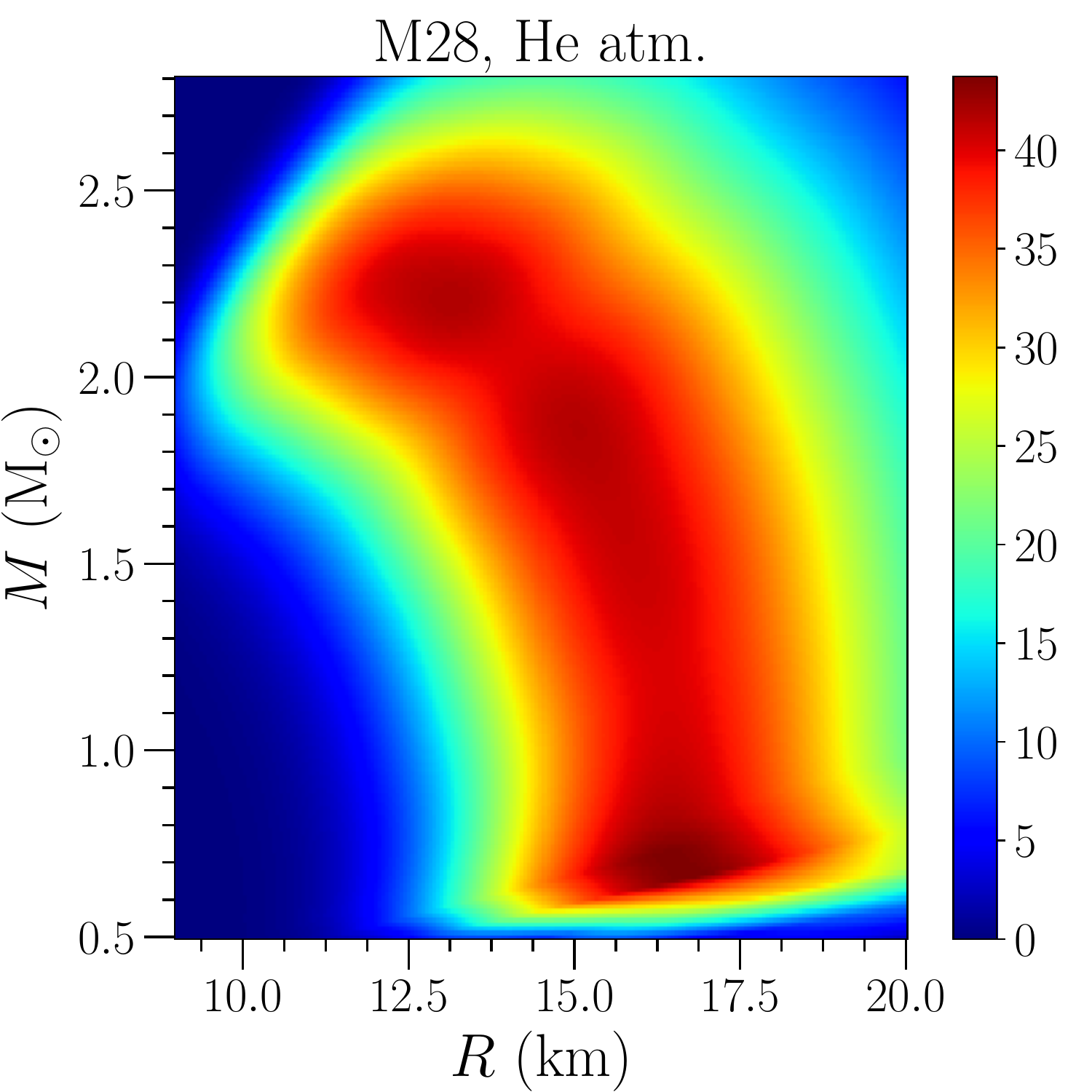}
  \includegraphics[width=1.6in]{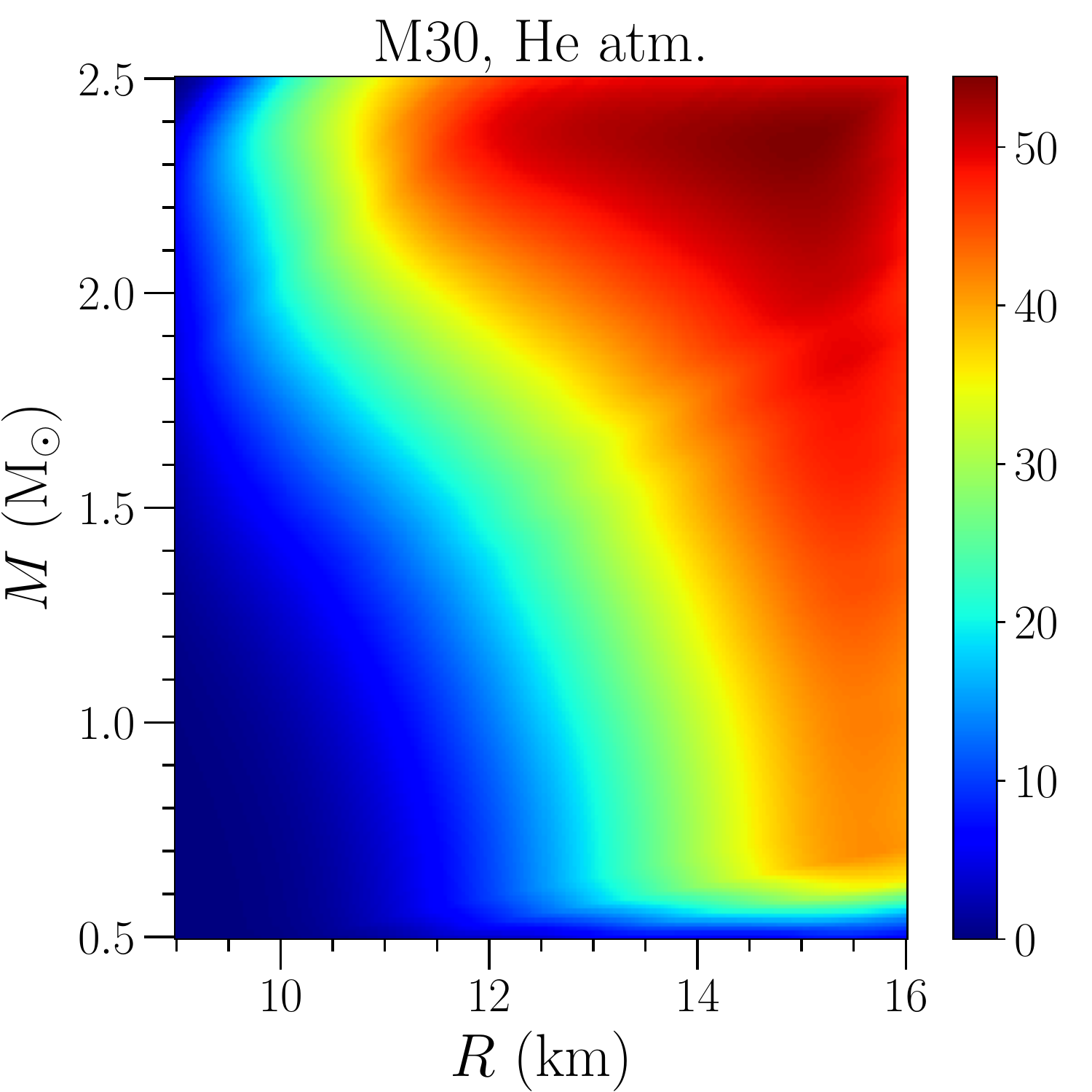}
  \includegraphics[width=1.6in]{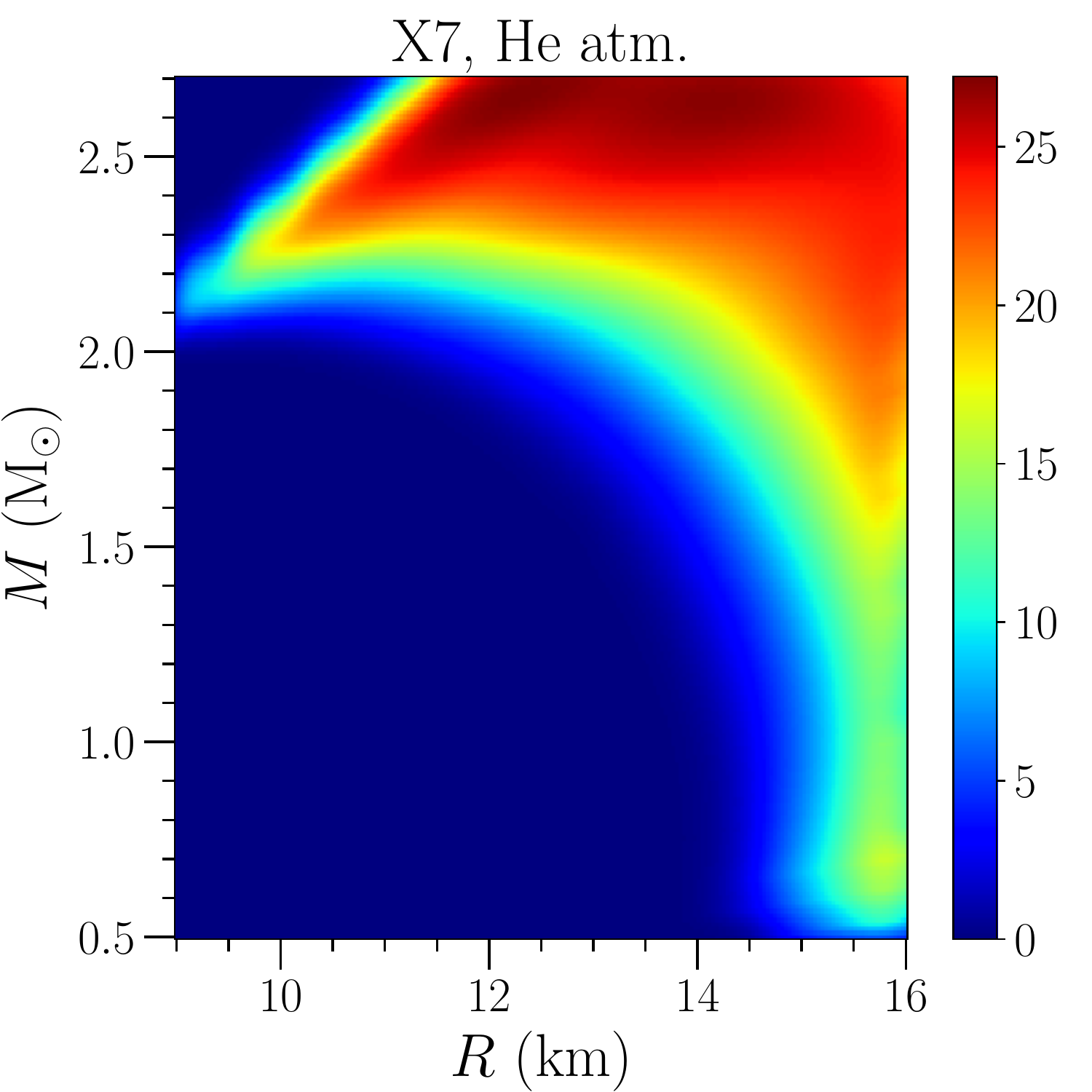}
  \caption{Left panel: The mass and radius constraints for the neutron
    stars in our data set when a hotspot and a He atmosphere is
    assumed (compare with Fig.~\ref{fig:data1}). Our baseline model
    includes He atmospheres for all neutron stars except those in
    $\omega$ Cen and X5.}
  \label{fig:hot2}
\end{figure}

\citet{Elshamouty16} found that the presence of temperature
inhomogeneities on the neutron star surface (hot spots) can bias the
radii inferred from X-ray spectral fits, leading to underestimates of
the radius by up to 28\%. We handle this by including an additional
nuisance parameter which increases $R_{\infty}$ by a fixed percentage
in order to compensate for this effect. We presume that this parameter
has a uniform prior distribution and take its value to be between 0\%
and 28\%. As an example, the probability distribution for the neutron
star in NGC 6304 after having made this correction is given in the
lower right panel of figure~\ref{fig:demo_Rz}. The final data set
presuming a possible hotspot with hydrogen atmospheres is in
Fig.~\ref{fig:hot1} and with helium atmospheres (where appropriate)
is in Fig.~\ref{fig:hot2}.

\subsection{Equation of State}

Quantum Monte Carlo calculations of neutron matter
\citep{Gandolfi12mm} provide an excellent description of matter up to
the nuclear saturation density ($\rho \approx 2.8 \times
10^{14}~\mathrm{g}/\mathrm{cm}^3$). As in \citet{Steiner15un}, we
assume that neutron star matter near the saturation density is
described by results from quantum Monte Carlo. We also assume the
neutron star has a crust as described in \citet{Baym71tg} and
\citet{Negele73ns}.

High-density matter may contain a strong phase transition. For the
purposes of this work, a strong phase transition may be defined as a
region of energy density for which the pressure is nearly flat. If
matter is unlikely to have a strong phase transition, then a
polytropic form is a good description of high-density matter. We
employ Model A as presented in \citet{Steiner13tn} and updated in
\citet{Steiner15un}. On the other hand, strong phase transitions are
not well-described by polytropes. In pressure-energy density space a
polytrope of the form $P=K \varepsilon^{\Gamma}$ is not a good
description of a nearly flat EOS because it requires $\Gamma$ to be
very small and $K$ to be anomalously large. This kind of behavior is
disfavored by uniform or weakly-varying priors in $\Gamma$. We thus
also use Model C~\citep{Steiner13tn,Steiner15un} which uses
line-segments in pressure and energy density space and makes stronger
phase transitions more likely. It is important to emphasize that the
ability of the parameterization to describe a generic EOS is not the
only consideration (as both polytropic models and those based on line
segments can reproduce almost all EOSs). Once a parameterization and a
prior for the parameters is specified, changes in the likelihood with
which various EOSs are selected can make a significant change in the
results. We present results from these two high-density EOSs
separately and assign an equal prior probability to each. We use the
open-source code from \citet{Steiner14oo,Steiner14ba} to perform the
simulations which have been updated to use affine-invariant
sampling~\citep{Goodman10es}.

\section{Results}

Ideally, one decides the prior distribution before any calculations
are performed. We assign all the different models and interpretations
of the data (described in detail below) equal probability. Thus our
full posteriors can be viewed as the sum over the results presented
for each individual model and/or data interpretation (after having
been properly weighted by their evidence). 

\begin{figure}
  \includegraphics[width=1.6in]{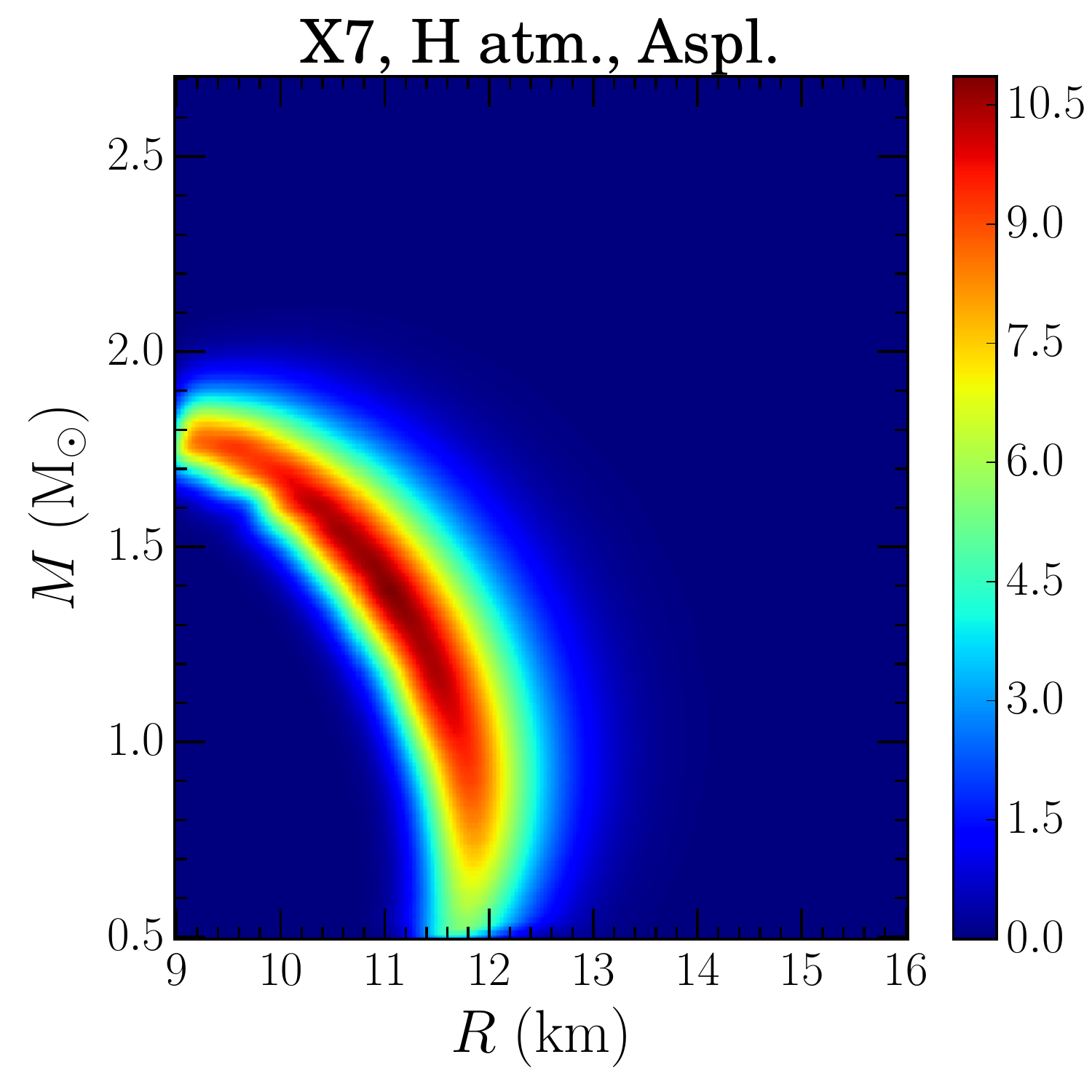}
  \includegraphics[width=1.6in]{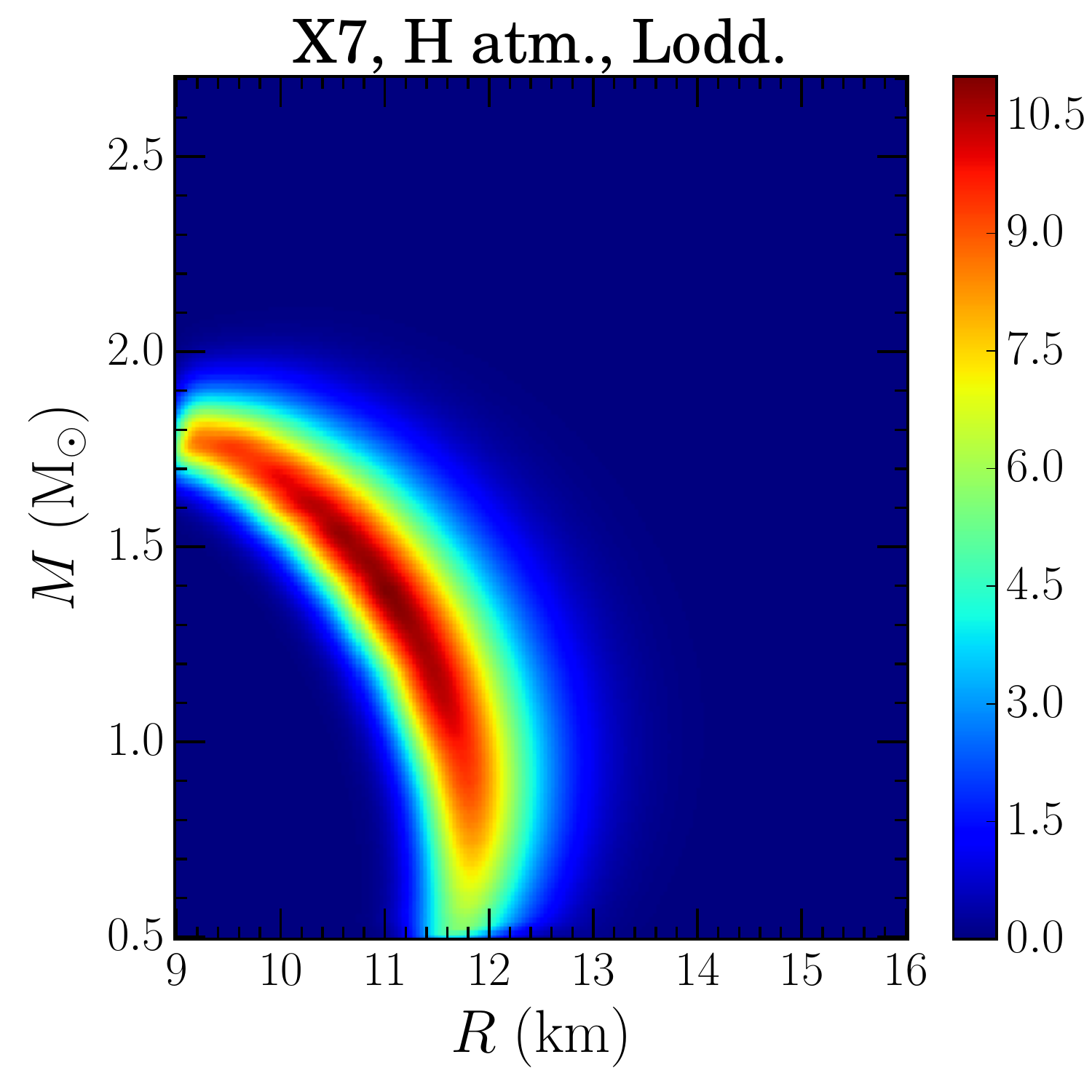}
  \caption{Left panel: The mass and radius constraints for the neutron
    star 47 Tuc in X7 when a H atmosphere is assumed and
\citet{Asplund09} abundances are used (compare with the lower right panel in
    Fig.~\ref{fig:data1}). Right panel: The mass and radius
    constraints for the neutron star 47 Tuc in X7 when a H atmosphere
    is assumed and \citet{Lodders03} abundances are used.}
  \label{fig:data3}
\end{figure}

For our baseline results, we include all neutron stars
except X5, and assume the polytropic model for high-density matter. We
include both H and He atmospheres for the objects in the baseline data
set, except $\omega$ Cen, where the atmosphere composition is known to
be H \citep{Haggard04}. We initially assume the abundances of elements
in the interstellar matter to be those given by \citet{Wilms00}, since
that paper collated the best available evidence on the abundance of
elements in the local interstellar medium. In order to test the change
in the inferred radius due to different abundance models, we compare
our mass and radius distributions derived using the \citet{Wilms00}
abundance model to those derived using the abundance models of
\citet{Asplund09} and \citet{Lodders03}. These abundance models,
produced using studies of the Sun and meteorites, respectively,
suggest a plausible range of uncertainty for the interstellar
abundances. This is in contrast to significantly older abundance
models such as those of \citet{Anders89}, which are quite different,
and would lead to significant changes to radius estimates
\citep{Heinke14}. Fig.~\ref{fig:data3} shows the mass and radius
constraints for X7 in 47 Tuc using these alternate abundance models
and demonstrates that the results are only slightly different
(differences in inferred radius are less than 1\%, in agreement with
\citealt{Bogdanov16} from those using \citet{Wilms00} abundances. This
holds for all of the objects in our data set, so we only present
results using \citet{Wilms00} abundances below. We have also tested
the effects of the updated {\tt tbnew} code by J\"{o}rn
Wilms\footnote{http://pulsar.sternwarte.uni-erlangen.de/wilms/research/tbabs/},
and find that the code improvements of {\tt tbnew} (mostly fine
structure around edges, affecting high-resolution spectroscopy) affect
radius estimates on the order of 0.1\%.

\begin{figure}
  \includegraphics[width=1.6in]{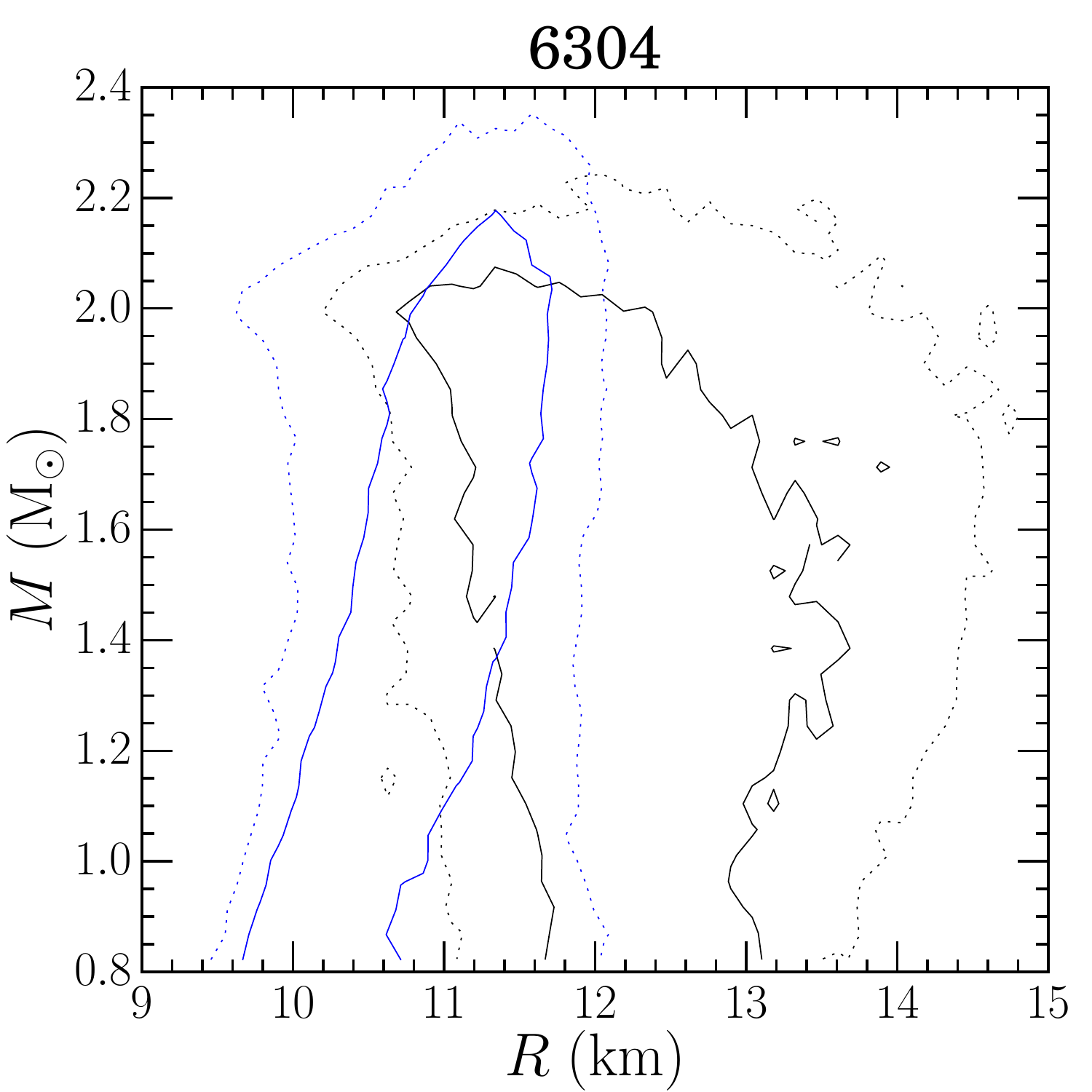}
  \includegraphics[width=1.6in]{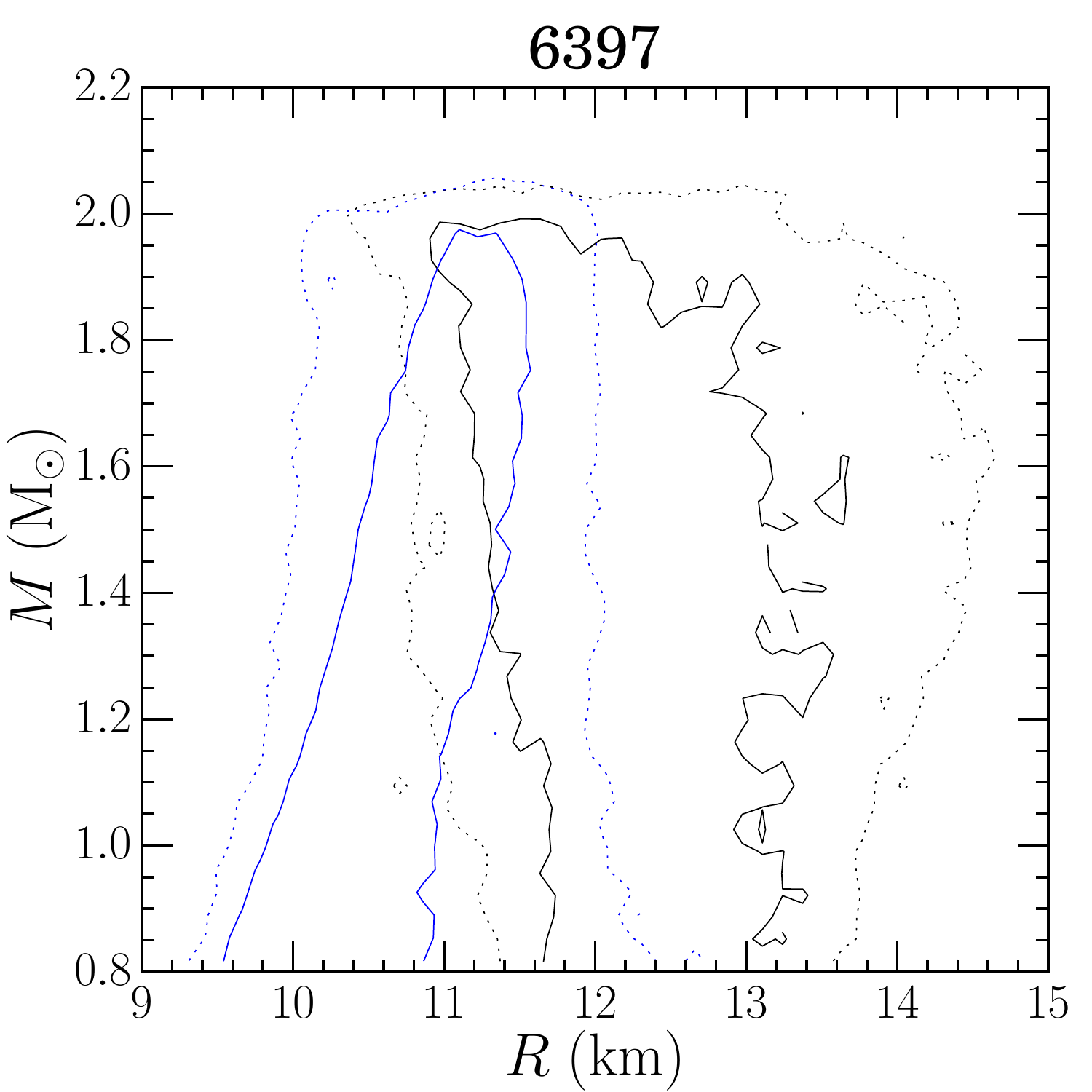}
  \includegraphics[width=1.6in]{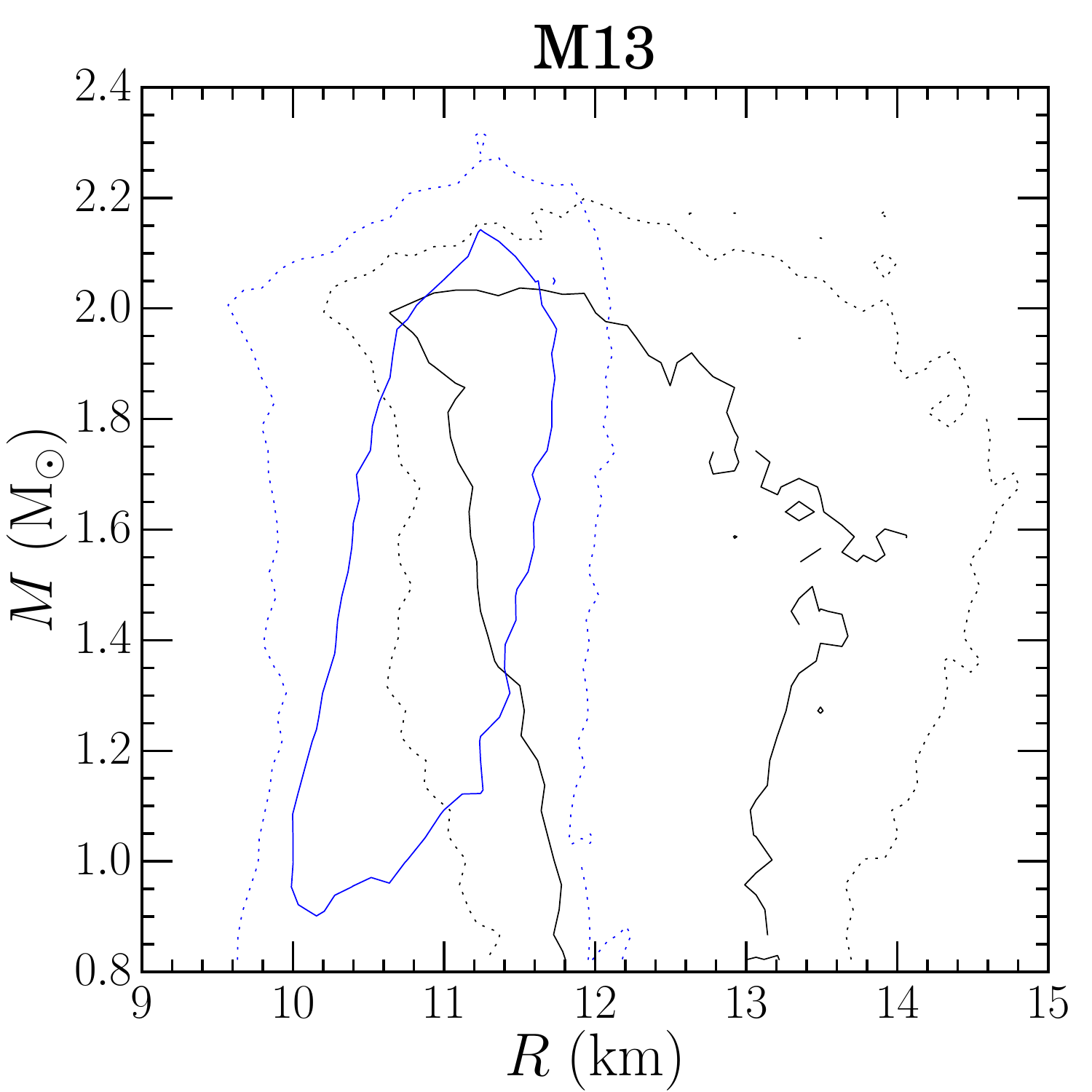}
  \includegraphics[width=1.6in]{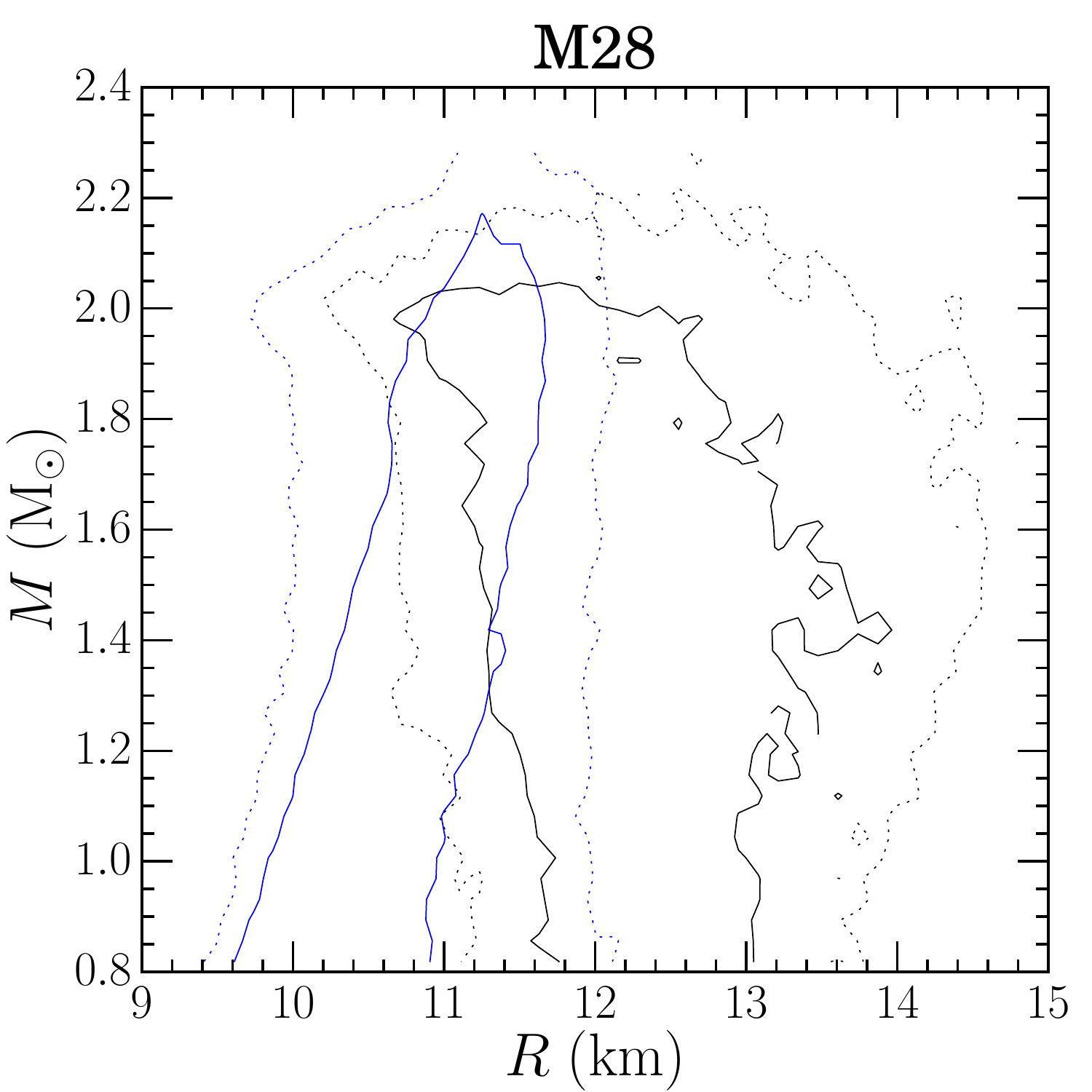}
  \includegraphics[width=1.6in]{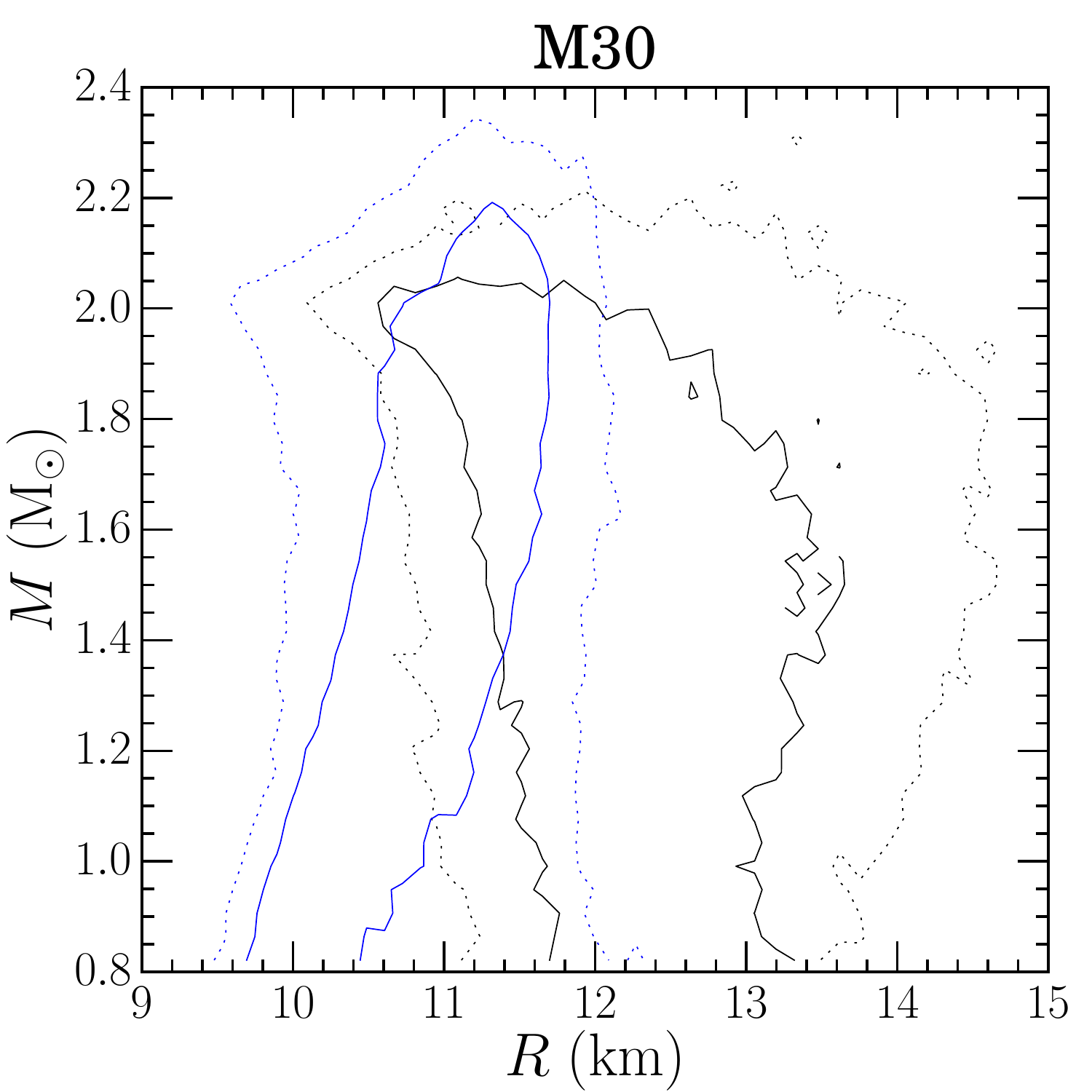}
  \includegraphics[width=1.6in]{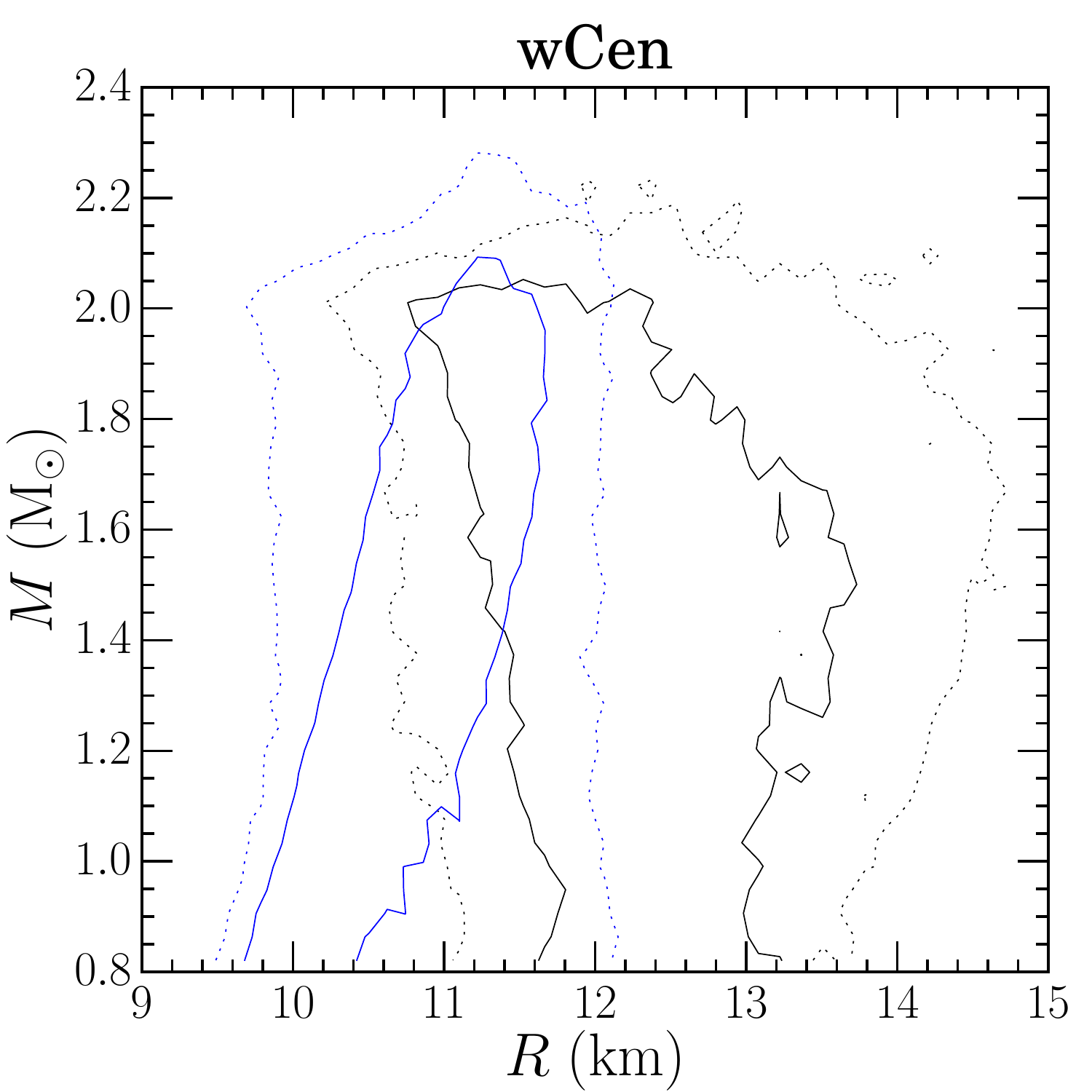}
  \includegraphics[width=1.6in]{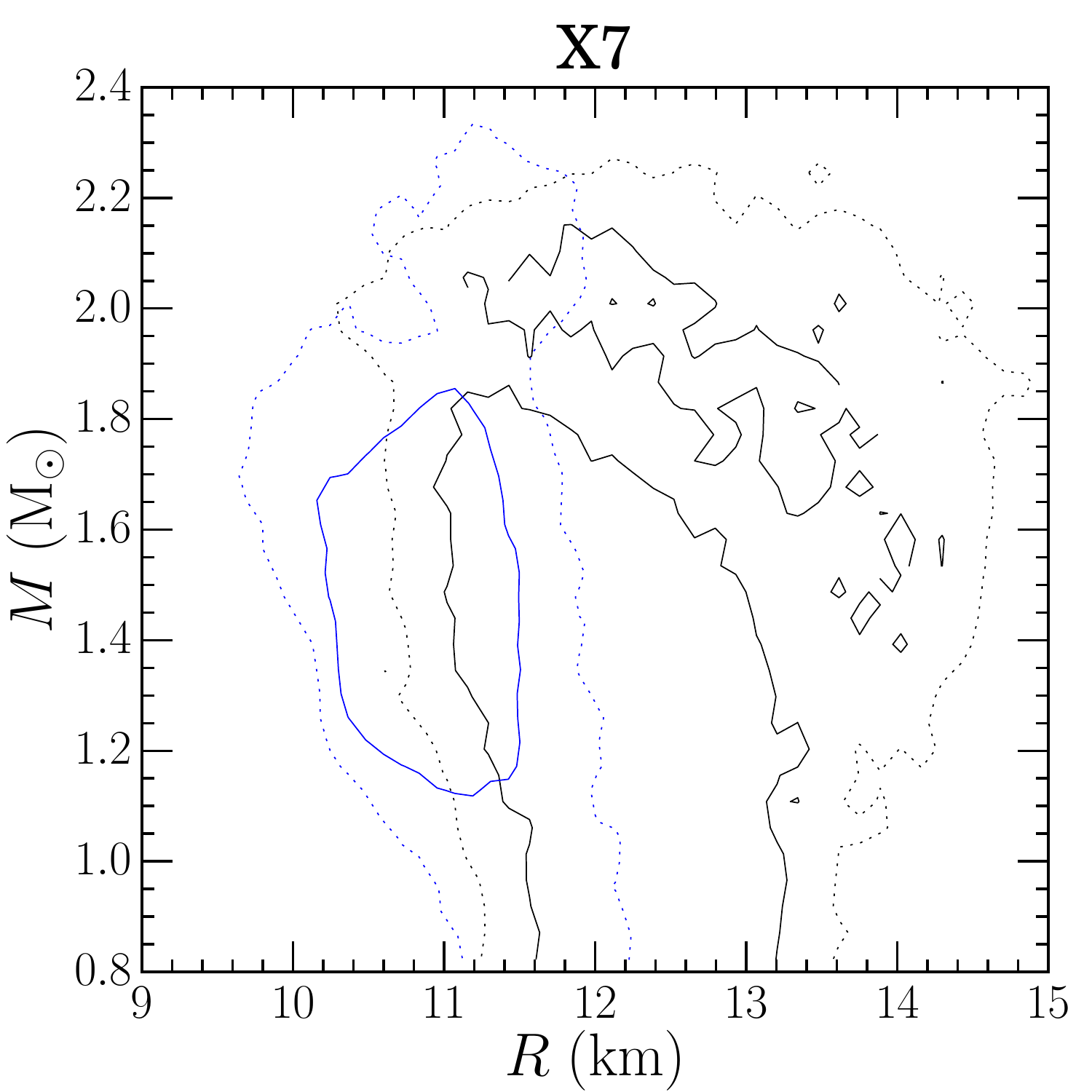}
  \caption{Contour lines representing 68\% and 95\% confidence limits
    for the 7 objects in the baseline model (black curves) and with
    Model C (blue curves). Most of the results appear similar except
    for X7 which has a bimodality resulting from the choice between H
    and He atmospheres which is only evident in this neutron star
    because the radius is strongly constrained. }
  \label{fig:post}
\end{figure}

The posterior mass and radius distributions for the neutron stars in
the baseline scenario are given in Fig.~\ref{fig:post}, together with
the results presuming Model C is used for the EOS. The mass posterior
distributions are relatively broad, with the sole exception for X7.
The mass of X7 must be larger than $1.1~\mathrm{M}_{\odot}$ in Model C
because smaller mass stars have central densities too low to allow for
the phase transition to decrease the radius sufficiently to match the
mass and radius implied by the observations.

\begin{figure}
  \includegraphics[width=1.6in]{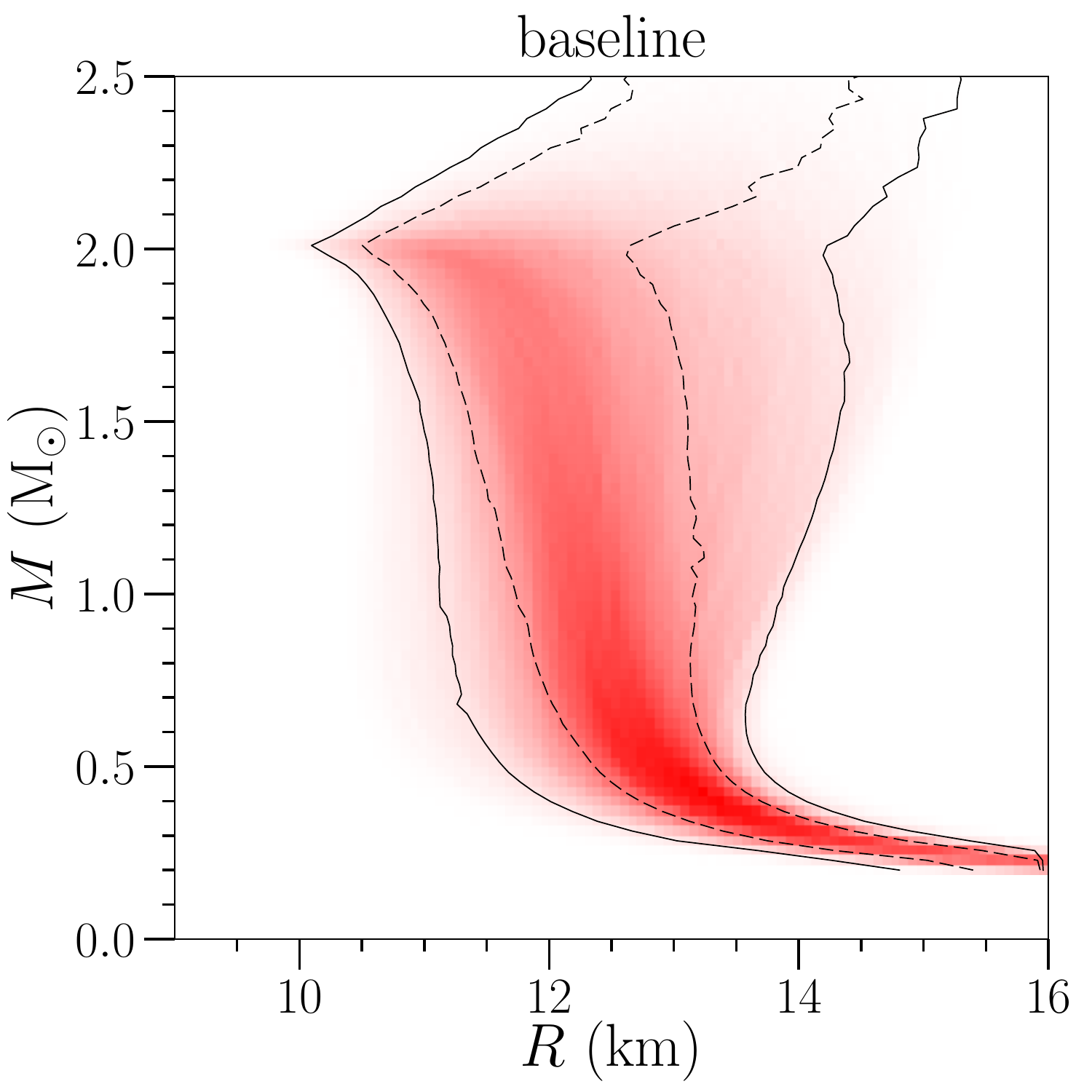}
  \includegraphics[width=1.6in]{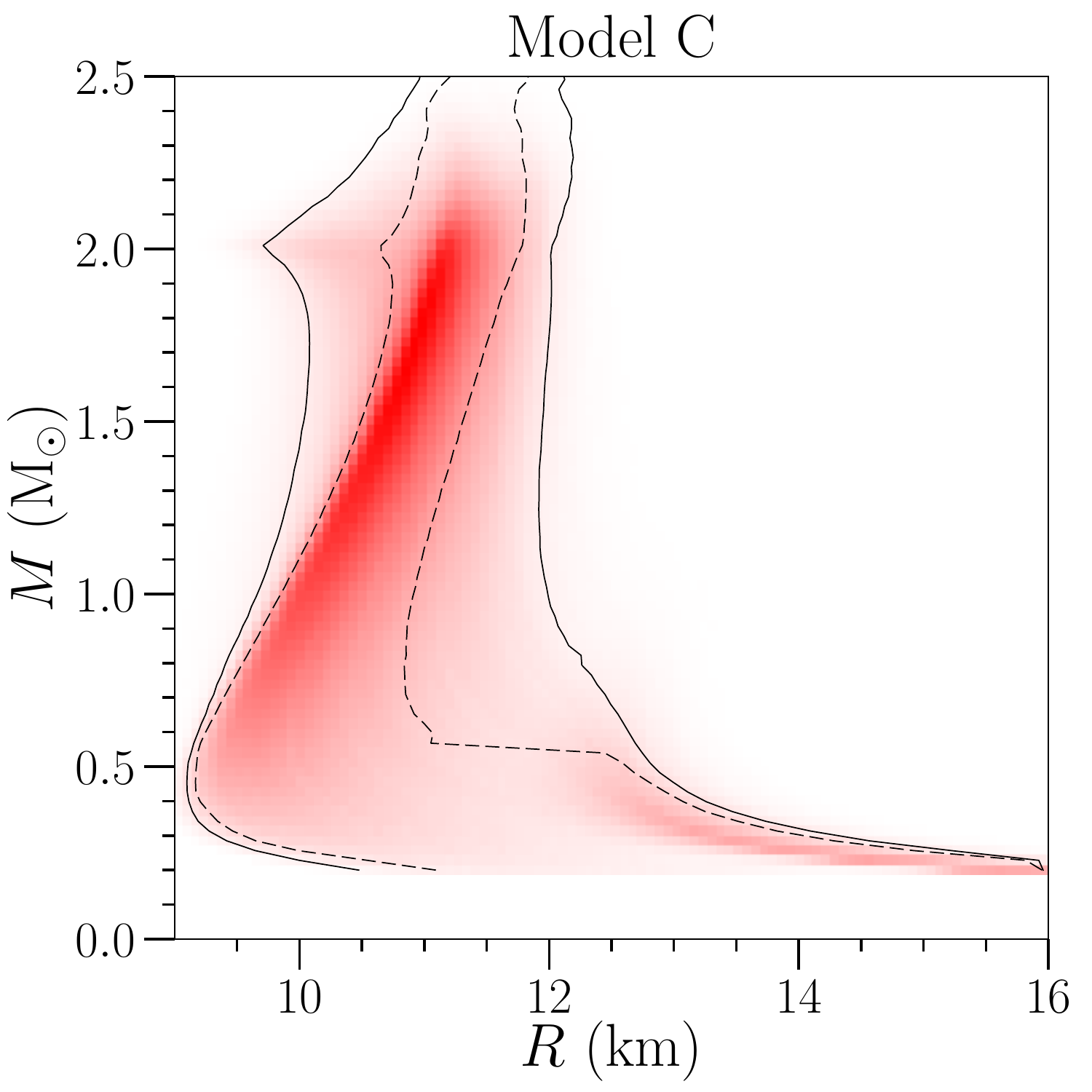}
  \includegraphics[width=1.6in]{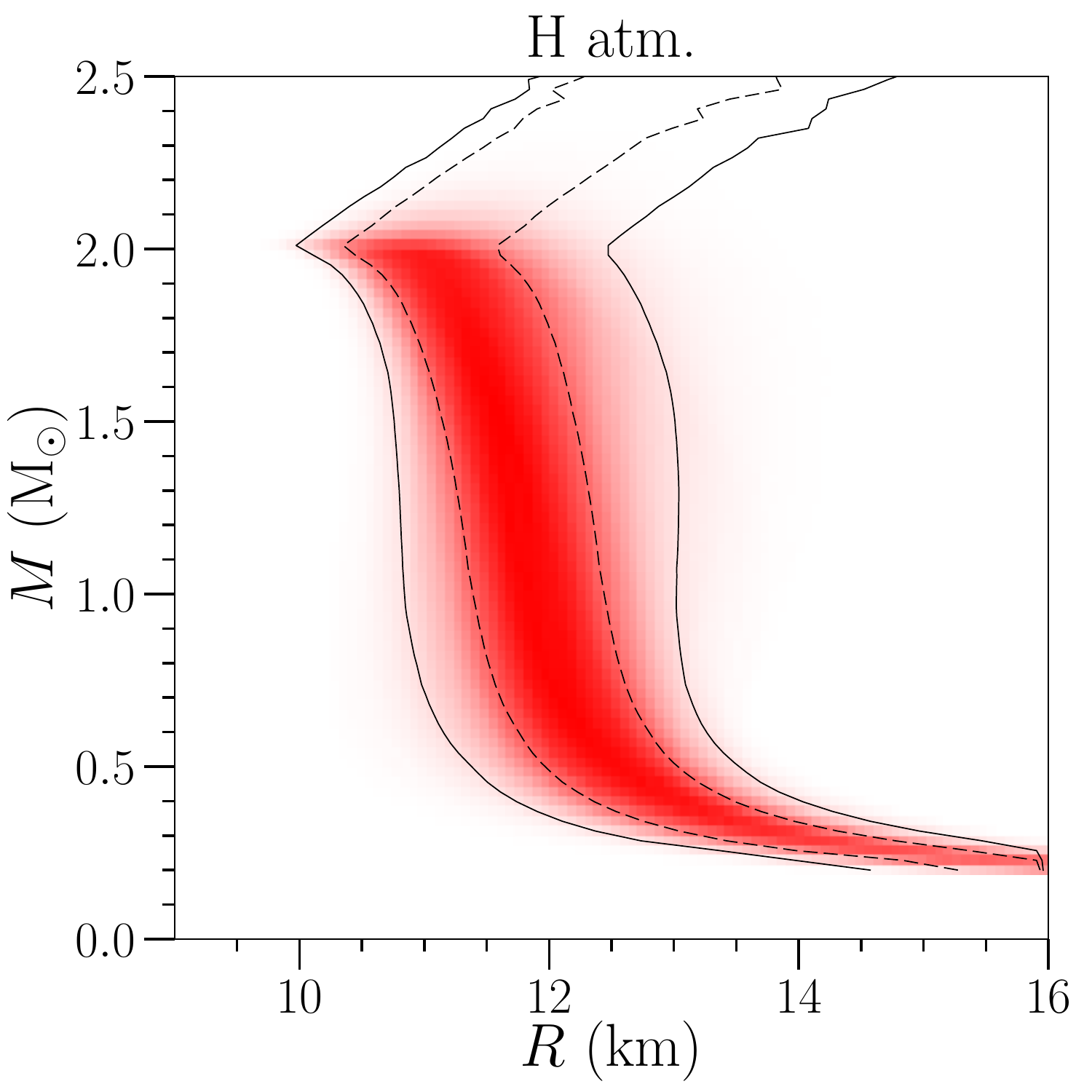}
  \includegraphics[width=1.6in]{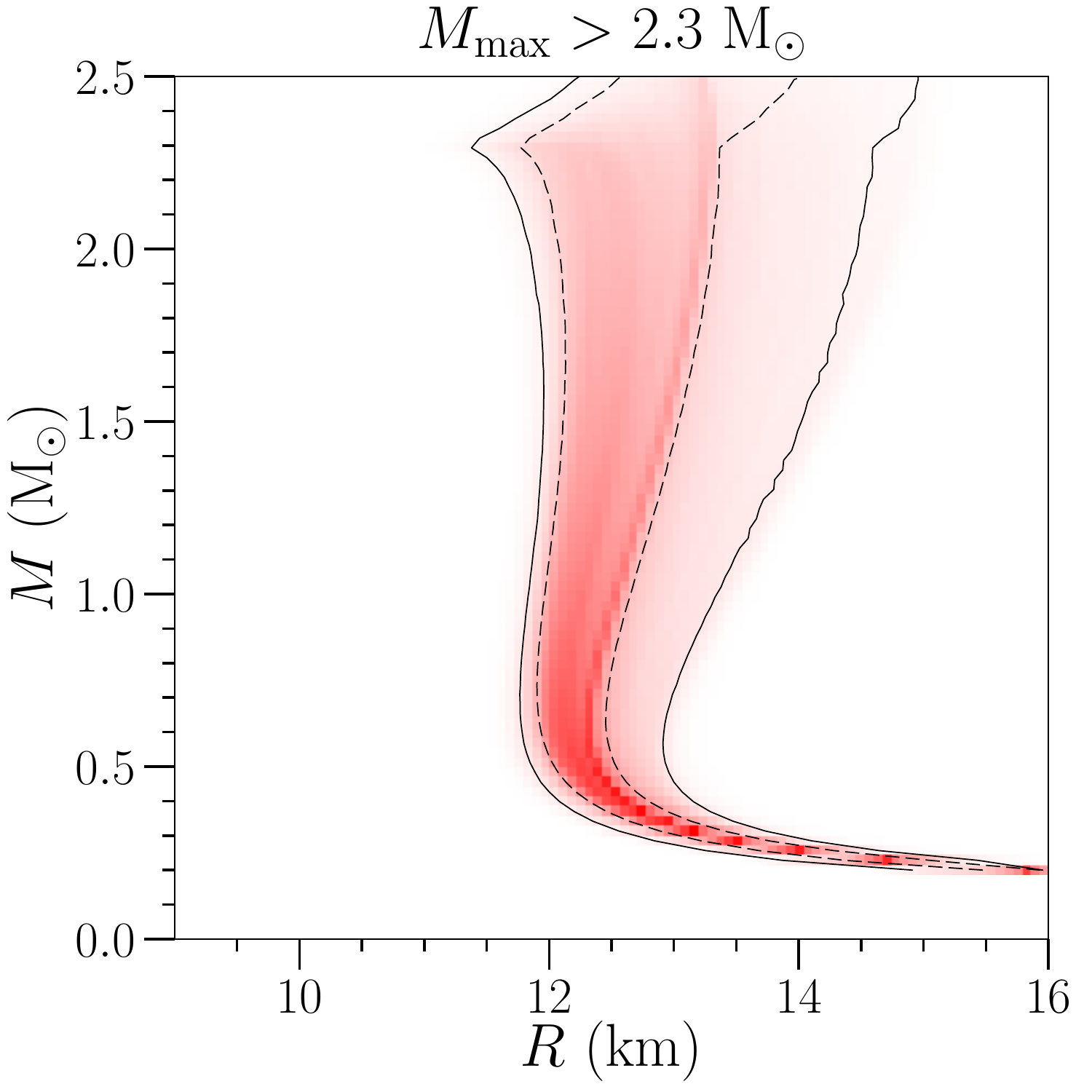}
  \caption{Probability distributions for radii as a function of
    mass for the baseline data set and baseline model (upper left
    panel), for the baseline data set with Model C (upper right),
    the baseline model and assuming H atmospheres (lower left),
    and the baseline model and baseline data set requiring
    $M_{\mathrm{max}}>2.3~\mathrm{M}_{\odot}$ (lower right).}
  \label{fig:MR_base}
\end{figure}

The upper left panel of
Fig.~\ref{fig:MR_base} shows an ensemble of one-dimensional radius
histograms for a fixed mass. The results imply that the radius for a
$M=1.4~\mathrm{M}_{\odot}$ neutron star is between 11.0 and 14.3 km
(to 95\% confidence; see Table~\ref{tab:radii}). The figure gets less
dark at higher mass because the area under a radius histogram at fixed
mass is normalized to the probability that the maximum mass is larger.
Note that the probabilities shift towards larger radii for masses
above 2 $\mathrm{M}_{\odot}$ because larger mass stars require a
larger maximum mass, which in turn requires a larger pressure at lower
densities and thus a larger radius. The probabilities for a helium
atmosphere for each source are given in Table~\ref{tab:atms} and
strongly prefer a He atmosphere for the neutron star in NGC 6397 and a
H atmosphere for the neutron star X7 in 47 Tuc. Note that the
probability of a He atmophere hovers around 1/3 for objects like the
neutron star in NGC 6304 and that in M30 since the probability
distributions for those two objects are too broad to
allow a strong constraint on the atmosphere composition (as shown in
Fig.~\ref{fig:demo_Rinfz} and Fig.~\ref{fig:data1}).

A significant decrease in the radius comes from using a
model of high-density matter which allows for strong phase
transitions, as found in \citet{Steiner13tn,Steiner15un}. This causes
likely radii for $M=1.4~\mathrm{M}_{\odot}$ neutron stars to drop by 1
to 2 km (see also upper right panel of Fig.~\ref{fig:MR_base}). The
probabilities for helium atmospheres drop significantly for some
objects, including the neutron star in NGC 6397, the neutron star in
M13, and X7 in 47 Tuc. Assuming all neutron stars must have a hydrogen
atmosphere (lower left panel of Fig.~\ref{fig:MR_base}) also decreases
the 95\% confidence limits in the radius by 0.2 (lower bound) and 1.3
km (upper bound). Requiring the neutron star maximum mass to lie above
2.3 $\mathrm{M}_{\odot}$ increases the lower limit for the radius by
0.7 km (lower right panel of Fig.~\ref{fig:MR_base}), similar to the
result found in \citet{Steiner13tn,Steiner16ns}.

\begin{figure}
  \includegraphics[width=1.6in]{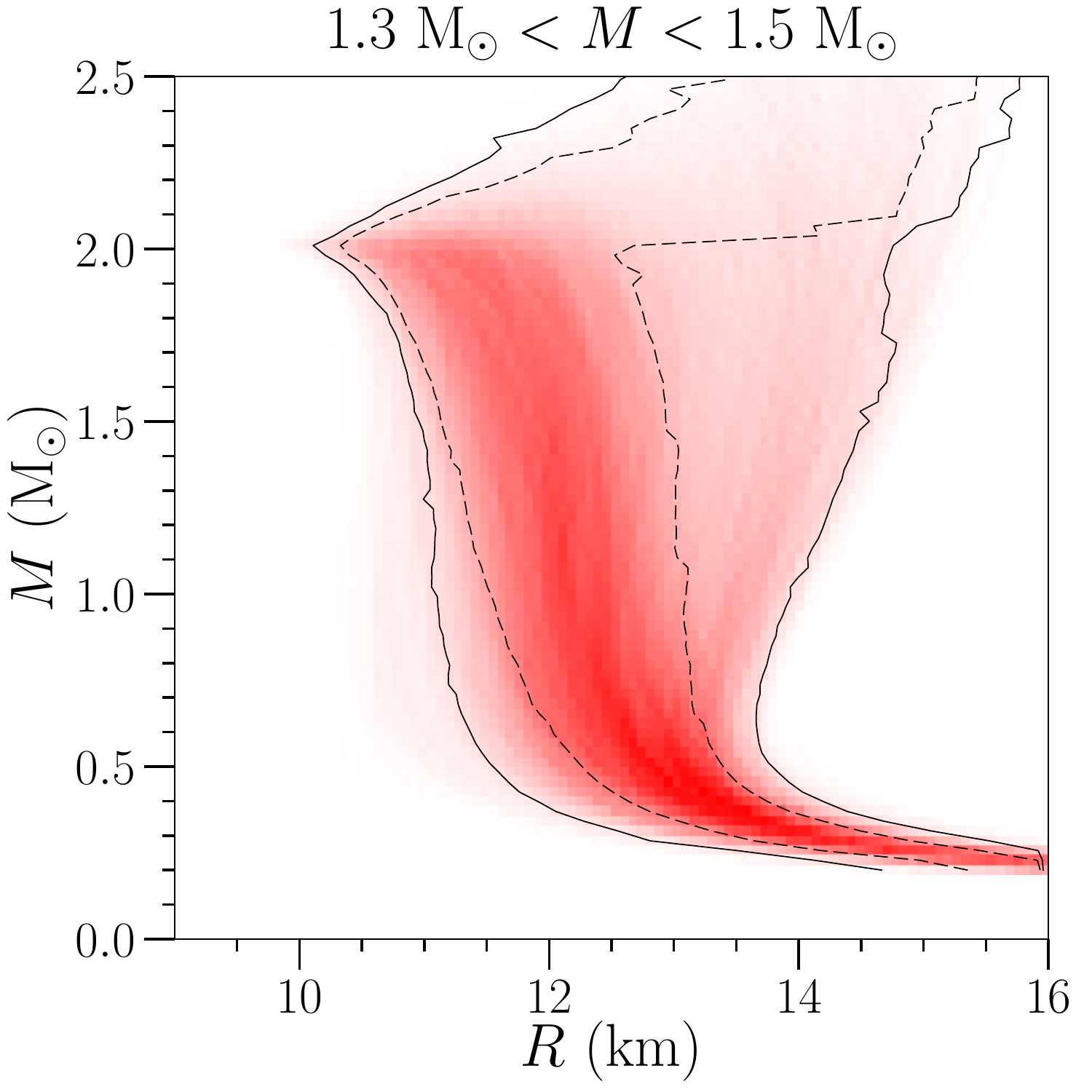}
  \includegraphics[width=1.6in]{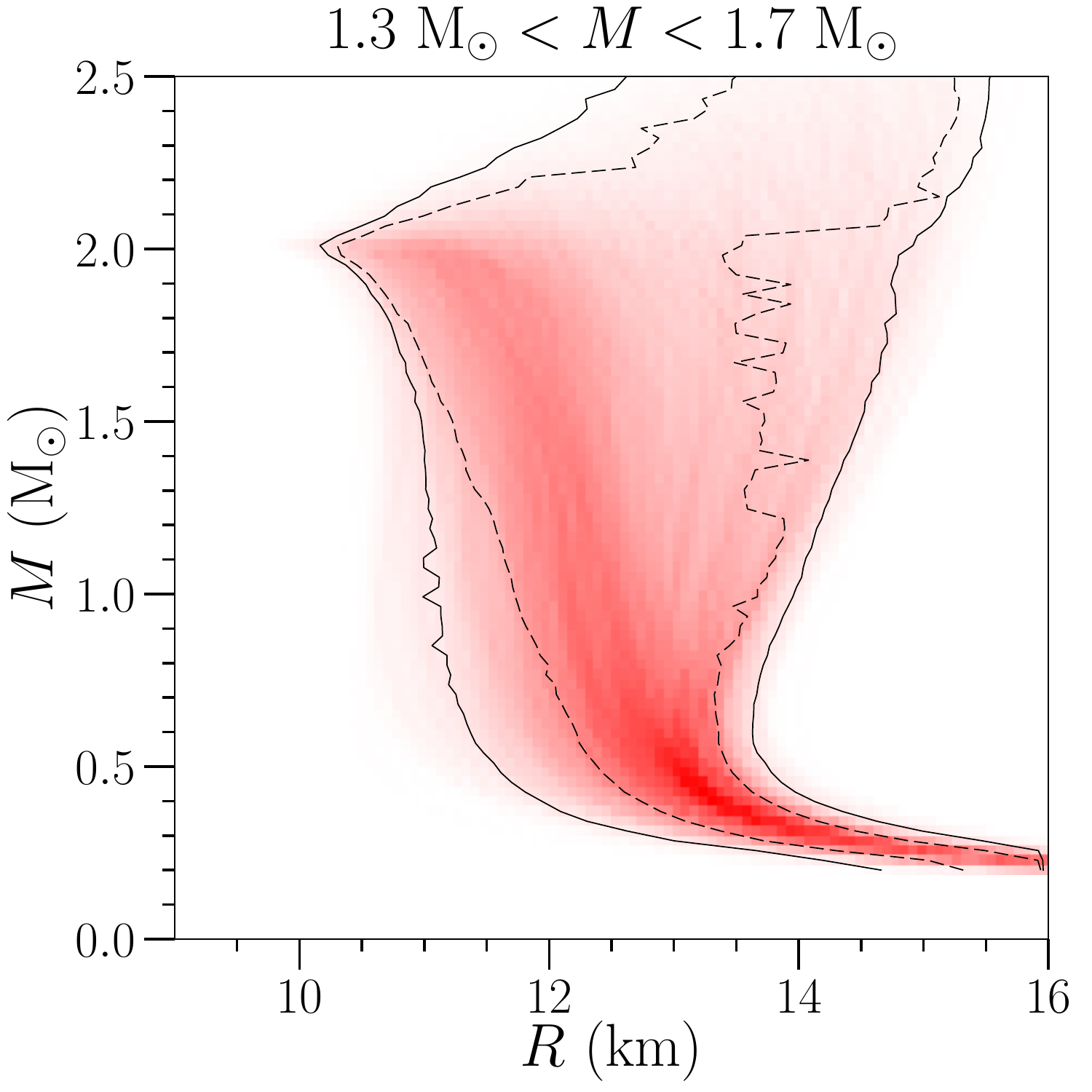}
  \includegraphics[width=1.6in]{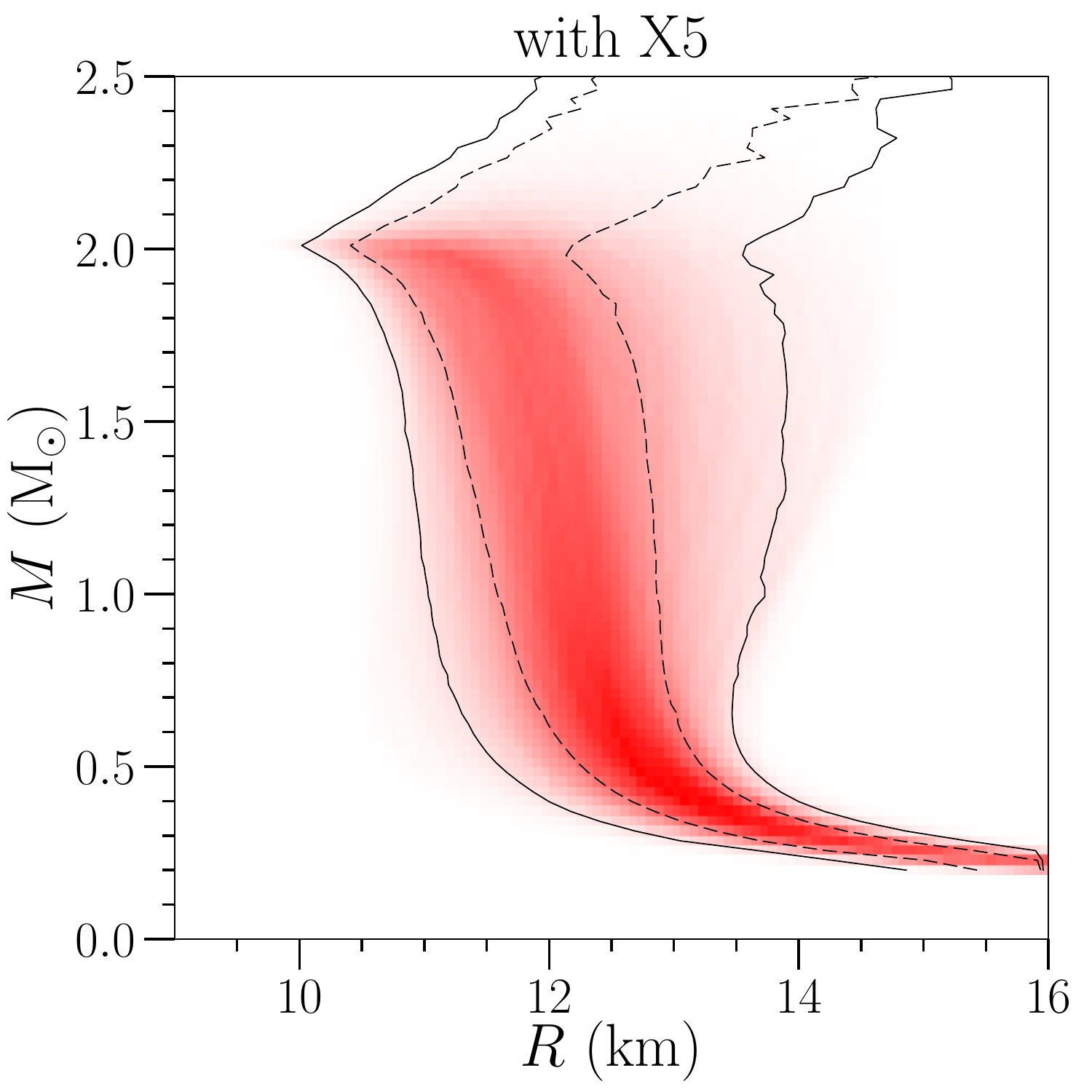}
  \includegraphics[width=1.6in]{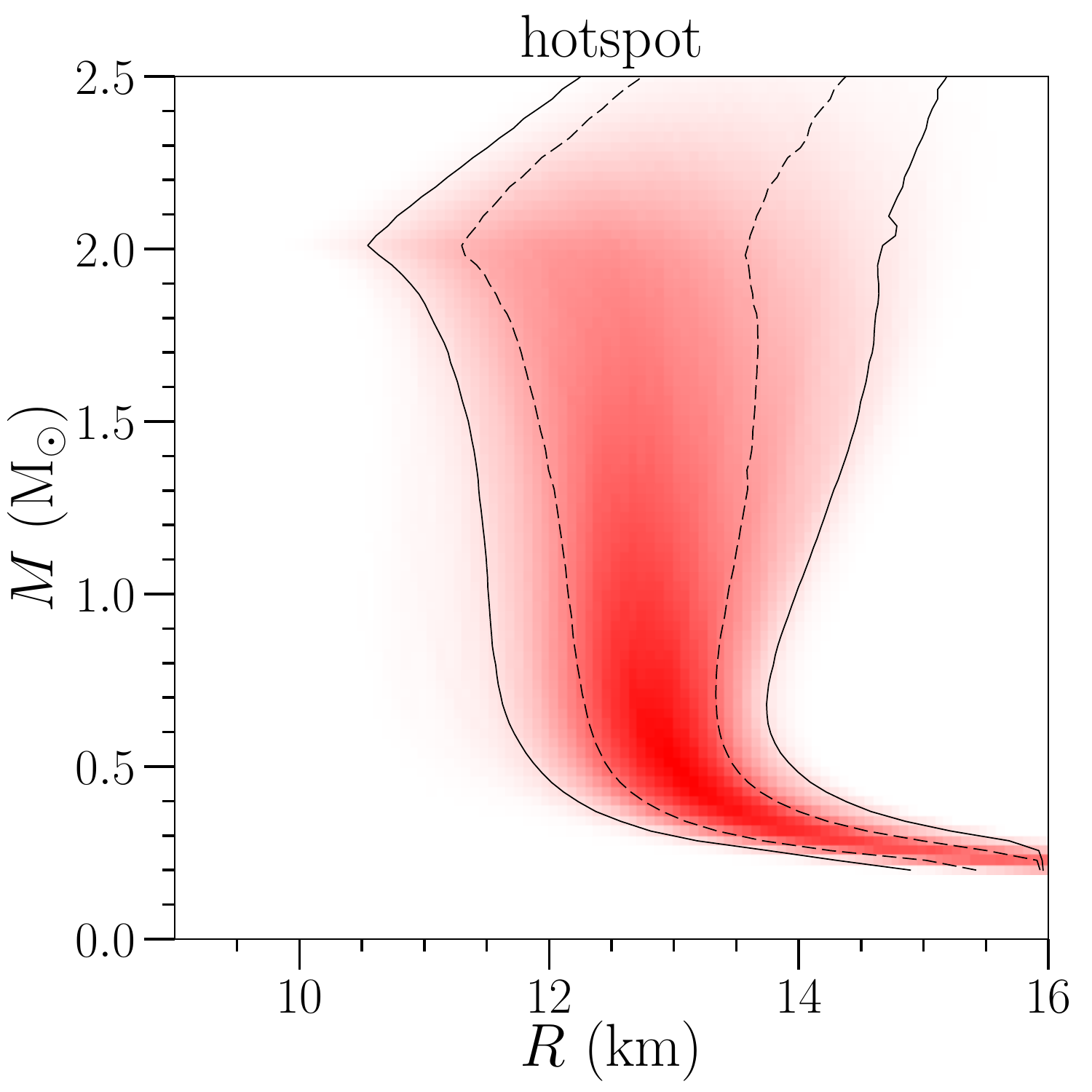}
  \caption{Probability distributions for radii as a function of mass
    for the baseline data set requiring all neutron stars to have
    masses $1.3~\mathrm{M}_{\odot} < M < 1.5~\mathrm{M}_{\odot}$
    (upper left panel), $1.3~\mathrm{M}_{\odot} < M <
    1.7~\mathrm{M}_{\odot}$ (upper right), with the baseline model and
    the baseline data set with the neutron star in X5 (lower left) and
    the baseline data set presuming a hotspot (lower right).}
  \label{fig:MR_base2}
\end{figure}

Constraining the mass of the neutron stars (to either 1.3$-$1.7
$\mathrm{M}_{\odot}$, or to 1.3$-$1.5 $\mathrm{M}_{\odot}$) also has
relatively little effect on the inferred radii as shown in the upper
left and upper right panels of Fig.~\ref{fig:MR_base2}. If we include
X5 in our data set (lower left panel of Fig.~\ref{fig:MR_base2}), then
radii above 13.9 km for a $M=1.4~\mathrm{M}_{\odot}$ neutron star are
strongly ruled out. The lower limit for the radius decreases, but only
slightly, as radii smaller than about 11 km require a strong phase
transition which are disfavored in polytropic models. In this case,
the preference for a helium atmosphere decreases in objects for which
the probabilities are not dominated by the prior choice (which tends
to be those stars which have posterior probabilities not near 33\%).
The model permitting hotspots significantly increases the 1$\sigma$
range for the radius (lower right panel of Fig.~\ref{fig:MR_base2}),
but has little effect on the 2$\sigma$ limits (Table~\ref{tab:radii}).
The effect of the hotspot on the posterior probability for the
atmosphere is most dramatic for the neutron star in NGC 6397 (see
table \ref{tab:atms}). In this case, the hotspot effectively increases
the inferred radius, and this makes the H atmosphere more probable,
thus decreasing the posterior probability for a He atmosphere by
almost a factor of two.

In the baseline model, we assume a 2/3 prior probability that each
neutron star has a hydrogen atmosphere. Increasing this prior to 90\%
decreases the posterior probability as seen in the last column of
table \ref{tab:atms}, and the effect of this prior choice on the
posterior probability is stronger than our other model choices. The
effect on the neutron star radius is more modest, as shown in the last
row of Table~\ref{tab:radii}.

\begin{table}
  \begin{tabular}{ccccc}
    & \multicolumn{4}{c}{$R(M=1.4~\mathrm{M}_{\odot})$ (km)} \\
    & \multicolumn{2}{c}{Lower limits} &
    \multicolumn{2}{c}{Upper limits} \\
    Model & 95\% & 68\% & 68\% & 95\% \\
    \hline
    baseline & 11.03 & 11.40 & 13.11 & 14.28 \\
    Model C & 9.995 & 10.40 & 11.23 & 11.93 \\
    H atm. & 10.78 & 11.20 & 12.26 & 13.03 \\
    $M_{\mathrm{max}}\geq 2.3$ & 11.95 & 12.09 & 12.98 & 13.94 \\
    $1.3<M<1.5$ & 11.02 & 11.22 & 13.04 & 14.44 \\
    $1.3<M<1.7$ & 11.01 & 11.30 & 13.68 & 14.41 \\
    with X5 & 10.88 & 11.32 & 12.78 & 13.87 \\
    hotspot & 11.04 & 11.97 & 13.62 & 14.40 \\
    90\% H & 10.85 & 11.28 & 12.82 & 14.03 \\
  \end{tabular}
  \caption{Two-sigma confidence limits for the radius of a
    $1.4~\mathrm{M}_{\odot}$ neutron star (in km) for the different
    combinations of data sets and model assumptions used in this
    work.}
  \label{tab:radii}
\end{table}

\begin{table}
  \begin{tabular}{crrrrr}
    Source & \multicolumn{5}{c}{Probability of He} \\
    & baseline & w/X5 & Model C & hotspot & 90\% H \\
    \hline
    NGC 6304 & 32 \% & 32 \% & 30 \% & 31 \% & 9.1\% \\
    NGC 6397 & 81 \% & 79 \% & 47 \% & 41 \% & 34 \% \\
    M13      & 28 \% & 26 \% & 19 \% & 26 \% & 7.0\% \\
    M28      & 52 \% & 48 \% & 31 \% & 33 \% & 16 \% \\
    M30      & 31 \% & 30 \% & 27 \% & 31 \% & 8.7\% \\
    X7       & 16 \% & 2.5\% & 3.4\% & 13 \% & 2.0\% \\
  \end{tabular}
  \caption{Probability of a Helium atmosphere for each neutron star
    depending on data set and model assumptions. The statistical
    uncertainties in these probabilities are about 3\%.}
  \label{tab:atms}
\end{table}

\begin{table}
  \begin{tabular}{lc}
    Data and model selection & Evidence \\
    \hline
    baseline & 2.4 \\
    H atm. & 0.68 \\
    $1.3<M<1.5$ & 0.88 \\
    $1.3<M<1.7$ & 1.17 \\
    $M_{\mathrm{max}}\geq 2.3$ & 1.0 \\
    Model C & 20.2 \\
    hotspot & 1.4 \\
    90\% H & 1.8 \\
  \end{tabular}
  \caption{The evidence, as computed by the properly
    normalized integral under the posterior distribution,
    for the different combinations of data sets and
    model assumptions used in this work.}
  \label{tab:evi}
\end{table}

In order to compare models, we employ Bayes factors, defined as the
ratio of the evidence. The evidence is the integral, over the full
parameter space, of the posterior distribution. The normalization of
the evidence here requires some care. When the two models which are
being compared have the same dimensionality and their parameters have
the same units, the Bayes factor needs no extra normalization.
However, this is not the case here as Model C and our baseline model
have fundamentally different parameters, so the evidence is computed
by rescaling the distinct model parameters which have dimensionful
units so that they are in the range $[0,1]$. The neutron star mass
parameters need no rescaling; it is important that they are not
rescaled in order to properly evaluate the Bayes factor for the models
which constrain the neutron star mass more than the baseline model.
Bayes factors between 1/3 and 3 are generally regarded as relatively
weak, and in this case no definitive statement can be made about the
two models. 

Requiring all neutron stars to have a H atmosphere is ruled out with a
Bayes factor of 1/3, but again as described above this result will
strongly depend on the prior probability of a H atmosphere.
Constraining the neutron star mass to between 1.3 and 1.7
$\mathrm{M}_{\odot}$ drops the evidence by a factor of 2 and further
constraining it to between 1.3 and 1.5 $\mathrm{M}_{\odot}$ drops the
evidence by an additional 30 percent. Assuming the presence of
hotspots drops the evidence by a factor of two. In agreement with
previous work, we find that the choice of EOS model has a strong
impact (see e.g. \citet{Steiner15un}). We find that the Model C is
strongly preferred over the baseline result (a Bayes factor of 8.4).
One way of diagnosing which object contributes most strongly to this
improved fit is by looking at the ratio of the average posterior
probability for each object between Model C and the baseline result.
These ratios are 1.03, 2.24, 1.03, 1.39, 0.94, 1.29, and 1.86 for the
neutron stars in NGC 6304, NGC 6397, M13, M28, M30, $\omega$ Cen, and
for X7 in 47 Tuc, respectively. (The product of these ratios is not
exactly equal to 8.4 because of correlations between the weight from
each neutron star). Thus the neutron star in NGC 6397 most strongly
pushes the results towards smaller radii, followed by X7 and then by
the neutron star in M28. There is no standard approach to computing
the Bayes factor when the data sets are different, so we cannot
evaluate whether or not including X5 is more or less consistent with
our model assumptions.

\begin{figure}
  \includegraphics[width=1.6in]{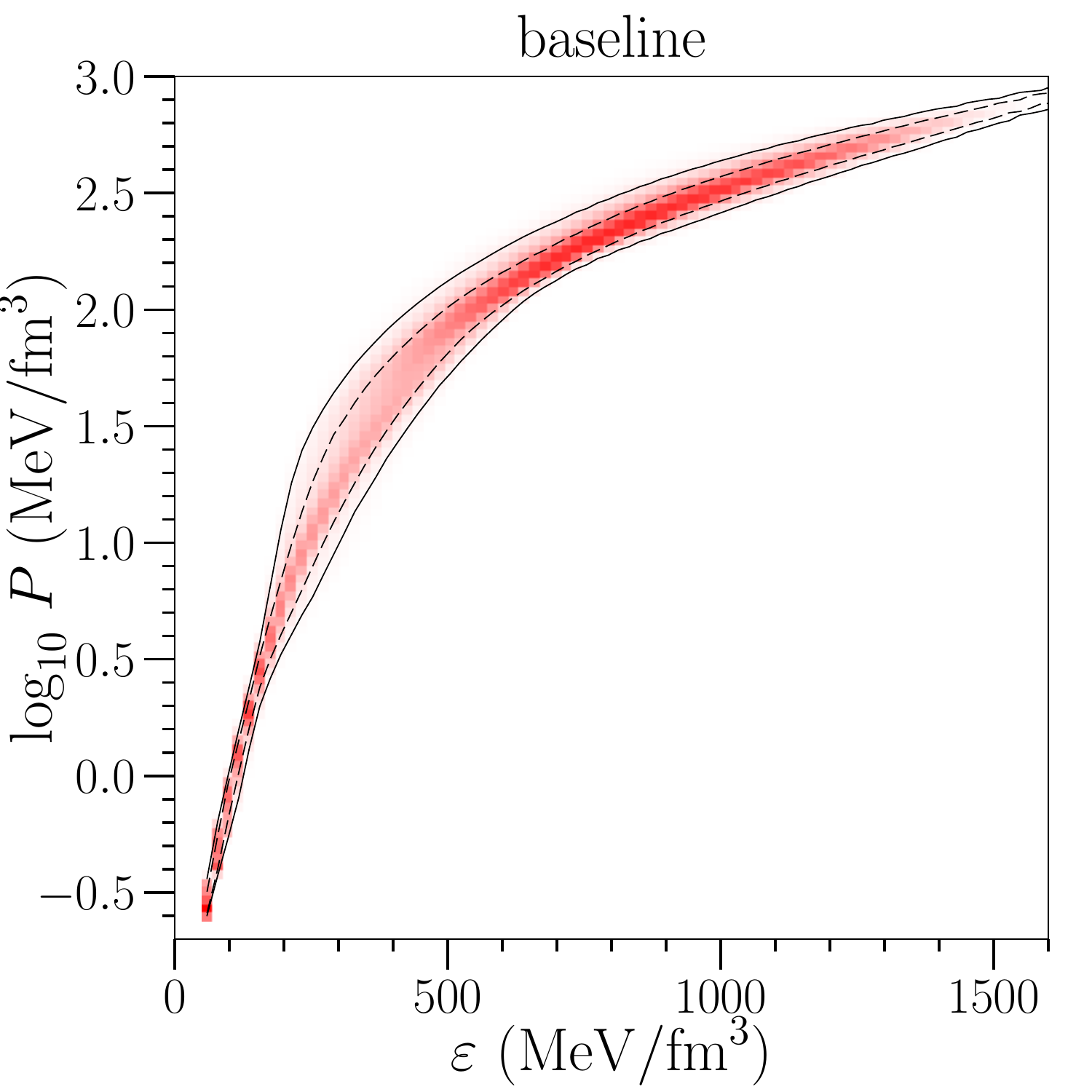}
  \includegraphics[width=1.6in]{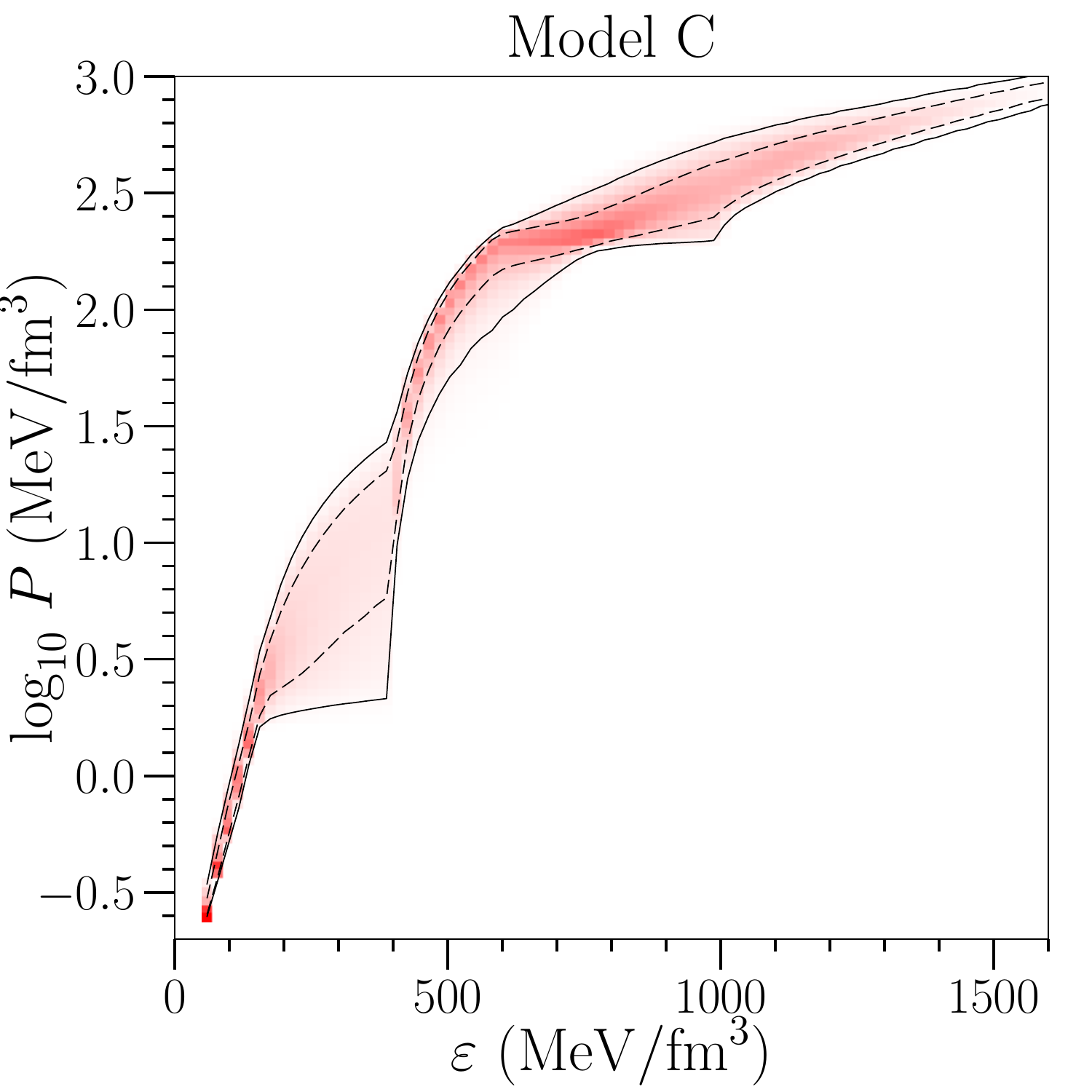}
  \includegraphics[width=1.6in]{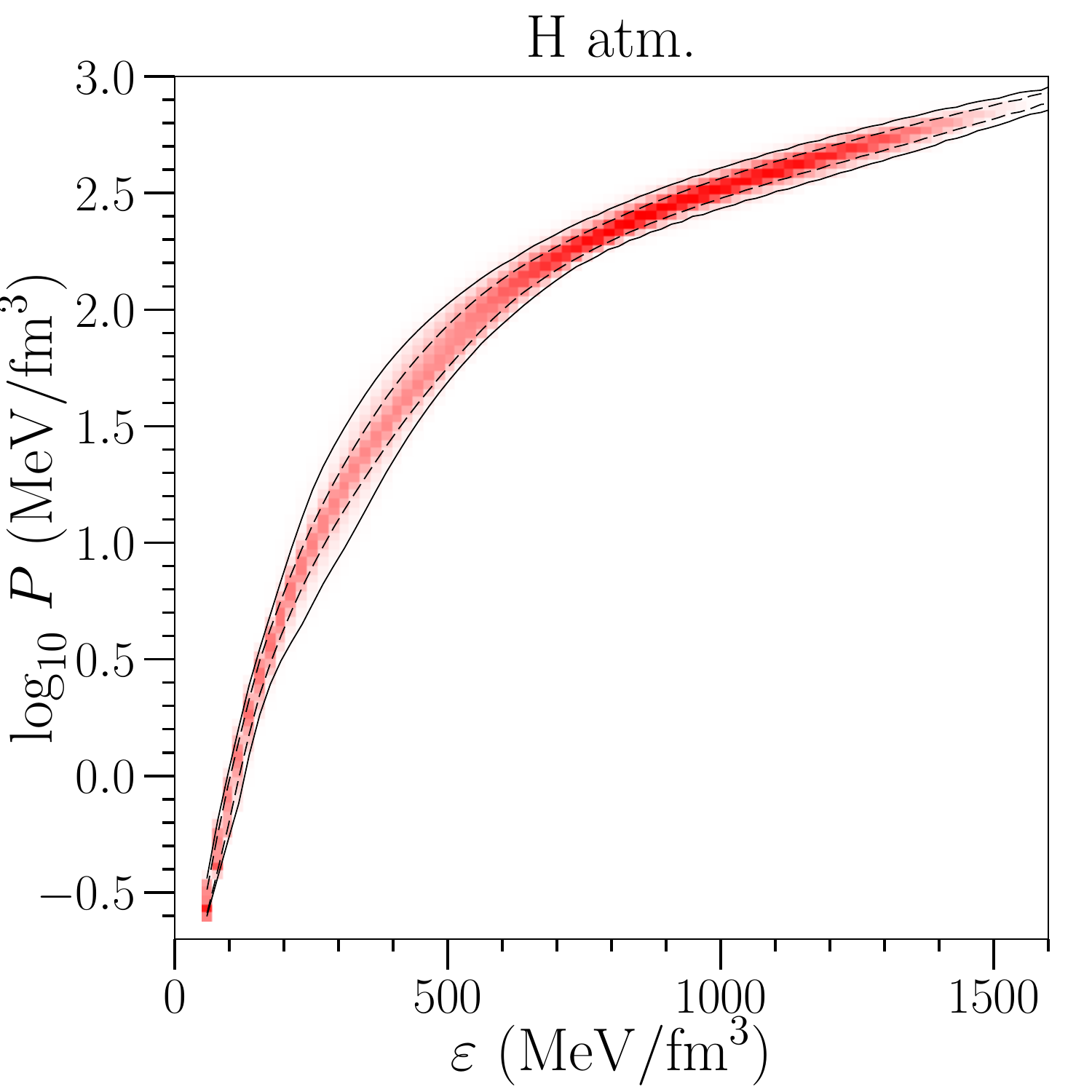}
  \includegraphics[width=1.6in]{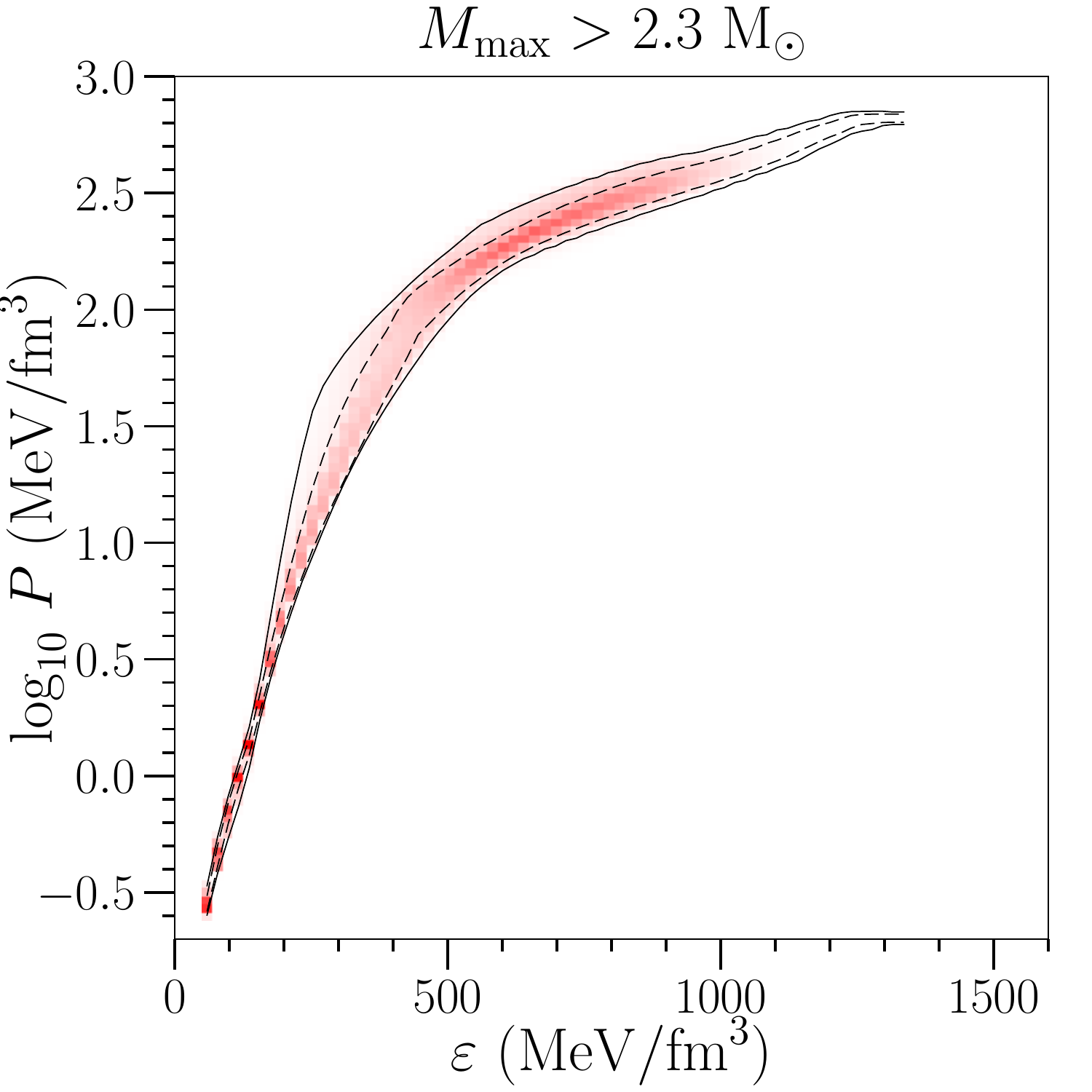}
  \caption{A set of posterior distributions for the pressure at
    fixed energy density over a range of energy densities for
    some of the models used in this work. The upper left panel shows
    results for our baseline model, the upper right shows the
    results for Model C, the lower left assumes
    H atmospheres for all sources and the lower right assumes
    a maximum mass larger than 2.3 $\mathrm{M}_{\odot}$.}
  \label{fig:eos}
\end{figure}

The posteriors for the relation between energy density and pressure
are presented in Fig.~\ref{fig:eos}. This figure shows a set of
pressure histograms at fixed energy density, normalized to the
probability that the central energy density in the maximum mass star
is larger than the specified energy density. Thus the plot becomes
less dark towards higher energy densities because they are often
higher than the maximum. The upper left panel shows our baseline
results. The plot is darker near 800 $\mathrm{MeV}/\mathrm{fm}^3$
because that energy density is more strongly constrained than near the
$\varepsilon=400~\mathrm{MeV}/\mathrm{fm}^3$. The strongest deviation
from the baseline model is for Model C, where the EOS parameterization
allows for large regions where the pressure is flat. These phase
transitions occur at rather low densities to match the stars which
have smaller inferred values of $R_{\mathrm{\infty}}$ and the pressure
increases strongly above the phase transition in order to ensure that
the maximum mass is above $2~\mathrm{M}_{\odot}$. The lower left panel
assumes H atmospheres, leading to a smaller pressure in the range of
energy densities from 300 to $500~\mathrm{MeV}/\mathrm{fm}^{3}$.
Increasing the maximum mass, shown in the lower right panel, has the
opposite effect. In the lower right panel, the central energy density
of the maximum mass star decreases significantly, as one expects when
increasing the maximum mass. Note that, since the posterior mass for
the neutron stars in our data set is not often near
$2~\mathrm{M}_{\odot}$, our constraint on the EOS at high densities
will be more depending on the prior distribution and the
parameterization of the EOS. The lower and upper limits for the
pressure over a range of energy densities for all nine scenarios is
given in Table~\ref{tab:summ_eos}.

\begin{table}
  \begin{tabular}{cccccc}
    & \multicolumn{2}{c}{Lower limits} & &
    \multicolumn{2}{c}{Upper limits} \\
    Model & 95\% & 68\% & Most prob. & 68\% & 95\% \\
    \hline
    \multicolumn{6}{c}{$P(\varepsilon=300~\mathrm{MeV}/\mathrm{fm}^3)$
      ($\mathrm{MeV}/\mathrm{fm}^3$)} \\
    \hline
    baseline & 7.202 & 11.38 & 15.81 & 23.42 & 38.17 \\
    with X5 & 7.200 & 11.41 & 15.63 & 20.55 & 33.36 \\
    Model C & 1.830 & 2.039 & 2.851 & 8.621 & 14.47 \\
    H atm. & 7.233 & 11.36 & 14.10 & 17.33 & 23.49 \\
    $M_{\mathrm{max}}>2.3$ & 14.12 & 15.46 & 17.20 & 25.05 & 48.72 \\
    $1.3<M<1.5$ & 6.265 & 10.44 & 15.45 & 21.61 & 48.92 \\
    $1.3<M<1.7$ & 6.427 & 9.766 & 15.55 & 23.58 & 47.80 \\
    hotspot & 8.625 & 13.22 & 17.35 & 27.87 & 44.01 \\
    90\% H & 7.083 & 11.24 & 15.52 & 20.32 & 34.02 \\
    \hline
    \multicolumn{6}{c}{$P(\varepsilon=450~\mathrm{MeV}/\mathrm{fm}^3)$
      ($\mathrm{MeV}/\mathrm{fm}^3$)} \\
    \hline
    baseline & 32.64 & 43.97 & 49.78 & 75.26 & 98.22 \\
    with X5 & 32.88 & 41.75 & 57.95 & 69.45 & 86.33 \\
    Model C & 23.05 & 37.57 & 52.91 & 60.70 & 69.57 \\
    H atm. & 31.83 & 38.87 & 45.68 & 59.46 & 76.96 \\
    $M_{\mathrm{max}}>2.3$ & 59.92 & 66.99 & 73.30 & 113.6 & 132.0 \\
    $1.3<M<1.5$ & 30.12 & 39.68 & 48.93 & 74.57 & 117.5 \\
    $1.3<M<1.7$ & 30.37 & 38.95 & 47.83 & 75.55 & 118.1 \\
    hotspot & 40.14 & 49.62 & 71.14 & 86.97 & 114.6 \\
    90\% H & 33.03 & 41.12 & 48.65 & 70.25 & 87.11 \\
    \hline
    \multicolumn{6}{c}{$P(\varepsilon=600~\mathrm{MeV}/\mathrm{fm}^3)$
      ($\mathrm{MeV}/\mathrm{fm}^3$)} \\
    \hline
    baseline & 89.84 & 104.7 & 119.0 & 140.4 & 173.8 \\
    with X5 & 88.63 & 104.7 & 117.5 & 135.5 & 157.4 \\
    Model C & 71.67 & 137.0 & 197.1 & 208.1 & 217.1 \\
    H atm. & 87.19 & 97.74 & 113.4 & 131.1 & 150.3 \\
    $M_{\mathrm{max}}>2.3$ & 147.4 & 157.8 & 184.7 & 202.1 & 254.3 \\
    $1.3<M<1.5$ & 86.88 & 98.81 & 119.6 & 144.7 & 196.3 \\
    $1.3<M<1.7$ & 86.03 & 97.40 & 118.6 & 143.8 & 201.1 \\
    hotspot & 97.57 & 108.9 & 125.4 & 157.8 & 197.5 \\
    90\% H & 88.30 & 103.6 & 117.8 & 135.7 & 158.0 \\
    \hline
    \multicolumn{6}{c}{$P(\varepsilon=1000~\mathrm{MeV}/\mathrm{fm}^3)$
      ($\mathrm{MeV}/\mathrm{fm}^3$)} \\
    \hline
    baseline & 248.9 & 292.9 & 319.3 & 372.6 & 438.6 \\
    with X5 & 251.7 & 294.9 & 315.2 & 360.2 & 404.4 \\
    Model C & 206.9 & 240.8 & 310.0 & 402.7 & 505.6 \\
    H atm. & 273.9 & 303.6 & 326.8 & 362.2 & 399.9 \\
    $M_{\mathrm{max}}>2.3$ & 299.3 & 342.2 & 389.4 & 460.8 & 530.8 \\
    $1.3<M<1.5$ & 247.0 & 298.8 & 331.2 & 382.9 & 497.2 \\
    $1.3<M<1.7$ & 243.7 & 298.1 & 328.3 & 382.5 & 494.7 \\
    hotspot & 243.3 & 292.7 & 341.0 & 407.5 & 487.1 \\
    90\% H & 254.3 & 296.8 & 316.6 & 364.6 & 408.5 \\
  \end{tabular}
  \caption{Constraints on the pressure at four energy densities in the
    various model and data set choices used in this work. Note, in
    particular, the strong variation in the pressure at
    $\varepsilon=300~\mathrm{MeV}/\mathrm{fm}^3$ which is about twice
    the nuclear saturation density.}
  \label{tab:summ_eos}
\end{table}

\begin{figure}
  \includegraphics[width=1.6in]{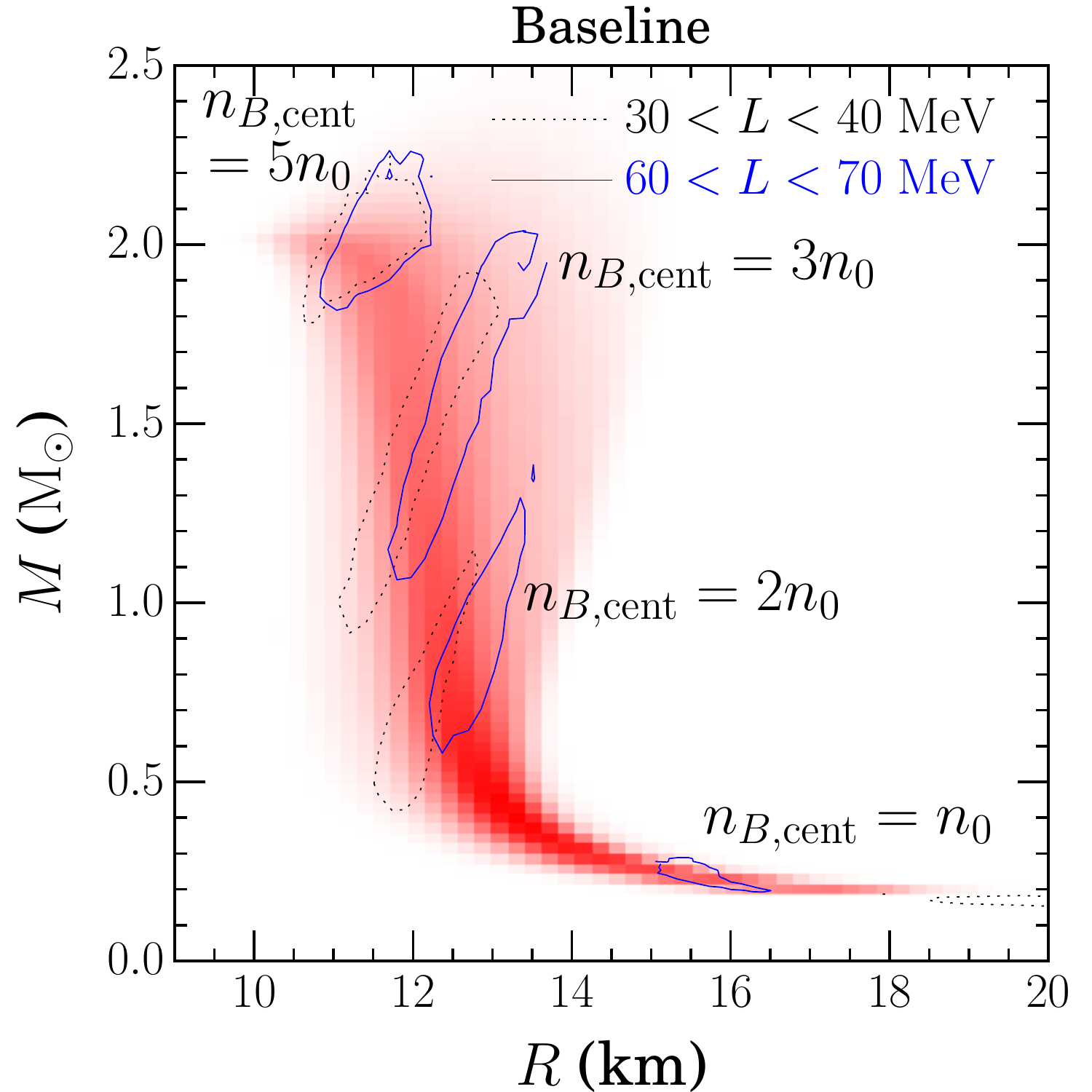}
  \includegraphics[width=1.6in]{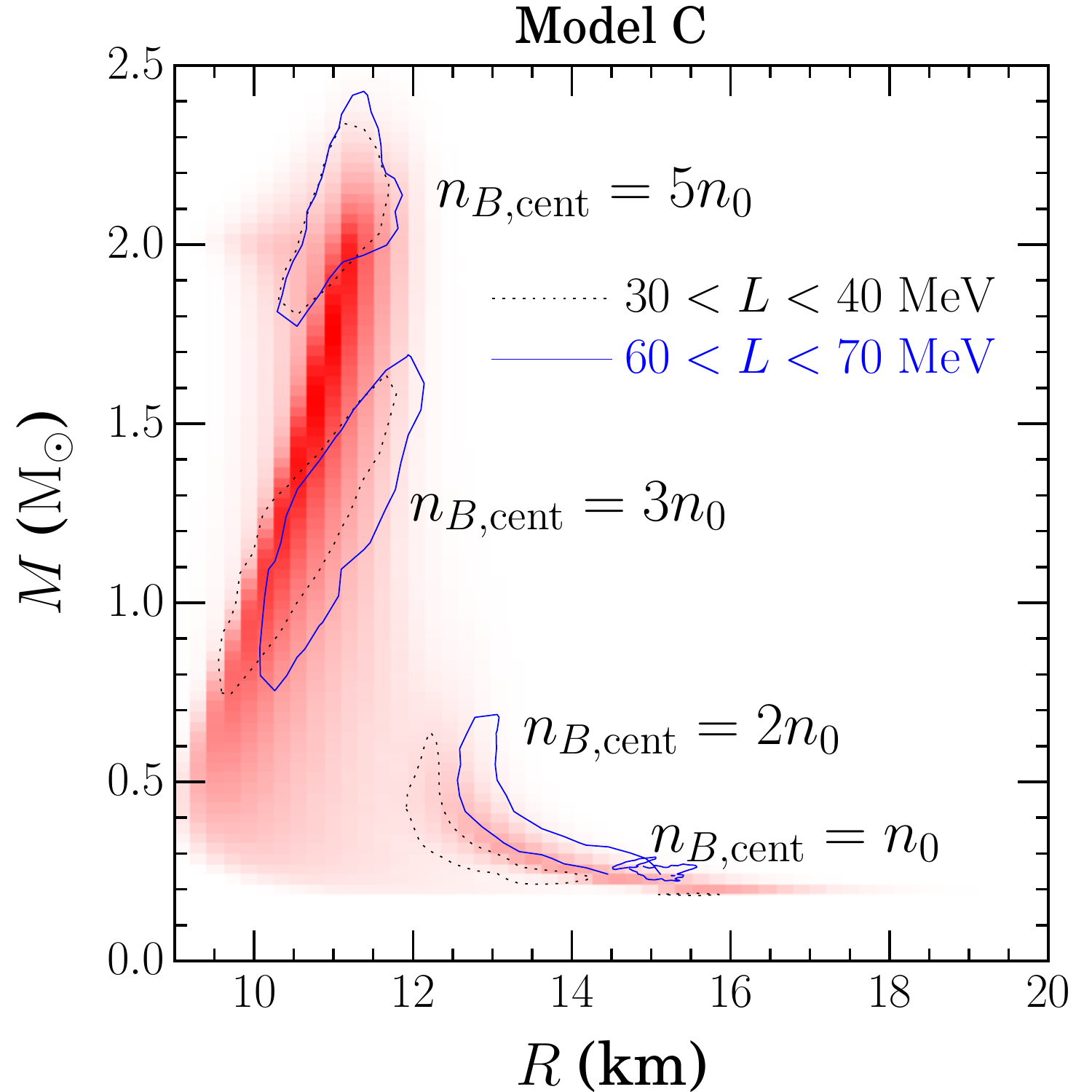}
  \caption{A plot showing the mass and radii of neutron stars which
    have a central baryon density equal to an integer multiple of the
    nuclear saturation density, $n_0 = 0.16~\mathrm{fm}^{-3}$. The
    results are separated into different assumptions about the slope
    of the symmetry energy, $L$. The left panel shows the baseline
    model and the right panel shows the results from Model C.}
  \label{fig:mrnb}
\end{figure}

The neutron star central baryon density is an important parameter for
understanding the extent to which nuclear physics can play a role in
neutron star structure. We find that the central baryon density of the
maximum mass star could be as low as 4 times the nuclear saturation
density, $n_0 = 0.16~\mathrm{fm}^{-3}$ or as large as $8 n_0$, in
agreement with \citet{Steiner15un}. The masses and radii for neutron
stars which have central baryon densities equal to integer multiples
of $n_0$ are shown in Fig.~\ref{fig:mrnb}. The left panel shows our
results for the baseline model and the right panel shows the results
for Model C. This figure is an updated version of Fig.~2 from
\citet{Gandolfi12mm} which properly quantifies the uncertainty in the
result by removing the unwarranted assumption that matter at high
densities can be described by the same Hamiltonian which is used at
the nuclear saturation density. The relationship with the density
derivative of the symmetry energy, $L$ is also shown. It is standard
to write the energy density $\varepsilon(n_B,\delta)$ as a function of
the neutron and proton number densities, $n_n$ and $n_p$, using the
definitions $n_B=n_n+n_P$ and $\delta=1-2 n_p/n_B$. Then the quantity
$L$ is defined by
\begin{equation}
L \equiv 3 n_B \frac{\partial}{\partial n_B}
\left( \frac{1}{2 n_B} \frac{\partial^2 \varepsilon}
     {\partial \delta^2}\right) \, .
\end{equation}
\citet{Lattimer01} showed that for a set of typical models $L$ is
correlated with the radius of a 1.4 solar mass neutron star. However,
this correlation is weak if one assumes a prior probability which
allows for strong phase transitions in the EOS~\citep{Steiner16ns} as
also shown in Fig.~\ref{fig:mrnb}. The lower and upper limits for the
value of $L$ and for the radii of higher mass stars for all nine
scenarios is given in Table~\ref{tab:summ_LR}.

\begin{table}
  \begin{tabular}{cccccc}
    & \multicolumn{2}{c}{Lower limits} & &
    \multicolumn{2}{c}{Upper limits} \\
    Model & 95\% & 68\% & Most prob. & 68\% & 95\% \\
    \hline
    \multicolumn{6}{c}{$L$ (MeV)} \\
    \hline
    baseline & 34.00 & 41.66 & 51.40 & 57.68 & 66.39 \\
    with X5 & 34.33 & 41.99 & 49.43 & 57.70 & 65.61 \\
    Model C & 30.26 & 30.52 & 30.79 & 45.00 & 58.61 \\
    H atm. & 30.55 & 38.94 & 49.08 & 58.09 & 64.89 \\
    $M_{\mathrm{max}}>2.3$ & 30.21 & 30.31 & 30.72 & 33.66 & 40.23 \\
    $1.3<M<1.5$ & 30.59 & 40.56 & 49.79 & 59.87 & 69.41 \\
    $1.3<M<1.7$ & 30.97 & 40.88 & 50.97 & 58.87 & 69.41 \\
    hotspot & 34.47 & 40.91 & 48.19 & 61.62 & 69.55 \\
    90\% H & 32.50 & 41.77 & 51.85 & 57.59 & 64.48 \\
    \hline
    \multicolumn{6}{c}{$R(M=1.7~\mathrm{M})$ (km)} \\
    \hline
    baseline & 10.82 & 11.19 & 11.79 & 13.02 & 14.37 \\
    with X5 & 10.74 & 11.12 & 11.72 & 12.65 & 13.88 \\
    Model C & 10.08 & 10.66 & 10.91 & 11.48 & 11.98 \\
    H atm. & 10.67 & 10.99 & 11.39 & 12.07 & 12.89 \\
    $M_{\mathrm{max}}>2.3$ & 11.95 & 12.13 & 13.06 & 13.16 & 14.25 \\
    $1.3<M<1.5$ & 10.80 & 10.99 & 11.76 & 12.81 & 14.64 \\
    $1.3<M<1.7$ & 10.76 & 10.88 & 11.90 & 13.85 & 14.62 \\
    hotspot & 11.18 & 11.78 & 12.69 & 13.67 & 14.58 \\
    90\% H & 10.71 & 11.06 & 11.73 & 12.59 & 13.99 \\
    \hline
    \multicolumn{6}{c}{$R(M=2.0~\mathrm{M})$ (km)} \\
    \hline
    baseline & 10.10 & 10.50 & 11.30 & 12.65 & 14.24 \\
    with X5 & 10.01 & 10.41 & 11.07 & 12.15 & 13.59 \\
    Model C & 9.702 & 10.65 & 11.23 & 11.79 & 12.03 \\
    H atm. & 9.958 & 10.35 & 10.91 & 11.59 & 12.48 \\
    $M_{\mathrm{max}}>2.3$ & 11.84 & 12.07 & 13.21 & 13.30 & 14.50 \\
    $1.3<M<1.5$ & 10.08 & 10.32 & 11.16 & 12.57 & 14.81 \\
    $1.3<M<1.7$ & 10.09 & 10.20 & 11.08 & 12.72 & 14.85 \\
    hotspot & 10.54 & 11.28 & 12.45 & 13.58 & 14.67 \\
    90\% H & 9.996 & 10.40 & 11.00 & 12.12 & 13.62 \\
  \end{tabular}
  \caption{Constraints on the derivative of the symmetry energy and
    the radii of 1.7 and 2.0~$\mathrm{M}_{\odot}$ neutron stars for
    the nine scenarios explored in this work.}
  \label{tab:summ_LR}
\end{table}

The one- and two-sigma limits for the radius of the maximum mass
neutron star are given in Table~\ref{tab:summ_max}. The most extreme
radii are observed if the EOS at high densities is assumed to have
strong phase transitions, in which case the two-sigma limit goes below
9.4 km. Larger lower limits occur if the maximum mass is greater than
2.3~$\mathrm{M}_{\odot}$ or if the neutron stars have uneven
temperature distributions, otherwise the lower two-sigma limit is
typically around 9.7 km. Table~\ref{tab:summ_max} also presents the
limits for the central baryon (ranging from 0.61 to
1.3~$\mathrm{fm}^{-3}$) and energy density (ranging from 740 to
1700~$\mathrm{MeV}/\mathrm{fm}^3$). Increasing the maximum mass tends
to decrease these central densities significantly.

\begin{table}
  \begin{tabular}{cccccc}
    & \multicolumn{2}{c}{Lower limits} & &
    \multicolumn{2}{c}{Upper limits} \\
    Model & 95\% & 68\% & Most prob. & 68\% & 95\% \\
    \hline
    \multicolumn{6}{c}{$R_{\mathrm{max}}$ (km)} \\
    \hline
    baseline & 9.743 & 10.01 & 10.39 & 11.64 & 13.00 \\
    with X5 & 9.721 & 9.959 & 10.35 & 11.22 & 12.58 \\
    Model C & 9.353 & 9.740 & 10.28 & 11.24 & 11.79 \\
    H atm. & 9.693 & 9.938 & 10.30 & 10.87 & 11.65 \\
    $M_{\mathrm{max}}>2.3$ & 11.04 & 11.44 & 11.83 & 12.77 & 13.79 \\
    $1.3<M<1.5$ & 9.696 & 9.891 & 10.37 & 11.51 & 13.72 \\
    $1.3<M<1.7$ & 9.679 & 9.839 & 10.33 & 11.59 & 13.76 \\
    hotspot & 9.974 & 10.43 & 11.23 & 12.31 & 13.52 \\
    90\% H & 9.720 & 9.956 & 10.38 & 11.21 & 12.58 \\
    \hline
    \multicolumn{6}{c}{$\varepsilon_{\mathrm{max}}$
      ($\mathrm{MeV}/\mathrm{fm}^3$)} \\
    \hline
    baseline & 858.9 & 1153 & 1453 & 1565 & 1631 \\
    with X5 & 933.2 & 1210 & 1504 & 1574 & 1642 \\
    Model C & 977.8 & 984.9 & 1000 & 1554 & 1704 \\
    H atm. & 1110 & 1305 & 1508 & 1586 & 1670 \\
    $M_{\mathrm{max}}>2.3$ & 743.4 & 873.7 & 935.2 & 1129 & 1232 \\
    $1.3<M<1.5$ & 751.3 & 1141 & 1439 & 1586 & 1638 \\
    $1.3<M<1.7$ & 752.9 & 1135 & 1509 & 1622 & 1641 \\
    hotspot & 779.1 & 959.1 & 1214 & 1405 & 1562 \\
    90\% H & 949.0 & 1212 & 1500 & 1574 & 1644 \\
    \hline
    \multicolumn{6}{c}{$n_{B,\mathrm{max}}$ ($\mathrm{fm}^{-3}$)} \\
    \hline
    baseline & 0.7079 & 0.9224 & 1.075 & 1.198 & 1.246 \\
    with X5 & 0.7956 & 0.9795 & 1.169 & 1.207 & 1.253 \\
    Model C & 0.8341 & 0.8453 & 0.8687 & 1.203 & 1.300 \\
    H atm. & 0.9120 & 1.044 & 1.165 & 1.221 & 1.268 \\
    $M_{\mathrm{max}}>2.3$ & 0.6140 & 0.6983 & 0.8518 & 0.8986 & 0.9560 \\
    $1.3<M<1.5$ & 0.6296 & 0.9243 & 1.137 & 1.230 & 1.254 \\
    $1.3<M<1.7$ & 0.6168 & 0.9252 & 1.162 & 1.239 & 1.258 \\
    hotspot & 0.6509 & 0.7834 & 0.9674 & 1.096 & 1.198 \\
    90\% H & 0.7977 & 0.9841 & 1.162 & 1.212 & 1.259 \\
  \end{tabular}
  \caption{The constraints on the radius, central energy density,
    and central baryon density of the maximum mass star for the
    nine scenarios explored in this work.}
  \label{tab:summ_max}
\end{table}

\section{Discussion}

\citet{Steiner13tn} found radii between 10.4 and 12.9 km (to 95\%
confidence) for a 1.4 $\mathrm{M}_{\odot}$ neutron star, and our
results allow for a larger range of radii. This is principally
because, except for the neutron star in $\omega$ Cen and X5 in 47 Tuc,
we have relaxed the assumption that QLMXB atmospheres are composed of
hydrogen. Once we assume QLMXBs have hydrogen atmospheres, we
reproduce the previous result. Our model with the largest evidence,
Model C, suggests that the radius of a 1.4 $\mathrm{M}_{\odot}$
neutron star is less than 12 km to 95\% confidence. One could conceive
of many possible combinations among the model assumptions which we
have explored. For example, if the maximum mass were larger than 2
$\mathrm{M}_{\odot}$, and additionally QLMXBs all had uneven
temperature distributions, then their radii could be larger than 14
km, especially if strong phase transitions were ruled out by
theoretical work on the nucleon-nucleon interaction. Alternatively, we
would find even smaller radii than 12 km if we assumed that He
atmospheres are unlikely and the data from X5 was confirmed (a
scenario similar to that in explored in~\citet{Ozel16}).

Other works which suggest stronger constraints always employ
assumptions which we have relaxed. \citet{Hebeler13eo} and
\citet{Steiner16ns} have shown that neutron star radii are between 11
and 13 km, assuming that chiral effective theory approaches to neutron
matter can be employed above the nuclear saturation density. However,
it is difficult to fully quantify the uncertainties in chiral
effective theory above the saturation density. The same difficulty is
found in the quantum Monte Carlo model we have used above, but we do
not employ it at high densities. While \citet{Lattimer14} ruled out
radii larger than 13 km from a similar set of neutron stars, our
updated data set and more complete consideration of distance
uncertainties weakens the case for smaller radii, as can be seen by
comparing our Fig.~\ref{fig:data1} with Figure 5 in \citet{Lattimer14}
(particularly the change in the constraints for the neutron star in
NGC 6304). \citet{Guillot13} and \citet{Ozel16} have found smaller
radii by assuming that (all or most) QLMXB atmospheres must be
composed of H. \citet{Ozel16} also use data from photospheric
expansion X-ray bursts, but their small radii were strongly driven by
the QLMXB data.

Further progress in our understanding of neutron star structure will
come from more data which constrains neutron star masses and radii,
including additional QLMXB spectra, constraints from NICER and LIGO,
and from future missions such as ATHENA, Lynx, and/or STROBE-X. This
work has shown that one can approximately quantify the effect that
hotspots, atmosphere composition, assumptions about the mass
distribution and the maximum mass, and assumptions about the presence
of strong phase transitions have on neutron star radii and the
equation of state. \\

\noindent {\bf Acknowledgements}\\

\noindent AWS and SH are supported by grant NSF PHY 1554876. This work
was supported by U.S. DOE Office of Nuclear Physics. This project used
computational resources from the University of Tennessee and Oak Ridge
National Laboratory's Joint Institute for Computational Sciences. COH
is supported by an NSERC Discovery Grant and an NSERC Discovery
Accelerator Supplement. WCGH acknowledges support from the Science and
Technology Facilities Council (STFC) in the United Kingdom.

\bibliographystyle{mnras}
\bibliography{Comb_MR}

\end{document}